\documentclass{ar-1col}
\usepackage{natbib,amssymb,rotating,longtable,subfig}

\jname{Xxxx. Xxx. Xxx. Xxx.}
\jvol{AA}
\jyear{YYYY}
\doi{10.1146/((please add article doi))}

\makeatletter \newcommand{\ion}[2]{#1$\;$\textsc{\rmfamily\@roman{#2}}\relax} \makeatother
\newcommand{\HII}{{\ion{H}{2}}}
\newcommand{\HeII}{{\ion{He}{2}}}
\newcommand{\HeI}{{\ion{He}{1}}}
\newcommand{\HI}{{\ion{H}{1}}}

\newcommand{\SiIIIf}{[{\ion{Si}{3}}]}
\newcommand{\SiIII}{{\ion{Si}{3}}]}
\newcommand{\NIII}{{\ion{N}{3}}]}
\newcommand{\NI}{[{\ion{N}{1}}]}
\newcommand{\NeIV}{[{\ion{Ne}{4}}]}
\newcommand{\NeV}{[{\ion{Ne}{5}}]}
\newcommand{\ArIV}{[{\ion{Ar}{4}}]}
\newcommand{\ClIII}{[{\ion{Cl}{3}}]}

\newcommand{\OII}{[{\ion{O}{2}}]}
\newcommand{\AlII}{[{\ion{Al}{2}}]}
\newcommand{\AlIIint}{{\ion{Al}{2}}]}

\newcommand{\CII}{[{\ion{C}{2}}]}
\newcommand{\OI}{[{\ion{O}{1}}]}
\newcommand{\OIV}{[{\ion{O}{4}}]}
\newcommand{\OV}{[{\ion{O}{5}}]}

\newcommand{\OIIIHb}{[{\ion{O}{3}}]/H$\beta$}
\def\ratioR23{([\ion{O}{2}]~$\lambda$3726 +[\ion{O}{3}]~$\lambda\lambda$4959,5007)/H$\beta$}
\def\R23{${\rm R}_{23}$}

\newcommand{\Msun}{${\rm M}_{\odot}$}

\newcommand{\NII}{[{\ion{N}{2}}]}
\newcommand{\NeII}{[{\ion{Ne}{2}}]}
\newcommand{\OIIIOII}{[\ion{O}{3}]/[\ion{O}{2}]}

\newcommand{\CIIIf}{[\ion{C}{3}]}
\newcommand{\CIII}{\ion{C}{3}]}

\newcommand{\NIIOII}{[\ion{N}{2}]/[\ion{O}{2}]}
\newcommand{\NIIIOIII}{[\ion{N}{3}]/[\ion{O}{3}]}

\newcommand{\OH}{$\log({\rm O/H})+12$}
\newcommand{\NIISII}{[\ion{N}{2}]/[\ion{S}{2}]}

\newcommand{\SIVSIII}{[\ion{S}{4}]/[\ion{S}{3}]}
\newcommand{\NIIHa}{[\ion{N}{2}]/H$\alpha$}
\newcommand{\SIIHa}{[\ion{S}{2}]/H$\alpha$}
\newcommand{\OIHa}{[\ion{O}{1}]/H$\alpha$}

\newcommand{\NeIIIOII}{[\ion{Ne}{3}]/[\ion{O}{2}]}
\newcommand{\SII}{[{\ion{S}{2}}]}
\newcommand{\SIII}{[{\ion{S}{3}}]}
\newcommand{\AlIII}{[{\ion{Al}{3}}]}
\newcommand{\SiII}{[{\ion{Si}{2}}]}
\newcommand{\ArIII}{[{\ion{Ar}{3}}]}

\newcommand{\SIV}{[{\ion{S}{4}}]}
\newcommand{\SV}{[{\ion{S}{5}}]}
\newcommand{\NIV}{[{\ion{N}{4}}]}
\newcommand{\CIV}{[{\ion{C}{4}}]}

\newcommand{\NV}{{\ion{N}{5}}}
\newcommand{\Hb}{{H$\beta$}}
\newcommand{\OIII}{[{\ion{O}{3}}]}
\newcommand{\NeIII}{[{\ion{Ne}{3}}]}
\newcommand{\Ha}{{H$\alpha$}}

\newcommand{\rion}[2]{{\ensuremath{\mbox{\rm #1$\,${\sc\expandafter{\romannumeral#2\relax}}}}}}
\newcommand{\Te}{${\rm T_{e}}$}

\begin{document}

\markboth{Kewley et al.}{Galaxy Evolution through Emission Lines}

\title{Understanding Galaxy Evolution through Emission Lines}

\author{Lisa J. Kewley$^1$$^2$, David C. Nicholls$^1$$^2$, \& Ralph S. Sutherland$^1$
\affil{$^1$Research School for Astronomy \& Astrophysics, Australian National University, Canberra, Australia, 2611; email: lisa.kewley@anu.edu.au, 
$^2$ARC Centre of Excellence for All Sky Astrophysics in 3 Dimensions (ASTRO 3D)}}

\begin{abstract}
We review the use of emission-lines for understanding galaxy evolution, focusing on excitation source, metallicity, ionization parameter, ISM pressure and electron density.  We show that the UV, optical and infrared contain complementary diagnostics that can probe the conditions within different nebular ionization zones.  In anticipation of upcoming telescope facilities, we provide new self-consistent emission-line diagnostic calibrations for complete spectral coverage from the UV to the infrared.  These diagnostics can be used in concert to understand how fundamental galaxy properties have changed across cosmic time.  We describe new 2D and 3D emission-line diagnostics to separate the contributions from star formation, AGN and shocks using integral field spectroscopy.   We discuss the physics, benefits, and caveats of emission-line diagnostics, including the effect of theoretical model uncertainties, diffuse ionized gas, and sample selection bias.  Accounting for complex density gradients and temperature profiles is critical for reliably estimating the fundamental properties of \HII\ regions and galaxies. Diffuse ionized gas can raise metallicity estimates, flatten metallicity gradients, and introduce scatter in ionization parameter measurements.  We summarize with a discussion of the challenges and major opportunities for emission-line diagnostics in the coming years.  
\end{abstract}


\maketitle
\tableofcontents

\section{Introduction}\label{Introduction}

A galaxy spectrum contains a wealth of information on the fundamental physical processes occurring within the galaxy.  A single optical spectrum alone can tell us the chemical abundance, the amount of dust, the electron density, the age of the stellar population, the pressure of the interstellar medium, and the rate of star formation.  The same spectrum can reveal whether there is an actively feeding supermassive black hole in the centre of the galaxy, or whether there are shocks from massive stellar winds, or gas collisions due to mergers, jets, or other transformative processes.  If an actively feeding supermassive black hole resides in the galaxy, the spectrum can tell us about the accretion rate, the shape of the AGN radiation field, and the strength of that radiation field.  If the spectrum reveals shocks, the emission-lines can be used to gauge the shock velocity, the density of the gas in the shock, and the mechanical energy of the shock.  In this way, galaxy spectra have allowed us to understand the dominant power source of galaxies, the star formation history of galaxies, the chemical history of galaxies, and the prevalence of galactic-scale winds in galaxies.  
Specific spectral features or combinations of features used to infer galaxy properties are called ``diagnostics''.  
Diagnostics are calibrated either empirically or theoretically to allow estimates of fundamental galaxy properties. 

Our ability to model galaxy spectra has advanced rapidly over the past few decades.  The first models were based on individual two- or three-level model atoms, which were applied to \HII\ regions and planetary nebulae. The electron density and electron temperature were derived, with assumptions about the state and structure of the gas, including constant density and constant electron temperature.  Fortunately, we no longer need to rely on simple atoms and assumptions to derive galaxy properties. Thanks to sophisticated quantum mechanical modelling and laboratory measurements, we now have large databases of atomic data, including atomic energy levels, collision and excitation rates, collision cross-sections, as well as data on the composition and effects of astronomical dust.  At the same time, ground and space telescope observations have built enormous databases of stellar spectra, while physics laboratories have advanced our understanding of the processes that produce spectra, including photoionization, collisional excitation, and shock physics.  Now, galaxy spectral models include the latest atomic data and radiative transfer physics, including detailed dust processes, and shock processes. 

\begin{textbox}[b]
{\it ``To try to make a model of an atom by studying its spectrum is like trying to make a model of a grand piano by listening to the noise it makes when thrown downstairs.''} - Anonymous.\\
Therefore, making a model of a galaxy by studying its spectrum is like modelling an entire symphony orchestra from the noise it makes when falling downstairs.  As we model galaxy spectra, it is crucial to understand the limitations of the models and the conditions where the models are invalid.
\end{textbox}

Space-based infrared telescopes and efficient high-resolution optical and infrared spectrographs on the world's largest ground-based telescopes have recently unlocked the rest-frame infrared and UV emission-line spectra of galaxies.  With this new window, we can now track the evolution of fundamental galaxy properties across 12 billion years of cosmic time to 1/10th the age of the universe (i.e. to $z\sim 4$).  In the coming decade, new larger space and ground-based telescopes will reveal the first galaxies in the universe, $>13$ billion years ago.  Rest-frame UV emission-lines will become critical for understanding the early phases of galaxy evolution, while the full suite of UV, optical, and infrared lines will be required to track the evolution of fundamental galaxy properties across cosmic time.

In this review, we focus on the emission-lines in a galaxy spectrum from the UV to the infrared.  We do not cover the use of emission lines to measure star formation rates or dust extinction because both topics have already been widely reviewed \citep[see][]{Kennicutt98,Calzetti07,Calzetti10,Calzetti13,daCunha16}.   Here, we review and expand on the existing library of diagnostics for the electron density, ISM pressure, chemical abundance, and excitation source of galaxies.  For this purpose, we use the latest Pressure and Density Models from \citet{Kewley18}, calculated by combining Starburst99 and MAPPINGS v5.1 photoionization models.  In anticipation of upcoming telescope facilities, we use our new models to recalibrate some of the existing optical diagnostics, and for spectral regions where diagnostics are missing, we provide new diagnostics.  These UV-optical-IR diagnostics can be used in concert to understand how fundamental galaxy properties have changed across cosmic time.

\subsection{Definitions}

Atomic and ionic energy levels can be populated and depopulated by collisions with electrons and, less frequently, by collisions with heavier particles such as protons.  Here, we focus on electron collisions and radiative processes, giving a brief mathematical description of this process.  We refer the reader to \citet{Peimbert17} for an excellent tutorial of the heating and cooling processes occurring in gaseous nebulae.  A comprehensive overview of the physics of gaseous nebulae can be found in many existing papers and textbooks, including \citet{Stromgren39}, \citet{Stromgren48}, \citet{Seaton60}, \citet{Osterbrock89}, \citet{Aller84}, and \citet{Dopita03}.

For thermal electron energy distributions, the collisional excitation rate per unit volume, $R_{ij}^{\rm coll}$ (cm$^{-3}$\,s$^{-1}$) from energy level $i$ to into energy level $j$ is given by

\begin{equation}
R_{ij}^{\rm coll} = n_e N_i \,\alpha_{ij}\;,
\label{eq_R12}
\end{equation}

where $n_e$ is the electron density per unit volume and $N_i$ is the density of the ions with electrons in the lower level $i$. The collisional coefficient $\alpha_{ij}$  (cm$^3$s$^{-1}$) depends on the temperature, $T$, of the gas through an exponential thermal excitation term, $\exp(-E_{ij}/kT)$, and a power law term, $T^{-1/2}$, by

\begin{equation}
\alpha_{ij} = k_{\rm coll} \, \frac{\Omega_{ij}}{g_i}\, T^{-1/2} \, \exp\left( \frac{-E_{ij}} {kT}\right)\;,
\end{equation}
with $ k_{\rm coll} = \left[(2\pi \hbar^4)/(k m_e^3)\right]^{1/2} \approx 8.63\times10^{-6}$.
Here, $\Omega_{ij}$ is the collision strength for the transition from level $i$ to level $j$, $g_i$ is the statistical weight of level $i$, and $E_{ij}$ is the energy difference between levels $i$ and $j$.  The relative importance of each of the two temperature terms governs whether a line ratio pair is sensitive to the electron density or pressure of the gas.

The upper energy level $j$ can also be depopulated by collisional de-excitation by electrons. The collisional de-excitation rate, $R_{ji}^{\rm dex}$  is given by
\begin{equation}
R_{ji}^{\rm dex} = n_e N_j \beta_{ji}\;,
\label{eq_R21}
\end{equation}
with \begin{equation}
\beta_{ji} =  k_{\rm coll} \, \frac{\Omega_{ji}}{g_j}\; T^{-1/2}\; .\end{equation}
Here $\Omega_{ji}$ is the collision strength for the transition from level $j$ to level $i$, and $g_j$ is the statistical weight of level $j$.
The term $\beta_{ji}$ is similar to $\alpha_{ij}$, but does not contain an exponential thermal excitation term.

In the past, the collision strengths were approximated using \citet{Seaton58}, with a classical formula and a Gaunt Factor correction $G(T)$: 
\begin{equation}
\Omega_{i,j} = \frac{8\pi}{\sqrt{3}}\, E_{ij}^{-1}\, \mbox{gf}_{ij}\, G(T)\; .
\label{eq_Gaunt}
\end{equation}

For modern atomic data, the collision strengths are computed numerically from full quantum mechanical collision calculations.  In either case, $\Omega_{ij}$ retains a dependence on temperature.
This residual temperature dependence becomes important for the mid-IR fine structure lines. 

Finally, the radiative depopulation rate of level $j$ is given simply by $A_{ji}$, the spontaneous transition probability, and the density of ions with populations in level $j$.  The radiative depopulation rate is independent of temperature:
\begin{equation}
R_{ji}^{\rm rad} = N_j\,A_{ji}\;.
\end{equation}

\begin{marginnote}[120pt]
\entry{Electron Density ($n_e$)}{The number of electrons per unit volume in a nebula, in $cm s^{-1}$}
\entry{Critical Density ($n_{crit}$)}{The density in $cm s^{-1}$ where the collisional de-excitation probability equals the radiative de-excitation probability for the excited state.}
\entry{ISM Pressure ($\log(P/k)$)}{The pressure within the \HII\ region, in $cm^{-3} {\rm K}$.  The \HII\ region pressure will in general be at a higher pressure than the surrounding diffuse medium.}
\end{marginnote}

The electron density is often calculated using simple atom models, assuming a
constant nebular density and temperature. The nebular pressure, $P$, is related to the nebula temperature, $T$, and total particle density $n$ through the ideal gas assumption $P = nkT$.  Assuming that the nebular and the ISM pressure are related, the ISM pressure can be approximated using the mean nebular temperature $T_e$:

\begin{equation}
P_{\rm ISM} \approx n k T_e,
\end{equation} 
where the total density $n$ is approximated from the electron density through $n \sim 2 n_e(1 + \mathrm{He}/ \mathrm{H})$.  In a fully ionized plasma, the electron temperature is often assumed to be $\sim 10^4$~K, and $P_{\rm ISM}$ is assumed to be directly proportional to $n_e$.

A critical value in determining the state of the plasma in a nebula is the ionisation parameter:

\begin{equation}
 q = {\Phi_{H^0}}/{n_{\rm H}},
 \end{equation}

which is the ratio of the local ionising photon flux $\Phi$ (cm$^{-2}\;$s$^{-1}$) and the local hydrogen density $n_{\rm H}$ (cm$^{-3}$).  For a spherical geometry, the ionization parameter $q_s$ at radius $R$, can be defined to take into account the spherical divergence of radiation :

\begin{equation}
q=\frac{L_{{\rm H}^{0}}}{4 \pi {R}^2 n_{\rm H}}  \label{q_spherical}
\end{equation}

where $L_{{\rm H}^{0}}$ is the ionizing photon luminosity (s$^{-1}$) above the Lyman limit.   This dimensional ionization parameter is related to the dimensionless ionization parameter $U$ by dividing by the speed of light (i.e. $U\equiv q/c$).  The dimensionless ionization parameter is typically $ -3.2 < \log U < -2.9$ for local \HII\ regions \citep{Dopita00} and star-forming galaxies \citep{Moustakas06,Moustakas10}.  
The ionization parameter $q$ has units of velocity (cm/s) and to first order, can be considered the velocity of the ionization front that an ionizing radiation field is able to drive into the surrounding neutral medium.

Different ionization parameter calibrations from different authors cannot be used interchangeably.   
Some calibrations use the ionization parameter at the inner edge of a plane parallel nebula 
($U(R_{\rm in}$) \citep{Kewley02,Kobulnicky04,Levesque10}, while \citet{Stasinska15} derive a volume averaged ionization parameter, $\bar{U}$:

\begin{equation}
\bar{U} = \frac{\alpha_{\rm B}^{2/3}}{c} \,
                    \left(
                          \frac{3}{4 \pi} 
                           n_{\rm H} \,
                          \epsilon^2 \, 
                           L_{{\rm H}^{0}}
                    \right)^{1/3} \, 
                    \left[ 
                          \left( 
                              1+f_S^3 
                          \right)^{1/3} - f_S 
                    \right] \; ,
\end{equation}
where $\alpha_{\rm B}$ is the case B recombination coefficient, $\epsilon$ is the volume filling factor of the gas, and $f_S=R_{in}/R_{S}$ is the ratio of the model inner radius $R_{in}$ to the Str\"{o}mgren radius, where the Str\"{o}mgren radius is calculated assuming $R_{in}=0$.   \citet{Jaskot16} adopt $\alpha_{\rm B} = 2.6 \times 10^{-13}$ cm$^3$ s$^{-1}$, the value for 10$^4$ K gas from \citet{storey95}, to derive useful conversions for comparison amongst ionization parameter values in the literature.  They relate $\bar{U}$  to the ionization parameter at the Str\"{o}mgren radius, $U(R_S)$, and to the ionization parameter at the inner radius, $U(R_{\rm in})$:
\begin{eqnarray}
\bar{U} & = & 3 U(R_S) \, \left[(1+f_S^3)^{1/3}-f_S \right] \, , \\
\bar{U} & = & 3 U(R_{\rm in} )\, f_S^2 \, \left[(1+f_S^3)^{1/3}-f_S \right] \; .
\end{eqnarray}

The ionization parameter can be defined in other ways, depending on the models used and the application.  In theoretical star cluster models, the ionization parameter represents the ratio of the ionizing radiation pressure to the gas pressure \citep{Yeh12}.  Most different ionization parameter calibrations are based on different geometries. In practice, if spherical or plane parallel models have the same ionization parameter (defined at the same distance in the nebula), they will produce very similar spectra, assuming all other parameters are held constant \citep[e.g.,][]{Dopita00}.

The gas-phase metallicity strongly influences the emission from \HII\ regions and galaxies.  The gas-phase metallicity is usually calculated as the oxygen abundance relative to hydrogen, and is defined in units of $\log({\rm O/H})+12$. Oxygen is used to define the overall gas-phase metallicity because oxygen is the dominant element by mass in the universe, and is readily observable in the optical spectrum using temperature-sensitive collisionally excited lines.  These lines are sensitive to the oxygen abundance both through the amount of oxygen in the gas, as well as through the electron temperature of the gas.  The electron temperature is sensitive to the total gas-phase metal abundance (i.e. all the metals, not just oxygen) because metals act as coolants in a nebula.  As the nebula cools, there are less collisional excitations and the strength of collisionally excited lines becomes anti-correlated with the gas-phase metallicity.

\subsection{Theoretical Models}

Current galaxy spectral models are based on the combination of stellar evolution synthesis simulations and photoionization models.  These two components have evolved separately and are still used as stand-alone codes, depending on the application.  

\subsubsection{Stellar Evolutionary Synthesis Models}

The first stellar evolutionary synthesis models for galaxies were developed by \citet{Tinsley68}, who calculated a large population of stars with continuous star formation, in bins of stellar mass.  She placed the stellar population on the Hertzsprung-Russell (H-R) diagram, and evolved the population according to evolutionary tracks.  Stars of different mass evolve along different tracks, and the spectrum of an entire population can be calculated by stopping the tracks at a given age. By adjusting the rate of stellar births and the age of the stellar population, Tinsley's models successfully reproduced the colors, mass-to-light ratio, relative gas mass, and types of galaxies.  

Stellar evolutionary synthesis models developed substantially over the subsequent decades. Major advances include chemical evolution calculations and massive star evolution \citep{Talbot71,Truran71}, stellar and gas-phase abundances \citep{Tinsley72}, heavy element yields \citep{Tinsley73}, changing star formation histories \citep{Larson78}, secondary nucleosynthesis \citep{Arimoto86}, nebular emission and internal extinction \citep{Guideroni87}, and stellar mass loss \citep{Alongi93}.  New observational datasets have also been incorporated, including giant branch luminosity functions \citep{Tinsley76}, the UV spectra of stars \citep{Rocca-Volmerange81,Bruzual83}, stars in the horizontal, asymptotic giant, and post-asymptotic giant branches \citep{Buzzoni89}, near-infrared stellar spectra \citep{Bruzual93}, and stellar atmosphere opacities \citep{Bressan93}.  Modern stellar evolution synthesis models are still based on Tinsley's method, but now use isochrone synthesis, where isochrones are fit to the evolutionary tracks across different masses rather than discretely assigning stellar mass bins to specific tracks \citep[e.g.,][]{Charlot91}.

In the mid 1990s, stellar evolution synthesis codes diverged into two types: fixed metallicity, and chemical evolution.  
Many different stellar evolutionary synthesis models now exist, and each code has its advantages and disadvantages.  
Fixed metallicity stellar evolution models include a sophisticated treatment of stellar processes such as stellar rotation, mixing, metal opacities, and massive star evolution \citep{Leitherer99}.  Some models provide high spectral resolution for modeling the contribution from stellar atmospheres and winds to blended UV lines \citep{Vazdekis16}.   The chemical evolution spectral synthesis models include the evolution of the stellar and gas-phase metallicities over cosmic time \citep{Bressan94,Kotulla09,Fioc11}.  Models now include extinction and gas physics \citep{Fioc11}, and dust absorption and re-emission \citep{Piovan06a}.  

Many stellar evolutionary synthesis models are now available. Starburst99 includes detailed models for massive stars, including metal opacities, and is useful for modelling the global emission from starburst galaxies \citep{Leitherer99}.  SLUG treats stellar populations stochastically and is particularly useful for modelling \HII\ regions and star-forming dwarf galaxies \citep{Krumholz15}. P\'{E}GASE includes detailed chemical evolution of the stellar population, with less emphasis on modelling massive stars or their atmospheres \citep{Fioc11}, while P\'{E}GASE-HR is a high-resolution version of P\'{E}GASE and provides reliable fits to the continua of quiescent galaxies.  The latest version of GALAXEV \citep{Bruzual11} includes new stellar tracks and treatment of the EUV ionizing radiation field, as well as new TP-AGB star models based on observations in the LMC and SMC \citep[see][for a description]{Gutkin16}. \citet{Schaerer13} and \citet{Conroy13} give useful overviews of the full range of stellar evolution synthesis models available.

\begin{figure}
 \centering
  \begin{minipage}[b]{0.49\textwidth}
    \includegraphics[width=\textwidth]{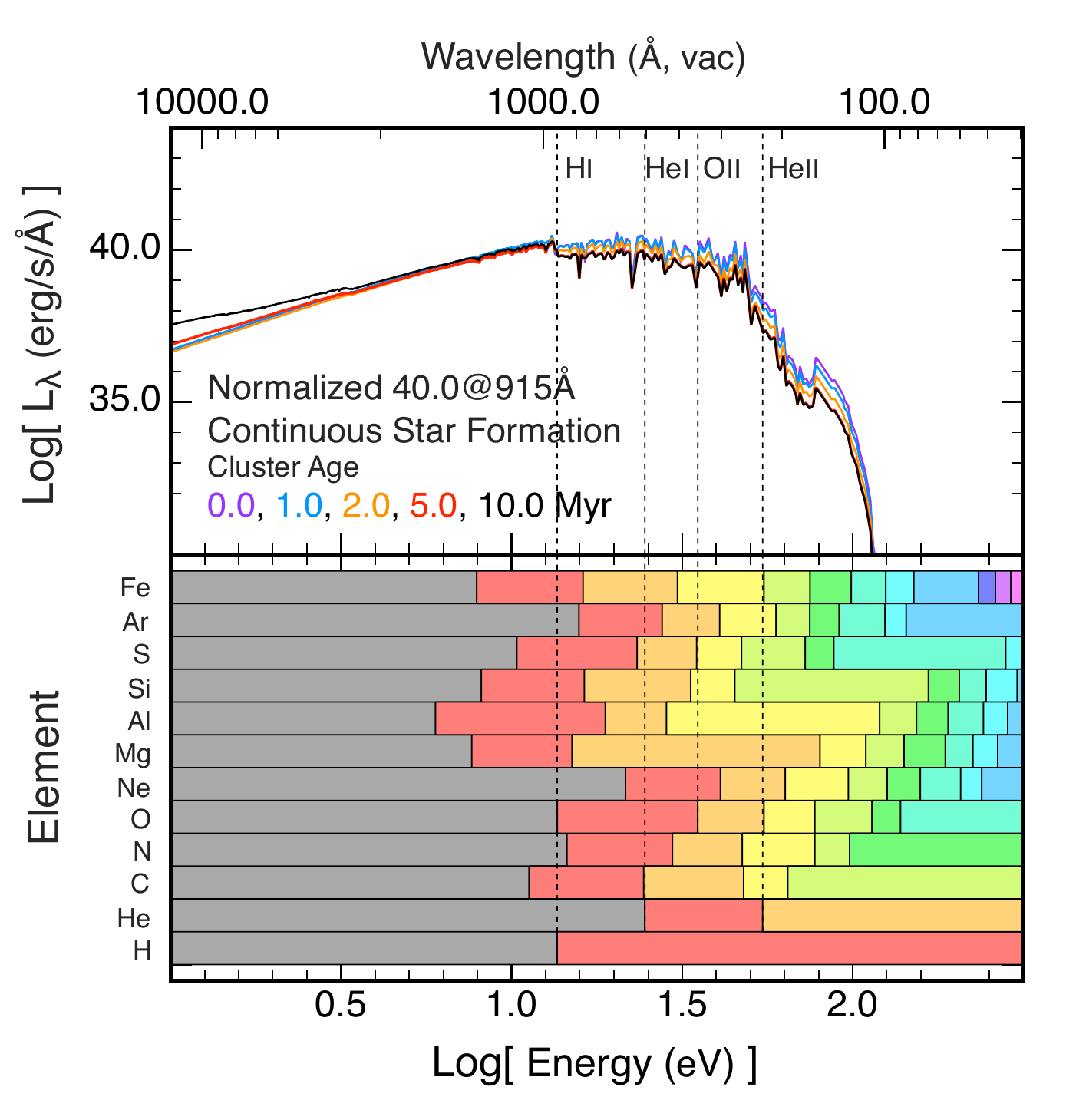}
  \end{minipage}
  \begin{minipage}[b]{0.49\textwidth}
    \includegraphics[width=\textwidth]{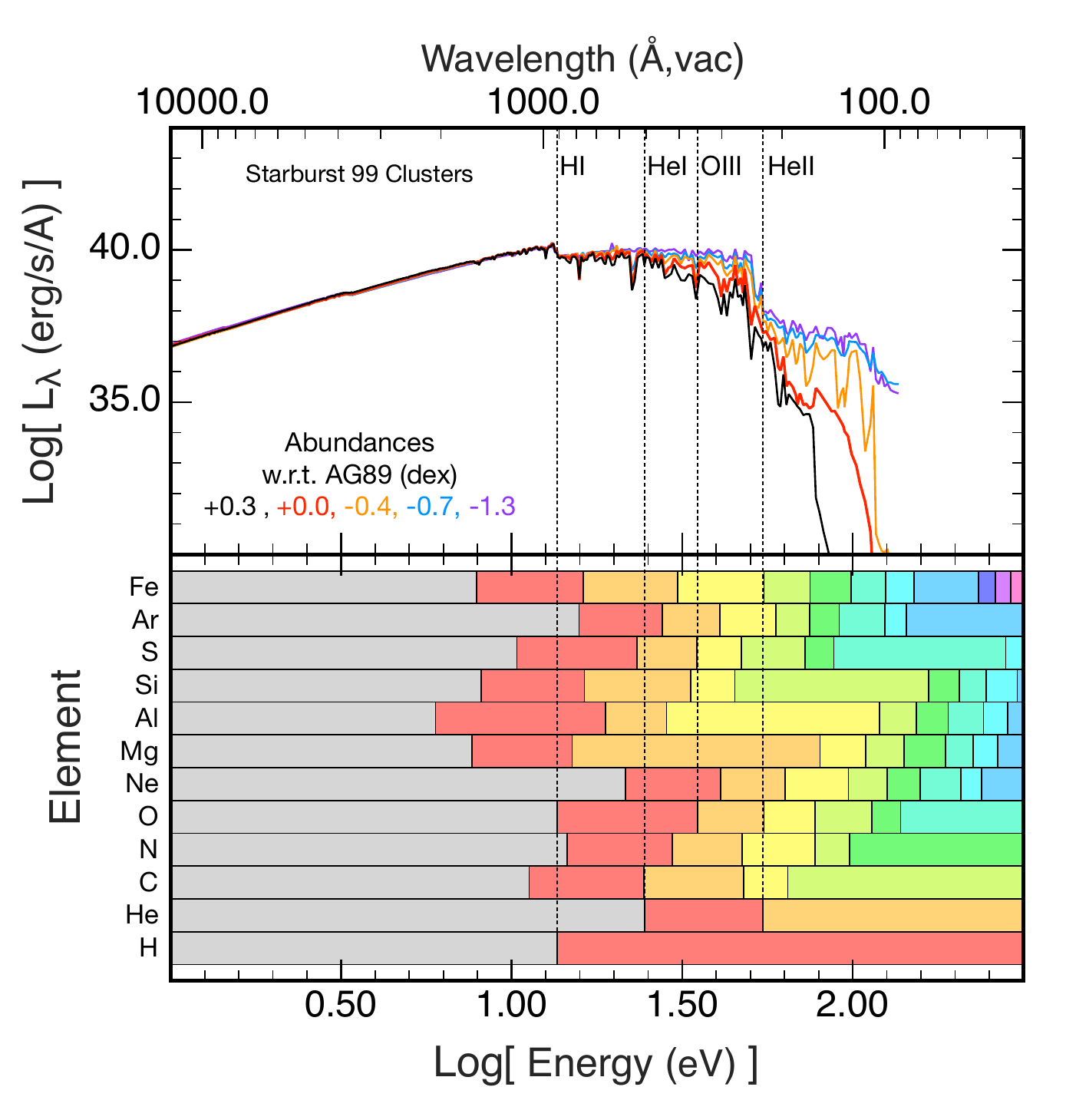}
  \end{minipage}
\caption[Kewley_figure1.pdf,Kewley_figure1b.pdf]{(Left) The spectral energy distribution for Starburst99 simulations of star clusters with ages of 0-10 Myr. (Right) The spectral energy distribution for Starburst99 simulations of star clusters for metallicities scaled with respect to \citet{Anders89} with log scaling factors of $+0.3, 0.0, -0.4, -0.7, -1.3$ (i.e. \OH$= 9.2, 8.9, 8.5, 8.2, 7.6$).  In the left and right panels, the ionization potentials of four key elements (\HI, \HeI, \ion{O}{2}, \HeII) are shown as vertical dashed lines, and the ionization potentials of 12 selected elements are shown in the lower panels for comparison.}
\label{population_age}
\end{figure}

The shape of the ionizing radiation field produced by stellar evolution models depends on the age of the stellar population and on the metallicity.  Certain diagnostics will therefore be very sensitive to, for example, the presence of Wolf-Rayet (W-R) stars or other features of the EUV radiation field. Figure~\ref{population_age} (left panel) shows how the shape of the ionizing radiation field changes for a continuously forming cluster as the stellar population ages from 0 Myr to 6 Myr, normalized at 915\AA.  The ionization potentials of key atomic elements are given in the lower panel for comparison.  The number of ionizing photons in the spectrum at the ionization potential of each atomic species will determine the ionization structure of the nebula.  Figure~\ref{population_age} shows that the change in age from 0 to 5 Myr causes a larger change in $He^{+}$ ionizations than $O^{+}$ ionizations due to the different ionization potentials of $He^{+}$ and $O^{+}$.  

The metallicity also strongly affects the shape of the EUV radiation field, as shown in Figure~\ref{population_age} (right panel).  The EUV radiation field is harder low metallicities because (1) there are less metals in the stellar atmosphere to absorb the stellar radiation field, (2) massive stars at low metallicity have hotter effective temperatures, and (3) the main sequence lifetimes are longer at low metallicity due to lower mass loss rates \citep[see][for a discussion]{Levesque10}.  The effect of W-R stars is also larger at high metallicities because W-R winds are stronger and have longer lifetimes in metal-rich environments, and a smaller mass is required to reach the W-R stage.  In general, species with ionization potentials $>1.5$eV are strongly affected by the metallicity of the EUV radiation field.  Therefore, prior to applying an emission-line diagnostic, it is important to check the shape of the ionizing radiation field relative to the ionization potential of the emission-lines of interest, to understand how uncertain the EUV radiation field is at those wavelengths, and its sensitivity to age and metallicity.

\subsubsection{Photoionization Models}

The physical conditions in photoionized regions are so complex that analytical solutions are impossible.  Instead, we must rely on photoionization models.  Early photoionization models were based on the observations by \citet{Struve38}, who showed that the gas around young O and B stars contain  \Ha\ and \OII~$\lambda 3727$ emission.  These observations inspired \citet{Stromgren39} to create the first model of these regions, based on spheres of gas.  A modified version of the Str\"omgren sphere is still used today to model \HII\ regions.

Major advances in photoionization modelling of \HII\ regions were made in the 1950s and 1960s.  \citet{Zanstra51} showed that the ionization state of the gas can be calculated if the gas is optically thick to the incident radiation field.  This assumption, known as the ``on-the-spot'' approximation, means that recombination of ions with electrons to the ground state produces diffuse radiation which immediately causes re-ionization of the neutral gas.    \citet{Williams67} considered the ionization and thermal balance within the gas surrounding a diffuse ionizing radiation field, and \citet{Rubin68} modeled an \HII\ region in thermal and ionization equilibrium with an arbitrary density distribution.  

AGN photoionization models were developed separately.  Early calculations of temperatures and densities in AGN focused on QSOs \citep{Osterbrock66,Mathez69}, including stratified models in which distinct zones contain ions at different stages of ionization \citep{Burbidge66,Bahcall69a,Bahcall69b}. Optically thin and optically thick X-ray nebulae were considered \citep{Tarter69a,Tarter69b}, and the physics of Auger ionization, Compton heating, and charge transfer collisions were subsequently included \citep{Hatchett76,Halpern80}.  \citet{Davidson72} calculated iterative photoionization models in which the gas has a central QSO ionizing source and spherical symmetry.  High temperature shocks were suggested as a possible cause of the forbidden lines of QSO and Seyfert galaxies \citep{MacAlpine74}, while \citet{Shields74} showed that a power-law radiation field can explain the emission from the radio-loud Seyfert 1 galaxy 3C 120.  Resonance line-trapping and collisional de-excitation can effect the emission-line spectrum
in the high gas densities associated with nebulae around QSOs, and these were included in subsequent models \citep{Krolik78,Ferland79,Kallman82}.

Over the past three decades, two large self-consistent photoionization codes emerged that include the physical developments discussed above:  MAPPINGS \citep{Binette85,Sutherland93,Sutherland17}  and CLOUDY \citep{Ferland98,Ferland17}.  MAPPINGS includes self-consistent treatment of nebular, dust, and shock physics, and can be applied to \HII\ regions, AGN, and regions shocked by supernovae, galactic winds, and jets.  CLOUDY includes self-consistent treatment of nebular, dust, and molecular physics, and can be applied to \HII\ regions, AGN, and photodissociation regions.  These photoionization models were originally used as stand-alone codes with blackbody and power-law radiation fields to simulate regions excited by star formation and AGN, respectively.  It is now known that a blackbody does not provide an accurate representation of the ionizing radiation field from a star forming region because the stellar radiation field is transported through the stellar atmosphere.  The Helium and metals in the photosphere systematically absorb radiation from the star (known as line blanketing), significantly altering the spectral shape \citep{Pauldrach01}.  Stellar mass-loss, rotation, and the effect of binary stars can also alter the shape of the ionizing radiation field \citep{Kurucz79,Pauldrach86,Schmutz89,Leitherer08,deMink09,Levesque12,Eldridge12,Pauldrach12}.

More sophisticated techniques developed in the early 2000s.  Stellar evolutionary synthesis and photoionization models were combined to analyse the optical spectra of \HII\ regions \citep{Dopita00} and star-forming galaxies \citep{Kewley01a,Moy01}.  More realistic AGN ionizing radiation fields were included in photoionization models to understand the properties of AGN spectra \citep{Groves04a,Groves04b,Thomas16}.  Ideally, the stellar continuum and nebular gas are coupled self-consistently to produce line intensities that scale with the stellar population in terms of age and metallicity, as in \citet{Byler17}.

\begin{figure}
\vspace{-3cm}
\includegraphics[width=9cm,angle=90]{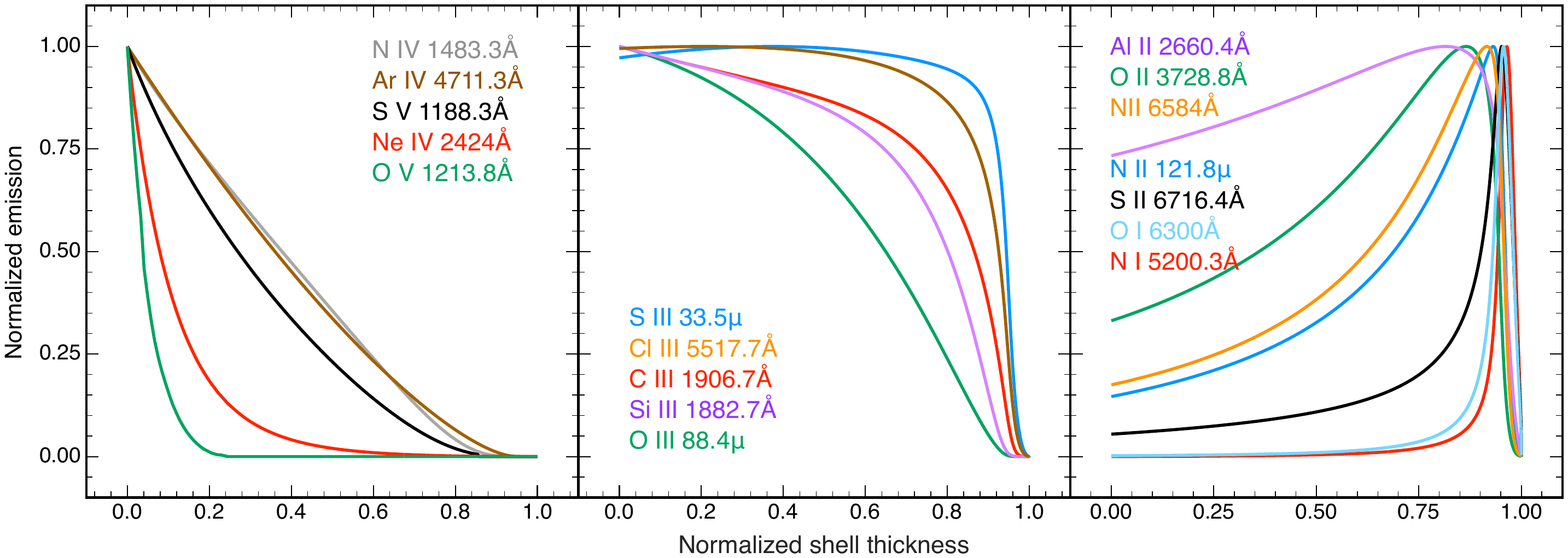}
\vspace{-3cm}
\caption[Ionization_structure2.pdf]{The ionization zones for selected strong emission-lines in the UV, optical, and infrared.  The different panels show how different levels of ionization probe different regions of the nebula. \label{Ionization_zones}}
\end{figure}

\subsection{Nebular structure}\label{nebular_structure}

Theoretical models assume an ionization structure, temperature structure, and density structure of the nebula.  It is important to assess whether particular diagnostics will be applicable to \HII\ regions or galaxy of interest by comparing the observed ionization, temperature, and density structure with the theoretical models that are used to interpret the diagnostic line ratios.

\subsubsection{Ionization Structure}\label{Ionization_structure}

Ionization lines of different species probe different zones of a nebula because they have different ionization potentials and different critical densities.  The X-ray (10 - 0.1 \AA) and the EUV spectrum (4000 - 10 \AA) cover up to 31 ionization stages\footnote{The National Institute of Standards and Technology (NIST) database gives the ionization potential of the elements \citep{Kramida15}}. 
Figure~\ref{Ionization_zones} gives the MAPPINGS v5.1 ionization zones for selected strong emission-lines.  Only a few sets of lines are produced throughout the nebula. The \OIII\ and \ion{C}{3}] lines are both produced throughout the nebula, and derived properties may be considered as broadly representative of the overall ISM within the nebula.  The \NII\ and \OII\ lines trace similar regions of the nebula, and may be used to trace intermediate-zone pressures and metallicities.  Higher ionization species, such as \NeIV, \OV, \ArIV, \NIV, and \SV, trace regions of high ionization, close to the ionizing source.   On the other hand,  the \SII, \OI, and \NI\ lines trace the very outer edge of the nebula, where the gas is only partially ionized. 

\begin{figure}
\includegraphics[width=10cm,angle=0]{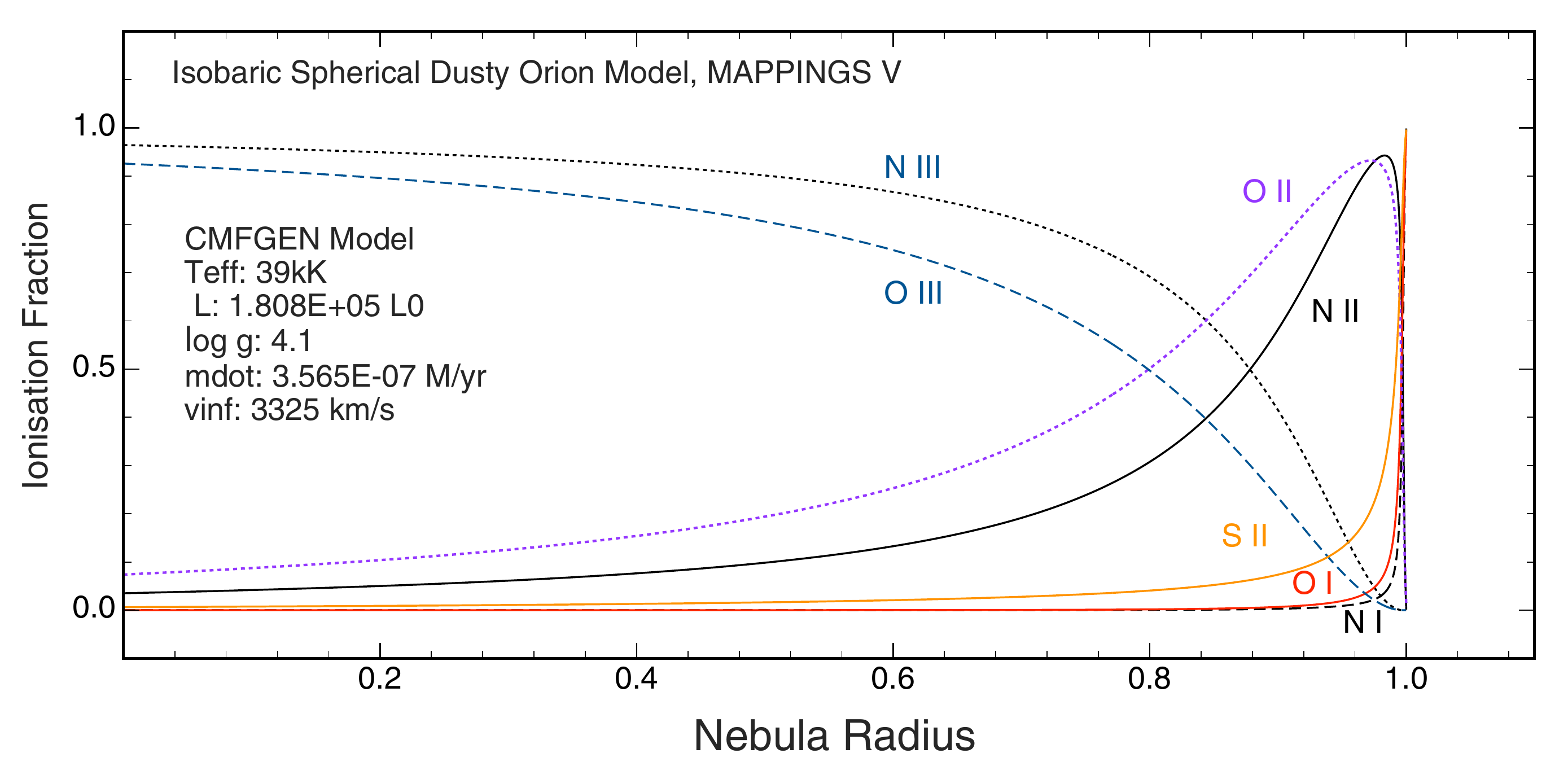}
\caption[Kewley_figure3.pdf]{Spherical Mappings V5.1 model of the ionization structure in Orion for isobaric conditions.  Model parameters for the dominant central star Theta 1C are shown.  Coloured curves correspond to the ionization fractions of commonly observed emission-lines of oxygen, sulphur, and nitrogen.}  
\label{Orion_ionization}
\end{figure}

Both Mappings and Cloudy assume plane parallel or spherical symmetry and cannot produce specific ionization structures that are seen in the Milky Way, such as the ionization bar in the Orion Nebula \citep[e.g.,][]{Simpson86,Rubin11}, but they can produce an overall equivalent spherical ionization structure (Figure~\ref{Orion_ionization}) that reproduces the observed spectrum.

\subsubsection{Temperature Structure}\label{Temperature_structure}

Optical density-sensitive line ratios have traditionally been calibrated assuming a simple model atom and a single electron temperature (typically $T_e = 10^4$~K) \citep[e.g.,][]{Osterbrock89,Rubin94}.  Single temperature models have also been assumed in AGN and starburst photoionization models for determining the power source of galaxies \citep[e.g.,][]{Filippenko85,Hill99,Groves04c}.

\begin{figure}
\centering
\includegraphics[width=\linewidth,angle=0]{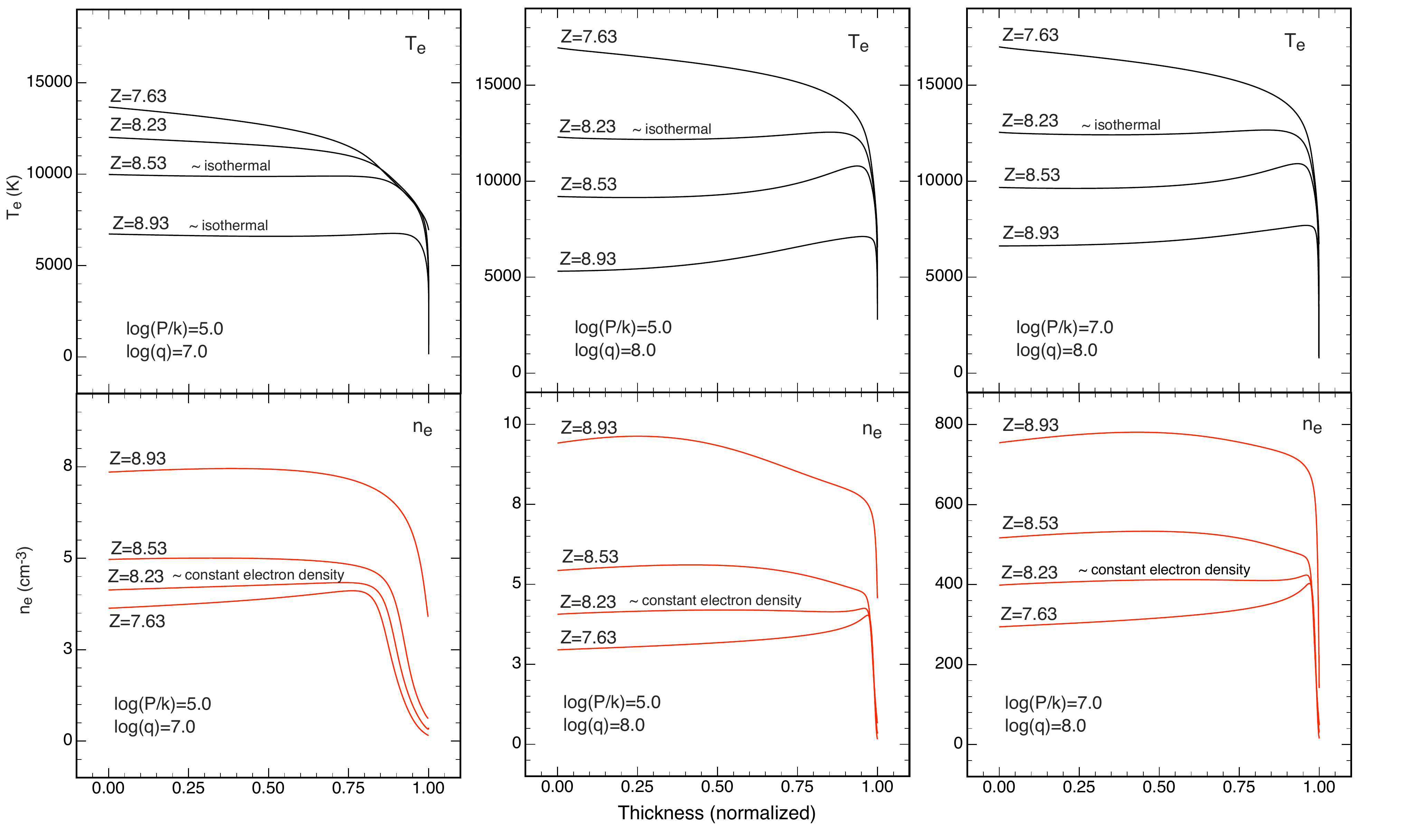}
\caption[Ionization_structure2.pdf]{(from Kewley et al. 2018) The electron temperature (upper panel) and electron density (lower panel) as a function of normalized model thickness for our Mappings V Pressure Models with $\log(P/k)=5.0$, $\log(q)=7.0$ (left),  $\log(P/k)=5.0$, $\log(q)=8.0$ (middle), and $\log(P/k)=7.0$, $\log(q)=8.0$ (right).}
\label{Model_TeNe}
\end{figure}

In current photoionization models, the electron temperature varies as a function of distance from the ionizing source.  Figure~\ref{Model_TeNe} (top) shows how the electron temperature varies through a theoretical nebula.  The electron temperature can only be approximated by isothermal conditions for some metallicities. At metallicities at or above \OH$\sim8.5$, the nebula becomes hotter towards the outer edges because the soft ionizing photons have already been absorbed by metals closer to the ionising source, leaving predominantly hard ionising photons, yielding more heat per ionisation.  In addition, the dominant coolant, \OIII, dominates the cooling in the inner nebula zone, but is not a significant coolant in the outer nebula.

\begin{figure}
\includegraphics[width=12cm]{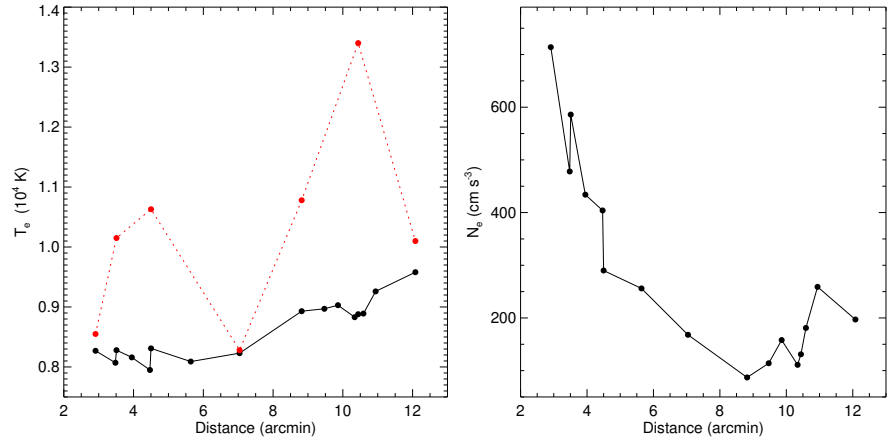} 
\caption[Kewley_figure5.eps]{The electron temperature (left) and electron density (right) as a function of distance from the centre (in arcmin) of the Orion nebula, as measured by \citet{Rubin11}.  The electron temperature is measured using the Auroral \NII~$\lambda 5755$ (black solid line) and \OIII~$\lambda 4363$ (red dashed) emission lines.  The electron density is measured using the \SII~$\lambda \lambda 6717,31$ doublet.}  
\label{Temden_Orion}
\end{figure}

The electron temperature structure in \HII\ regions has been notoriously difficult to measure due to the weakness of the emission-lines most sensitive to \Te\ \citep[e.g.,][]{Luridiana03}.  Integral field spectroscopic studies of local \HII\ regions have now made such measurements possible in just a few cases \citep[see e.g.,][for a discussion]{Wang04}.  However, detailed observations have been made of the temperature structure across $\theta^1$~Ori C in the Orion Nebula, shown in Figure~\ref{Temden_Orion} (left panel).  The electron temperature in the \NII\ (black solid line) and \OIII\ (red dashed line) zones is complex and varies across $\theta^1$~Ori C, rising with distance from the ionizing source, similar to the predictions of the metallicity theoretical photoionization models with \OH$>8.53$ and $\log(q)=8$.

Assuming a constant temperature becomes particularly problematic when fixed-size apertures capture the light from an ensemble of \HII\ regions, as is the case for global spectra of high redshift galaxies.  Fiber-based or slit-based spectra of nearby or distant galaxies represent the luminosity-weighted average of multiple (up to hundreds or thousands) of \HII\ regions, each of which may contain a complex electron temperature structure.  In addition, diffuse ionized gas (DIG) can contribute to electron temperature gradients and contaminate the emission-lines observed in fixed-size apertures \citep{Otte02}.  
Therefore, the spectra from an ensemble of \HII\ regions cannot be robustly modeled using a constant temperature model.   Models that allow a complex temperature structure or gradient will provide more realistic results.

\subsubsection{Density Structure}\label{Density_structure}

Current measurements of the electron density of \HII\ regions and galaxies are based on single atom models that assume a constant density across the \HII\ regions.  However, we know that this is not a realistic assumption. Ionized gas and radio continuum density measurements reveal complex radial gradients in many \HII\ regions \citep{Franco00,Perez01,Binette02,Luridiana03,McLeod16} and flat gradients in others \citep{Ramos-Larios10,Garcia-Benito10}.  These density gradients are often anti-correlated with \HII\ region size.  Ultracompact \HII\ regions typically have steep density gradients \citep{dePree95,Franco00,Kurtz02,Johnson03b,Phillips07}, while larger \HII\ regions have more shallow or flat density gradients \citep{Phillips08}.   

The Milky Way \HII\ region density structure has been studied extensively in the infrared. \citet{Simpson04} found significant density variations of $40-4000\,{\rm cm}^{-3}$ within a single \HII\ region in the inner region of the Milky Way, while \citet{Rubin11} found complex electron density structure in the Orion nebula, with density variations of $80-700 \,{\rm cm}^{-3}$.  In Figure~\ref{Temden_Orion} (right panel), we show the density structure of Orion with measurements from 
\citet{Rubin11} along $\theta^1$~Ori C.  The electron density is not a simple gradient from the ionizing source, and is likely the result of a blister \HII\ region and contamination by scattered light from a nearby more dense region. 

Photoionization models in pressure equilibrium calculate the electron density structure of the nebula based on the ionizing radiation field.  Figure~\ref{Model_TeNe} shows how the electron density varies through a theoretical nebula for differing pressure, ionization parameters, and metallicity.  At high metallicity, electron density gradients are seen, and at low metallicity, a more complex structure is seen, due to the relationship between electron density and ionization parameter, which is large in the outer edges of the nebula at low metallicity.

There are many potential causes for complex density structures.  \HII\ regions can trigger the formation of new stars through ``collect and collapse'' \citep{Elmegreen95} and ``radiation driven implosion'' \citep{Bertoldi89}.  In the ``collect and collapse'' model, the \HII\ region expands into a supersonically turbulent cloud, causing coagulation of gravitationally unstable clumps that can collapse and form new stars.  In the ``radiation driven implosion'' model, hot stars penetrate the ISM and heat the cold low-density gas.  The heating amplifies overdensities created by the turbulent ISM, which then collapse and form stars \citep{Gritschneder09,Dale12}.  Both of these processes can create globules of dense gas \citep[e.g.,][]{Tremblin13,Walch15,Schneider16}.   Unusual geometries can also be created by stellar wind-driven outflows, creating horse-shoe or other complex geometries \citep{Park10}.
 
Galaxies too have complex density structure and gradients.  Star-forming galaxies typically have electron densities that follow a $r^{-1/2}$ profile out to $\sim 10~{\rm kpc}$ \citep{Gutierrez10}.  The electron density can also exhibit structure, and is large at the edges of bars where gas collisions occur \citep{Herrera-Camus16}, as well as in some extended regions associated with galactic-scale winds \citep{Ho14,Rose18}.  Seyfert galaxies exhibit electron density gradients \citep{Kakkad18}.   In particular,  in Seyfert galaxies the gas density (a) increases with the ionization potential of the ions, and (b) is anti-correlated with the electron temperature \citep{Spinoglio15}.  

Emission-lines produced by different ions and different energy levels are sensitive to different density regimes, depending on the critical density of the transition.  Lines with low critical densities, like \OII\ and \SII, are affected by collisional de-excitation and are therefore weak in the high density regions, while \CIII, \ArIV\  trace higher density regions of a nebula.  Likewise, the \OIII\ fine structure lines have a significantly lower critical density than the \ArIV\ or the \ClIII\ lines, and therefore trace lower density regions than the \ArIV\ or \ClIII\ lines. The high ionization line ratios like \ArIV\ and \ClIII\ therefore trace the ISM pressure conditions within a luminosity-weighted average of the high density clumps.

Where density gradients or clumpiness is expected, we recommend the use of models that include complex density gradients and profiles, such as models that assume pressure equilibrium.  If constant density diagnostics are applied to \HII\ regions or galaxies that contain density gradients or clumps, the properties derived will likely be dominated by the regions that produce the largest emission-line strengths, and will not necessarily represent the galaxy average.

\section{ISM Pressure and Electron Density Diagnostics}

Electron density diagnostics are based on UV, optical, or infrared emission-line ratios of the same species.  The first calibrations were developed for diagnosing the electron density in planetary nebulae using UV or optical spectroscopy \citep[e.g.,][]{Aller61,Dopita76,Stanghellini89,Keenan92,Copetti02}.   Optical and UV density-sensitive line ratios were subsequently calibrated for \HII\ regions \citep{Esteban99,Wang04,Park10} using simple model atoms.  UV line ratios have also been used to derive the electron density in the gas around quasars \citep{Nussbaumer79,Negrete12}.   The infrared fine structure lines were first used to measure the electron density in planetary nebulae over a decade ago \citep[e.g.,][]{Liu01}, and are now accessible as density diagnostics in local \HII\ regions and galaxies, as well as at high-redshift \citep[e.g.,][]{Maiolino05,Maiolino09,Ferkinhoff10,Ivison10,Valtchanov11,Spinoglio15}.  

Direct measurements of the electron density using the same methods are now being made for increasing numbers of galaxies, particularly at high redshift.  Some studies find larger electron densities \citep{Hainline09,Bian10,Brinchmann08b,Shirazi13b} on average than in local galaxies, while others find similar electron densities to local galaxies \citep[see][]{Rigby11,Bayliss13}.  However, these large electron densities may be a selection effect.  \citet{Kaasinen17} matched high-z and local galaxies in stellar mass, star formation rate (SFR), and specific SFR.  They showed that larger electron densities in high-z galaxies seen in previous work disappear when the samples are matched in star formation rate, implying that the observed large electron densities are a result of selecting samples of the most luminous galaxies at a given epoch.  

Many electron density estimates assume a single model atom with a constant electron temperature through the nebula. As discussed in Section~\ref{Temperature_structure}, constant temperature is not a realistic assumption for most \HII\ regions and galaxies.  Full photoionization model grids with complex temperature structures have been developed to derive more realistic electron density estimates. \citet{Proxauf14} use CLOUDY models to calibrate the  \ArIII, \ArIV , \SII, \OII\ line ratios.  \citet{Kewley18} uses photoionization models to calculate a suite of 13 density calibrations for various line ratios from the UV-FIR, with minor corrections for metallicity and ionization parameter. 

All electron density estimates assume that the electron density is constant across the \HII\ region or galaxy.  It is clear that a constant density is not a valid assumption for many \HII\ regions and galaxies.  For these cases, the ISM pressure is a more realistic physical parameter to derive from most \HII\ regions or ensembles of \HII\ regions.  The ISM pressure is a critical parameter in fully self-consistent models that include detailed nebular temperature and density structures. In reality, the pressure is determined by the combination of the mechanical energy produced by the stellar population, as well as the strength and shape of the radiation field. Models with a constant pressure are appropriate when the sound crossing time is less than the heating and cooling timescales, which occurs in the majority of \HII\ regions \citep{Field65,Begelman90}. \citet{Gutierrez10} showed observationally that to a first approximation, the \HII\ regions can be considered in pressure equilibrium with their surroundings.

Many pressure diagnostics depend strongly on the gas-phase metallicity.  Different metallicity calibrations exhibit extremely large systematic discrepancies (up to 1 dex in \OH) \citep[see][for a review and discussion]{Kewley08}.  Due to these discrepancies, theoretical ISM pressure calibrations can only be used with metallicity calibrations that have been constructed using consistent theoretical models.  For the pressure calibrations summarized here, many suitable consistent metallicity calibrations exist, from simple fits to MAPPINGS photoionization models given here, to the more sophisticated Bayesian methods of \citet{Blanc15} and \citet{Thomas18} that simultaneously calculate the ionization parameter and metallicity probability distributions given the observed emission-lines in a spectrum.  


The UV contains several pressure and density sensitive emission-line ratios, which are described in detail  in \citet{Kewley18}.  Some of these ratios, like the \SiIIIf~$\lambda 1883/$\SiIII~$\lambda 1892$, \CIIIf~$\lambda 1907$/\CIII~$\lambda 1909$, and \AlII~$\lambda 2660/$\AlIIint~$\lambda 2669$ ratios, have been previously calibrated and successfully applied to planetary nebulae, to the gas around quasars, and to nearby \HII\ regions \citep[e.g.,][]{Nussbaumer79,Dufton84,Clegg87,Keenan92,Negrete12}.

\begin{figure*}
\includegraphics[width=12cm]{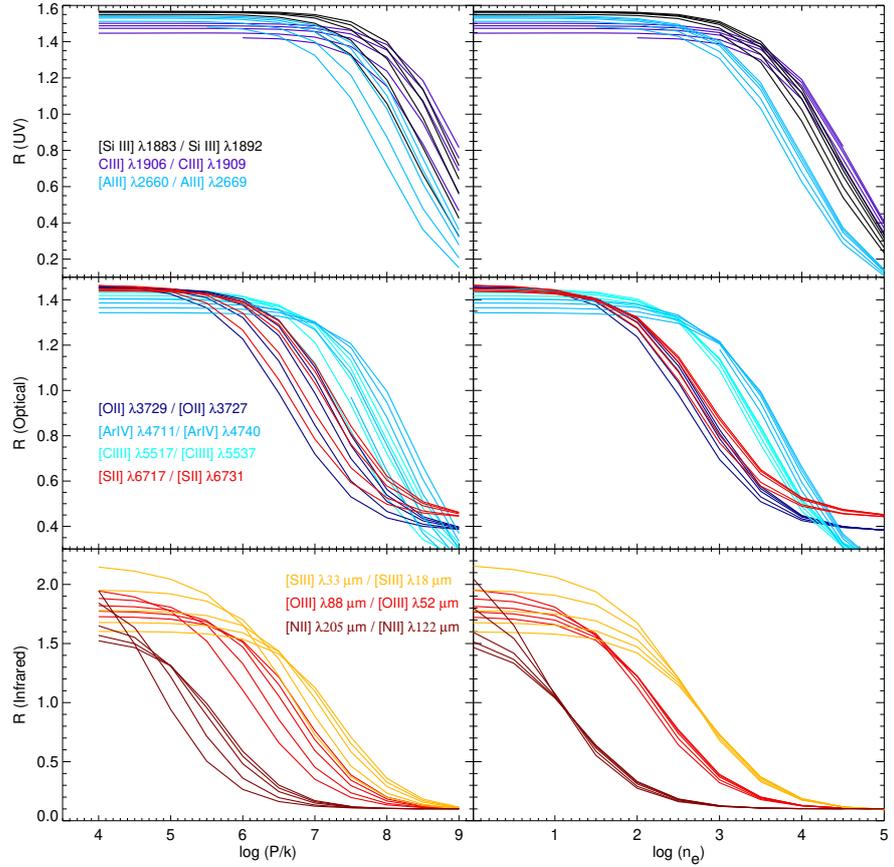}
\caption[Kewley_figure6.eps]{UV (upper panel), Optical (middle panel) and IR (lower panel) pressure (left panel) and density (right panel) diagnostics from \citet{Kewley18}.  For each ratio, models of constant metallicity are shown as lines from left to right corresponding to \OH$=[7.63, 8.23, 8.53, 8.93, 9.23]$}
\label{Density_diagnostics}
\end{figure*}

Figure~\ref{Density_diagnostics} (top panels) shows the relationship between the UV pressure-sensitive ratios, ISM pressure and electron density, as a function of metallicity.  Calibrations of these diagnostics in terms of metallicity, as well as a minor secondary dependence on ionization parameter can be found in \citet{Kewley18}.  The strong \SiIIIf/\SiIII\ and \CIIIf/\CIII\ ratios offer the most promise as UV electron density and pressure diagnostics for star-forming galaxies.  The wavelengths in each of these ratios are sufficiently close that flux calibration and extinction correction is not necessary.   The metallicity dependence of these ratios arises from the sensitivity of the C++ and S++ transitions to the electron temperature of the gas through their recombination coefficients.  Both ratios are important tracers of dense environments ($\log(\frac{\rm N_{e}}{\rm cm s^{-1}}) > 3.5$) and high pressures ($\log(\frac{{\rm P/k}}{\rm cm^{-3} k}) > 7$).  Both ratios have a residual dependence on temperature when used as a density diagnostic; temperature variations in the nebula may cause a scatter in the measured density of up to 0.3 dex. 

High ionization lines of neon and nitrogen can also be used as pressure and density diagnostics in the UV, but these ratios are weak in galaxies with high metallicity and low ionization parameter because the low temperature and low ionization parameter prevent ionization of neon and nitrogen to their higher states.  Care must be taken when using these two sets of lines because they may contain significant contributions from AGN or shocked gas.

The optical contains many emission-lines which together make density-sensitive line ratios, including the commonly used \OII~$\lambda 3729/$\OII~$\lambda 3726$ and \SII~$\lambda 6731$/\SII~$\lambda 6717$ ratios (Figure~\ref{Density_diagnostics}, middle panels).  These calibrations have been used for decades to diagnose the electron density in planetary nebulae, \HII\ regions, and galaxies \citep[see e.g.,][and references therein]{Osterbrock89,Kaasinen17}.
Weaker density-sensitive ratios such as the \ArIV~$\lambda 4711$/\ArIV~$\lambda 4740$, \NI~$\lambda 5198$/ \NI~$\lambda 5200$, and \ClIII~$\lambda 5517/$\ClIII~$\lambda 5537$ ratios have been applied to planetary nebulae and local \HII\ regions 
\citep[e.g.,][]{Aller70,Esteban99,Lee13}.
The \CIII\ ratio traces high density, high pressure regions of the gas, and is less sensitive to metallicity than the other diagnostics.  The \ArIV\ and \NI\ lines are weak for metallicities and ionization parameters of most star-forming galaxies, and the \NI\ ratio depends strongly on metallicity above \OH$>8.5$.  This behavior is primarily caused by collisional effects between the upper levels in the pair, as well as transitions from higher levels \citep[see][for a full explanation]{Kewley18}.

The infrared contains three extremely useful electron density diagnostics; the \SIII\ $33\mu$m/\SIII~$18 \mu$m, \OIII~$52 \mu$m/\OIII~$88 \mu$m, and \NII~$205 \mu$m/\NII~$122 \mu$m ratios (Figure~\ref{Density_diagnostics}, lower panels).  These ratios cover a broad range of densities, providing a complementary suite of diagnostics that can be used to build a comprehensive picture of the density structure of the ionized gas at a given redshift.  These line ratios depend strongly on the metallicity, with a secondary dependence on ionization parameter. 

The ionization parameter dependence of the pressure-sensitive line ratios is shown in Figures 5-17 of \citet{Kewley18}.  To derive accurate ISM pressures, we recommend the interpolation of the theoretical model data given in \citet{Kewley18} with an estimated ionization parameter and metallicity.


\section{Ionization Parameter in Galaxy Evolution Studies}

The ionization parameter across normal star-forming galaxies is remarkably uniform. The dimensionless ionization parameter is typically $ -3.2 < \log U < -2.9$ for local \HII\ regions and star-forming galaxies \citep{Dopita00,Moustakas10,Poetrodjojo18}. 
The largest ionization parameters in the local universe are found in the largest star clusters, called super star clusters.  The super star clusters in M82 have ionization parameters up to $\log(U) \sim -2.3$ \citep{Forster01,Smith06}.  Super star clusters in luminous infrared galaxies have similar ionization parameters \citep{Snijders07,Indebetouw09}. 
Ionization parameters have not been observed above $\log(U) \sim -2.3$, suggesting that some mechanism or mechanisms moderates or limits the ionization parameter.  Multiple mechanisms have been proposed, including expanding wind bubbles, radiation pressure confinement, and the effects of dust \citep[see][for an excellent overview]{Yeh12}. 

The global ionization parameter in galaxies is usually anti-correlated with the gas-phase metallicity, such that low metallicity galaxies or \HII\ regions have large ionization parameters \citep{Dopita86}.  \citet{Bresolin99} found that the ionization parameter in metal-poor disc \HII\ regions is $\sim 4 \times$ that in \HII\ regions with solar metallicity.  Similar results were reported by \citet{Maier06} and \citet{Nagao06}.   The cause of this anti-correlation between ionization parameter and metallicity is unknown.  \citet{Dopita06a} propose that the anti-correlation is caused by stellar atmospheres.  At high metallicity, the stellar wind has a larger metal opacity, and absorbs a larger fraction of the ionizing photons, leaving less EUV photons to ionize the surrounding \HII\ region.  Metal-rich stellar atmospheres also scatter photons from the photosphere more efficiently than metal-poor atmospheres.  This process allows luminous energy to be converted to mechanical energy more efficiently in the stellar wind region, reducing the number of ionizing photons incident on the \HII\ region. Intriguingly, the metallicity-ionization parameter relation does not necessarily hold for spatially resolved data.  In samples of \HII\ regions within single galaxies, ionization parameter generally does not correlate with metallicity \citep{Garnett97,Dors05,Dors11,Poetrodjojo18}.  

The ionization parameter, measured directly using the \OIIIOII\ lines, was larger in the past than today \citep{Kaasinen17}.  Over the past 6 billion years (from $z=0$ to $z=0.4$), the ionization parameter continued to fall by 0.1 to 0.25 dex, independent of metallicity and stellar mass \citep{Kewley15}.  This change is likely to be related to the fraction of young stars per unit volume, referred to as the cosmic star formation density \citep{Hirschmann17}.  


Despite the importance of the ionization parameter in understanding the properties of ionizing sources and their influence on the surrounding ISM, very few ionization parameter calibrations existed until recently.  The ionization parameter is typically measured by comparing two emission-lines from the same atomic species that are in different ionization states.  Many different species can be used to calculate the ionization parameter including carbon, sulphur, silicon, neon, nitrogen, and oxygen.  The most sensitive ionization parameter diagnostics usually come from two states with the largest difference in ionization potentials.  However the ionization parameter can also be calculated using ionization-sensitive line ratios of different species.  

The majority of high-ionization to low ionization sets of lines are separated far in wavelength, requiring accurate flux calibration and extinction correction.  Accurate flux calibration and/or extinction correction can be difficult at high redshift when ionization-sensitive emission-lines are detected in different wavebands, or when instrument flexure is severe.  To overcome this issue, \citet{Kobulnicky03a} proposed using emission-line equivalent widths in lieu of fluxes.  This method assumes that the continuum between the two ionization state lines is roughly constant between the wavelengths of interest.  Tests indicate that this assumption holds for the rest-frame blue continuum between the \OII~$\lambda \lambda 3726,9$ and  \OIII~$\lambda \lambda 4959,5007$ lines \citep{Liang07b}. However, we do not recommend the use of equivalent widths for UV ionization-parameter diagnostics due to the strong underlying stellar UV continuum. 

Below, we describe and provide new ionization parameter diagnostics from the UV through to the infrared. In Table~\ref{pres_diag_Pk5}, we provide bi-cubic surface fits to the ionization parameter in terms of log(R) and metallicity for $\log(P/k)=5.0$ and $\log(P/k)=7.0$, where R is the diagnostic line ratio. If the metallicity is unknown, our pressure diagnostics may be used with a metallicity estimated using the mass-metallicity relation, or by assuming a metallicity of \OH$\sim 8.7$, and accounting for the potential difference between this assumed metallicity and the potential metallicity.  Note that for the bi-cubic surface fits, some model data points at high metallicity and high ionization parameter have been removed from the fitting procedure, because the modelled behaviour is double-valued and cannot be fit by a simple function.  For ionization parameter estimates outside the modelled range, the full model data from \citet{Kewley18b} should be interpolated.  The fits in Table~\ref{pres_diag_Pk5} have average errors of \~3\% or less, and cover the range $-4\lesssim \log(U) \lesssim-2.5$ (or $6.5 \lesssim \log({\rm q})\lesssim 8$).  

Note that different authors have used different definitions of the ionization parameter, which can lead to confusion and errors when comparing ionization parameters calculated from different samples.  Ionization parameters can be defined on the inner edge of the nebula, or as an average throughout the nebula.  The ionization parameter is usually defined in terms of the hydrogen density of the gas, but sometimes the ion density or the electron density is used instead.  We recommend the application of consistent ionization parameter diagnostics when comparing \HII\ regions and galaxies from different samples.

\subsection{UV Ionization Parameter Diagnostics}

\begin{figure*}[t]
\includegraphics[width=12cm]{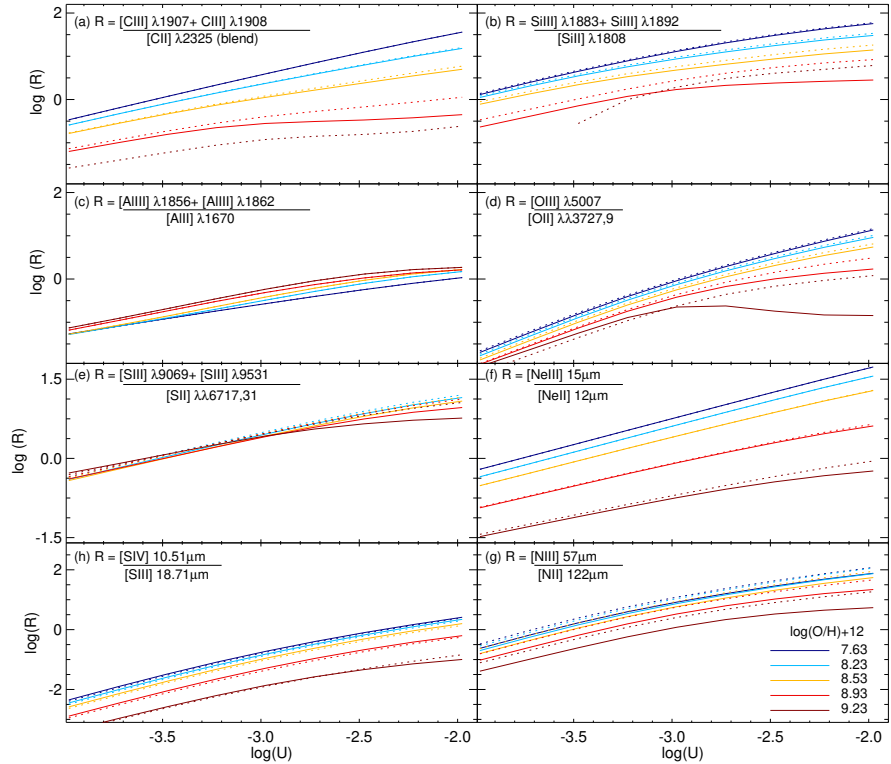}
\caption[Kewley_figure7.eps]{Useful ionization parameter diagnostic line ratios as a function of metallicity (colors blue, cyan, green, pink, grey correspond to \OH$=[7.63,8.23,8.53,8.93,9.23]$ respectively) (blue to red lines, respectively).  Solid and dotted lines correspond to ISM pressures $\log(P/K) = 5$ and $\log(P/K) = 7$, respectively.}
\label{ionization_diags}
\end{figure*}

The UV contains some excellent ionization parameter diagnostics, as shown in Figure~\ref{ionization_diags}.
Carbon, silicon, sulphur, nitrogen, aluminium, and iron all have multiple stages of ionization that produce emission-lines in the UV.  Most of these ratios are sensitive to metallicity and ISM pressure, however there are a few notable exceptions.  The (\CIIIf~$\lambda 1907$ + \CIII~$\lambda 1909$)/(\CII~$\lambda 2325$ blend) provides an ideal ionization parameter diagnostic for low metallicity galaxies (\OH$\lesssim 8.5$).  This diagnostic is based on the blend of lines at 2323.50\AA, 2324.69\AA, 2325.40\AA, 2326.93\AA, and 2328.12\AA, and is insensitive to ISM pressure and metallicity for this metallicity range \citep[see][]{Kewley18b}.  The \CIII$/$\CII~$\lambda 2325$ ratio is also insensitive to ISM pressure, but requires a correction for metallicity (Figure~\ref{ionization_diags}a).

The silicon ratios \SiIII~$\lambda 1206$/\SiII~$\lambda 1260$ and (\SiIII~$\lambda 1883$+\SiIII~$\lambda 1892$)/\SiII~$\lambda 1808$ are extremely stable to variations in pressure for metallicities \OH$\lesssim 8.5$.  Between $4<\log(P/k)<7$, both silicon ratios vary less than 0.1 dex with pressure (Figure~\ref{ionization_diags}b).   



The commonly-used optical \OIIIOII\ line ratio has an analog in the UV: (\OIII~$\lambda 1666$+\OIII~$\lambda 1660)/($\OII~$\lambda 2470.22$+\OII~$\lambda 2470.34$).  The UV \OIIIOII\ ratio depends strongly on the metallicity, and is stable to variations in ISM pressure between $4<\log(P/k)<7$.  This diagnostic should not be used at large pressures and low metallicities; for $\log(P/k)>7$, the \OIIIOII\ ratio can vary by up to 0.5 dex with pressure when the metallicity is \OH$\leq 8.23$.  Note that in the low density limit ($n_e \lesssim 20 \,{\rm cm}^{-3}$), the \OII~$\lambda \lambda 3727,9$ line can be used with the \OII~$\lambda 2470$ doublet as an electron temperature diagnostic \citep{Nicholls18}.

The Al32 ratio, (\AlIII~$\lambda 1856$+\AlIII~$\lambda 1862$)/\AlII$\lambda 1670$, is an ideal ionization parameter diagnostic, if the spectrum can be corrected for extinction (Figure~\ref{ionization_diags}c).  This ratio is relatively insensitive to metallicity and is stable for all ISM pressures spanned by our models ($4<\log(P/k)<9$) for metallicities \OH$\leq8.93$. The Al32 ratio is usually weaker than the other ionization-sensitive UV lines, and may not be observed in high redshift galaxies, except at the highest ionization parameters.   



\subsection{Optical Ionization Parameter Diagnostics}

The optical spectrum contains two strong-line ionization parameter diagnostics: O32, based on the \OIII~$\lambda 5007$/\OII~$\lambda \lambda 3727,29$ ratio, and S32, based on the (\SIII~$\lambda 9069$+\SIII~$\lambda 9531$)/\SII~$\lambda \lambda 6717,31$ ratio.   

The O32 ratio was first proposed by \citet{Aller42} as an excitation diagnostic.  This ratio was calibrated theoretically by \citet{Kewley02,Kobulnicky04} using a combination of stellar evolution and photoionization models.   Recent theoretical calibrations were made by \citet{Dors11} using single temperature photoionization models and \citet{Morisset16} using hybrid photoionization models with fits to optical line ratios of a large sample of \HII\ regions.  
The O32 ratio has a strong dependence on the the gas-phase metallicity, which led \citet{Kewley02} to recommend an iterative approach to solve for both metallicity and ionization parameter.  Figure~\ref{ionization_diags}d shows that in addition to metallicity, the O32 ratio is strongly influenced by the ISM pressure in metal-rich galaxies (\OH$>9.0$) at pressures $\log(P/k)<6$ due to collisional de-excitation of \OII.

The S32 ratio was proposed by \citet{Kewley02}, but has been difficult to measure in nearby galaxies because the far-red \SIII\ lines lie at the edge of optical bandpasses in a region of significant sky emission.  More recent S32 calibrations have been made by \citet{Dors11} and \citet{Morisset16}.  Figure~\ref{ionization_diags}e shows that the S32 ratio is very sensitive to ionization parameter, varying by three orders of magnitude across the full ionization parameter range of our models, with little variation with metallicity (varying only by 0.3 dex between $7.63 < $\OH$<8.93$).  The S32 ratio is also insensitive to the ISM pressure between between $4<\log(P/k)<7$. 

An important caveat with the use of the S32 ratio is that the \SII\ lines have been underestimated by photoionization models in the past, especially at low metallicities.  This underestimation has been attributed to the stellar evolution models producing a radiation field that is too soft, i.e. too few ionizing photons \citep{Kewley01a,Levesque10}, or the poorly known dielectronic recombination coefficients of sulfur \citep[e.g.,][]{Izotov09}. \citet{Dors11} cite the use of single temperature photoionization models as the culprit, but photoionization models with complex temperature structures also underestimate the \SII\ line strength \citep{Levesque10}. This issue is discussed further in Section~\ref{Models}.  Until photoionization models can reproduce the \SII\ line strengths across the full range of metallicities observed in \HII\ regions, the S32 ratio should be used with caution.

\subsection{Infrared Ionization Parameter Diagnostics}

Infrared ionization parameter diagnostics are less affected by extinction than optical or UV diagnostics and are now accessible using sensitive infrared and sub-mm instruments.  For example, \citet{Yeh12} investigate the use of mid-infrared ionization parameters to diagnose radiation and wind pressures in star-forming regions.  

The near-infrared contains the S32$_{\rm NIR}$ ratio of \SIII~$\lambda 9530$/\SII~$1\mu$m where the \SII~$1\mu$m line is a complex of the \SII\ lines at 1.029, 1.032, 1.034, 1.037$\mu$m lines.  The \SII~$1\mu$m complex is very weak ($0.02 - 0.00005 \times$\Hb), but it has been observed in low metallicity nearby galaxies \citep{Izotov09}.  The S32$_{\rm NIR}$ ratio has a complex relationship with pressure and metallicity, and is least sensitive to these properties for metallicities \OH$\leq  8.93$ and for pressures $4<\log(P/k)< 7$.   The S32$_{\rm NIR}$ ratio becomes large in high pressure environments because the \SII\ lines are collisionally de-excited under these conditions.  

The infrared \SIV~$10.51 \mu$m/\SIII~$18.71 \mu$m ratio was proposed by \citet{Yeh12}, who provide a diagram that calibrates \SIVSIII\ in terms of both ionization parameter and hydrogen density.  Figure~\ref{ionization_diags}h shows that the \SIVSIII\ ratio is a sensitive function of ionization parameter for all metallicities.  This ratio is insensitive to ISM pressure except at the highest pressure and metallicity (\OH$>9.0$ and $\log(P/k) = 9.0$).   The close spacing of the \SIV~$10.51 \mu$m and \SIII~$18.71 \mu$m lines allows the ratio to be used even in spectra with limited wavelength coverage, but the \SIV\ line is weak ($0.0001 \times$\Hb) and likely to be difficult to observe.

In the mid-infrared, the Ne32 ratio of \NeIII~$15\mu$m/\NeII~$12\mu$m is a useful diagnostic of the ionization parameter, as long as the metallicity is known (Figure~\ref{ionization_diags}f).  Although Neon is a noble gas, the collisionally-excited \NeIII\ and \NeII\ emission-lines are sensitive to the electron temperature of the gas (and therefore the metallicity), especially at high metallicities (\OH$>8.53$).  \citet{Thornley00} first calibrated the Ne32 ratio using starburst models that include time evolution. Later, \citet{Yeh12} calibrated the Ne32 ratio for ionization parameter and hydrogen density, showing that the N32 ratio is insensitive to the hydrogen density except at the highest densities.  Figure~\ref{ionization_diags}f indicates that 
the N32 ratio is insensitive to ISM pressure, except at the highest metallicities (\OH$>8.93$).

The far-infrared N32 ratio (\NIII~$57\mu$m/\NII~$122\mu$m) is less sensitive to metallicity but is more affected by the ISM pressure at high metallicities (\OH$>8.53$) than either the \SIV/\SIII\ or \NeIII/\NeII\ ratios (Figure~\ref{ionization_diags}h).

\subsection{Mixed line ratio diagnostics}

Mixed line ratios are primarily used when lines from two ionization states of the same species are unavailable in a spectrum.  This situation often occurs at redshift $z\sim 0.7$ where only part of the rest-frame blue spectrum is available in the optical.  Mixed ratios usually have a larger dependence on the metallicity than ratios of a single species. Mixed ratios should only be used if the gas-phase metallicity is known and is taken into account in the ionization parameter diagnostic.  

The optical \NeIII~$\lambda 3968$/\OII~$\lambda \lambda 3727,9$ ratio, was first proposed by \citet{Levesque10} for this situation.  This ratio is very sensitive to the ISM pressure and metallicity, and should only be used to determine ionization parameter if both the ISM pressure and metallicity can be reliably estimated.  

The \OIII~$\lambda 5007/$\Hb\ ratio is correlated with ionization parameter for metallicities \OH$<8.53$  (\OIIIHb$<0.5$). In this range the \OIIIHb\ ratio varies less than $0.4$~dex with metallicity, and can be used to provide a rough estimate of the ionization parameter.  The \OIIIHb\ should not be used to measure ionization parameter in super star clusters or luminous infrared galaxies because the \OIIIHb\ ratio becomes insensitive to ionization parameter at $\log(U)>-2.63$.

The infrared \OIII~$52\mu$m/\SIII~$33\mu$m ratio was used by \citet{Cotera05} to trace the ionization parameter in the Arches cluster.  Like \OIIIHb, this ratio is a strong function of ionization parameter but becomes insensitive to ionization parameter at larger ionization parameters ($\log(U)>-2.63$) \citep{Kewley18b}.  The \OIII/\SIII\ ratio is insensitive to ISM pressure variations between $4<\log(P/k)<6$.

\section{Metallicity in Galaxy Evolution Studies}

Measuring the chemical history of galaxies is critical to our understanding of galaxy formation and evolution. 
In a simple closed-box model of chemical evolution, metals increase over time through each generation of star formation.  
However, the closed-box model cannot explain the complex relationship between the chemical abundances and the luminosity or stellar mass of galaxies \citep{Tremonti04,Dalcanton07,Finlator07}, nor the evolution of these properties with time \citep{Kobulnicky04,Yuan13,Zahid14}.   

The metallicity history of galaxies is usually characterized as a function of stellar mass.  Theory predicts that as time progresses, the mean metallicity of galaxies increases with age as galaxies undergo chemical enrichment, while the stellar mass of a galaxy will increase with time as galaxies are built through merging and other accretion processes \citep[e.g.,][and references therein]{Somerville99,Nagamine01,DeLucia04,Dave11b}.  

The gas-phase metallicity strongly correlates with the stellar mass, and the galaxy bolometric luminosity.  A correlation between mass and metallicity naturally arises if low mass galaxies have larger gas fractions than higher mass galaxies, as observed in local galaxies \citep{McGaugh97,Bell00,Boselli01}, and recently produced in simulations \citep{Hunt16,Ma16,deRossi17}.  The detailed relationship between metallicity and mass depends on accretion of material from the IGM, as well as galactic-scale outflows driven by supernovae, stellar winds, or AGN \citep{Wild08}.  

The mass-metallicity (MZ) and luminosity-metallicity (LZ) relations were observed first in irregular and blue compact galaxies \citep{Lequeux79,Kinman81}, and later in disk galaxies  \citep[e.g.,][]{Rubin84,Wyse85,Vila92,Garnett02}.   The local mass-metallicity relation is now well-known, thanks to large local spectroscopic surveys such as the Sloan Digital Sky Survey (SDSS) and the 2 degree Field Galaxy Redshift Survey (2dFGRS) \citep[e.g.,][and references therein]{Baldry02,Tremonti04,Wu16}.   The local MZ relation is steep for masses  $\lesssim 10^{10.5}$~\Msun\ and flattens at higher stellar masses.    These characteristics have been interpreted in terms of efficient galactic scale winds that remove metals from low mass galaxies ($M \lesssim 10^{10.5}$~\Msun), and a saturation in the chemical yield \citep{Pettini02,Tremonti04,Zahid14a}.  Alternative scenarios include low star formation efficiencies in low-mass galaxies caused by supernova feedback \citep{Brooks07}, and a variable integrated stellar initial mass function \citep{Koppen07}.   

Most investigations into the metallicity history of star-forming galaxies focus on the change in the mass-metallicity relation with  redshift.  The mass-metallicity relation (or the related luminosity-metallicity relation) has now been measured to $z>3$ \citep[e.g.,][]{Shapley04,Savaglio05,Erb06,Maier06,Liang06,Zahid11,Zahid13,Yuan13,Ly16}.  
Some authors have reported a tight correlation between gas-phase metallicity, stellar mass, and star formation rate that does not evolve with redshift \citep{Lara-Lopez10,Mannucci10,Alaina13b}.  However, this relationship, its tightness, and observed evolution appear to be affected by systematic uncertainties in the derived metallicity, stellar mass, and star formation rates \citep{Yates12,Andrews13,Sanchez13,Ly14,Cullen14,Wuyts14,Maier14,Salim14,Kashino16}.  The relationship is also affected by environment \citep{Scudder12}, and may be caused by the more fundamental relation between gas mass, stellar mass, and star formation rate \citep{Lilly13,Zahid14a,Troncoso14,Ma16}.

Within galaxies, metallicity gradients vary depending on galaxy type, mass, and environment.  Isolated spiral galaxies like our Milky Way typically have steep metallicity gradients, consistent with inside-out disk formation, where early nuclear star formation enriches the central gas with metals \citep[e.g.,][]{Chiappini01}.  The gradient subsequently builds up over time as accretion from the surrounding IGM and satellites causes star formation to form further and further out in the disk \citep[e.g.,][]{Fu09,Kobayashi11,Cunha16}.  When normalized by galaxy size, isolated galaxies have very similar gradients, suggesting a common star formation and gas accretion history \citep{Sanchez14,Ho15}. Low mass galaxies have flatter, or even inverted gradients because the crossing timescale is short, allowing efficient mixing of the gas \citep[][and references therein]{Carton18,Wang18}

The gas-phase metallicity gradient is extremely sensitive to environment and gas infall.  Galaxy mergers drive pristine gas from the outer regions into the central regions, flattening metallicity gradients \citep{Kewley10,Rupke10,Rosa14,Torres-Flores14}. 
N-body merger models indicate that major gas infall occurs at two stages during the merger; at first close passage, and during final coalescence, both of which flatten metallicity gradients \citep{Perez11,Torrey12}.  Cosmological simulations support this scenario \citep{Bustamante18}. Late stage mergers have extremely flat metallicity distributions \citep{Rich12}.  If galactic-scale winds quench star formation at late merger stages, the post-merger galaxy may not ever fully recover a steep metallicity gradient.  Thus, the combination of observed metallicity gradients and N-body merger simulations has the potential to age date mergers based on recent gas infall, and to constrain the amount of star-formation quenching at late merger stages.

The gas-phase metallicity can be determined from a wide variety of emission-lines from the UV through to the IR.  In \HII\ regions, the abundance can sometimes be determined from recombination lines.  Recombination lines from Hydrogen and Helium are readily observed in spectra of \HII\ regions and galaxies, but the recombination lines of other elements, such as oxygen and carbon are weak and are only measurable in a few nearby regions.  Instead, three other techniques are commonly used.  The traditional method for deriving metallicity, known as the ``Direct Method" is based on the electron temperature from sensitive Auroral lines in the optical spectrum.  The Auroral lines are weak and are often not detected in the metal-rich galaxies.  To overcome this problem, ``Empirical" calibrations were created, providing relations between Auroral line metallicities and strong line flux ratios.  However, both methods rely on samples of \HII\ regions, simplified assumptions, and correction for unseen stages of ionization.  To overcome these issues, theoretical methods based on photoionization models were developed.  Each of these methods has its own strengths and weaknesses, and massive discrepancies exist amongst all of these methods \citep[see][for an overview]{Stasinska05,Kewley08,Peimbert17}.  Here, we discuss each method and associated caveats.  

\subsection{Auroral Line Metallicities}\label{Direct_method}

The use of temperature-sensitive Auroral lines to measure metallicities is commonly called the "Direct" method.  This is a misleading term because there are many steps and assumptions in the derivation of metallicity from an Auroral line \citep[see e.g.,][]{Dinerstein90,Lopez12}.  A particularly useful tutorial is given by \citet{PerezMontero17}.  An ionic abundance is calculated with a simple photoionization model of a five-level atom, assuming a homogeneous (constant density, constant temperature) medium.  The models in \citet{Aller84}, or {\it Temden} in {\it IRAF} \citep{deRobertis87,Shaw95} are commonly used, along with atomic data from \citet{Mendoza83}.  The atomic model is used in conjunction with the ratio of an Auroral line to a nearby line of the same species to calculate the electron temperature, $T_{e}$, of that ionic zone.  

The main Auroral line used to determine metallicities is the \OIII~$\lambda 4363$ line.
Both the \OIII~$\lambda 4363$ line and the nearby \OIII~$\lambda \lambda 4959, 5007$ lines are used to derive electron temperature in the O$^{2+}$ zone, $T_{e}($\OIII).   $T_{e}($\OIII) does not represent the average electron temperature, so the temperature in the unseen zones also need to be estimated.  Two ionization zones are usually assumed; a high-ionization (H$^{+}$+ He$^{+}$) zone and a low ionization (H$^{+}$+He$^{0}$) zone.  Some authors use average electron temperatures of several atoms to estimate the electron temperature in each zone \citep[e.g.,][]{Lopez12}.  Usually, the temperature in the O$^{+}$ zone, $T_{e}($\OII), is calculated based on a fit to the relation between $T_{e}($\OII) and $T_{e}($\OIII) derived from photoionization models.  An electron density needs to be included, which is calculated with the \OII\ or \SII\ doublets based on photoionization models of simple atoms.  The temperatures are then used to calculate ionic abundances, which are the abundances of each ion (i.e. O$^{+}$ and O$^{2+}$).  

To calculate the total oxygen abundance relative to hydrogen, the ionic abundances need to be corrected for the unseen stages of ionization.  Usually, an ionization correction factor (ICF) is used.  The ICF is based on photoionization models, sometimes with inputs from observed emission-lines, such as \OII~$\lambda 3727$ or \HeII~$\lambda 4686$ \citep{Izotov99}.   Models fits to sets of emission-line fluxes can be used to derive the ICF \citep[e.g.,][]{Stasinska80,Mathis91}.  These models assume a hydrogen density, the gas distribution, and the fraction of ionizing photons that are absorbed by the nebula.  Dust modifies the radiation field absorbed into the nebula, and the stellar radiation field needs to be accurately estimated using stellar atmosphere models.  \citet{Mathis85} gives a comprehensive and useful summary of the computation method and issues inherent in estimating ICFs, as well as tables of ICFs for various species. \citet{Lopez12} provide a fit to the relationship between oxygen abundance and electron temperature.  

Significant errors can be introduced into metallicity estimates through the ICF.  The calculation of an ICF usually assumes that the electron temperature is constant and that the \HII\ region is composed of atoms of only one, or a small number of species.  However, the electron temperature can differ by 2000-3000K from one \HII\ region to another \citep{Hagele06}, and we know that \HII\ regions are multi-zone ionized regions, composed of atoms of many species.  The ionization of elements depends strongly on the stellar radiation field, which is reliant on our understanding of stellar atmospheres.  \citet{Hagele08} show that an accurate \Te\ estimate requires the measurement of multiple faint Auroral lines (\OII, \SII, \OIII, \SIII, and \NII) and their accurate (better than 5 per cent) ratios to Balmer recombination lines.  Without these lines, the assumed ICF can differ significantly from the actual unseen stages of ionization  in low excitation \HII\ regions, causing the \Te\ metallicity to be underestimated by up to 0.2~dex \citep{Hagele08}.  

 In a chemically homogeneous medium, \Te\ lines originate from regions that are low density and metal-poor.  This bias is caused by the fact that the \Te\ line is predominantly produced in hot gas \citep{Dinerstein90}.  At the low temperatures typical of metal-rich regions, collisional excitation of the \OIII\ lines is small.  Therefore, integrated emission of the \OIII~$\lambda 4363$ line is weighted towards the hottest zones of an \HII\ region.   The standard Auroral \OIII~$\lambda 4363$ line is therefore rarely observed at metallicities \OH$>8.7$ \citep{Peimbert67,Rubin69,Stasinska05}, and is highly sensitive to temperature gradients and stratification \citep{Peimbert69}.  Multiple mechanisms for generating electron temperature fluctuations have been proposed, including shock waves and turbulence\citep{Peimbert91,O'Dell15}, mechanical energy from stellar winds produced by planetary nebulae \citep{Peimbert95} or Wolf-Rayet Stars \citep{Gonzalez94}.

This temperature effect causes a significant sample bias.  \citet{Hoyos06} analysed a large sample of \HII\ regions and analysed the differences between galaxies with and without detections of the \OIII~$\lambda 4363$ line.  They showed that \HII\ regions and galaxies without Auroral line detections are more metal-rich, more luminous, and have a lower ionization parameter than those where the Auroral line is detected.  Theoretical models suggest that temperature gradients can cause the \Te\ metallicity to be systematically underestimated by up to $0.9$~dex at high metallicity (\OH$\sim 9.0$) \citep{Stasinska05}.
  
Most Auroral methods assume a constant density throughout the nebula.  \citet{Copetti00} found that 50\% of their sample of galactic \HII\ regions have internal variations in electron density, while 60\% of giant extragalactic \HII\ regions have internal electron density variations \citep{Malmann02}.   \citet{Luridiana03} discusses the effect of density gradients on abundances calculated using \Te\ methods with various ionization correction factors.  They find that the \Te\ metallicity from oxygen is relatively insensitive to density gradients, as long as all of the different ionization stages of oxygen have been observed.  However, they suggest that nitrogen and sulphur abundances may become unreliable where large density gradients exist.

Recently, \citet{Nicholls18} used photoionization models to re-calibrate the relationship between metallicity and the \OIII~$\lambda \lambda 1660,66$, \OIII~$\lambda 2321$, and \OIII~$\lambda 4363$ Auroral lines.  These re-calibrations include the latest collisional, radiative excitation/de-excitation, and collisional cross-section data, as well as full cascade from higher energy levels.   Nicholls et al. point out that Auroral line calibrations give an {\it atomic equivalent temperature}, which may be used to estimate an average temperature and oxygen abundance, but do not take into account the heterogeneous structure of emission-line nebulae.

\begin{textbox}[b]
It is preferable to use \OH\ to refer to the gas-phase metallicity, rather than units in terms of solar, because solar abundances are not representative of local present day galactic abundances.  \citet{Asplund09} gives a bulk solar abundance of \OH$=8.72$ and a surface solar abundance of \OH$=8.69$.  The solar surface abundance shows evidence of evolution and mixing, and the solar bulk abundance is an estimate of the abundance when the Sun formed, 4.7 billion years ago. Local B stars (\OH$ = 8.76$) may be more representative of local galactic abundances \citep[see][for a discussion]{Nicholls17}. 
\end{textbox}

\begin{marginnote}[120pt]
\entry{Atomic Equivalent Temperature}{The electron temperature of a dust free, isolated assembly of oxygen atoms at a uniform electron density and a single electron temperature that generate the same emission line flux ratios as observed data.}
\end{marginnote}

\subsection{Auroral-Strong Line Calibrations}

Metallicity calibrations based on optical strong-line ratios were first developed for measuring metallicities in \HII\ regions where the Auroral lines are not observed.  These calibrations are derived by fitting the observed relationship between Auroral metallicities and strong-line ratios, and are often referred to as ``empirical methods".  The use of the term ``empirical" is misleading because it implies that the metallicities are directly observed, and neglects the theoretical atomic models and assumptions that the Auroral method is based on.  

\begin{figure*}[!t]  
\includegraphics[width=\linewidth]{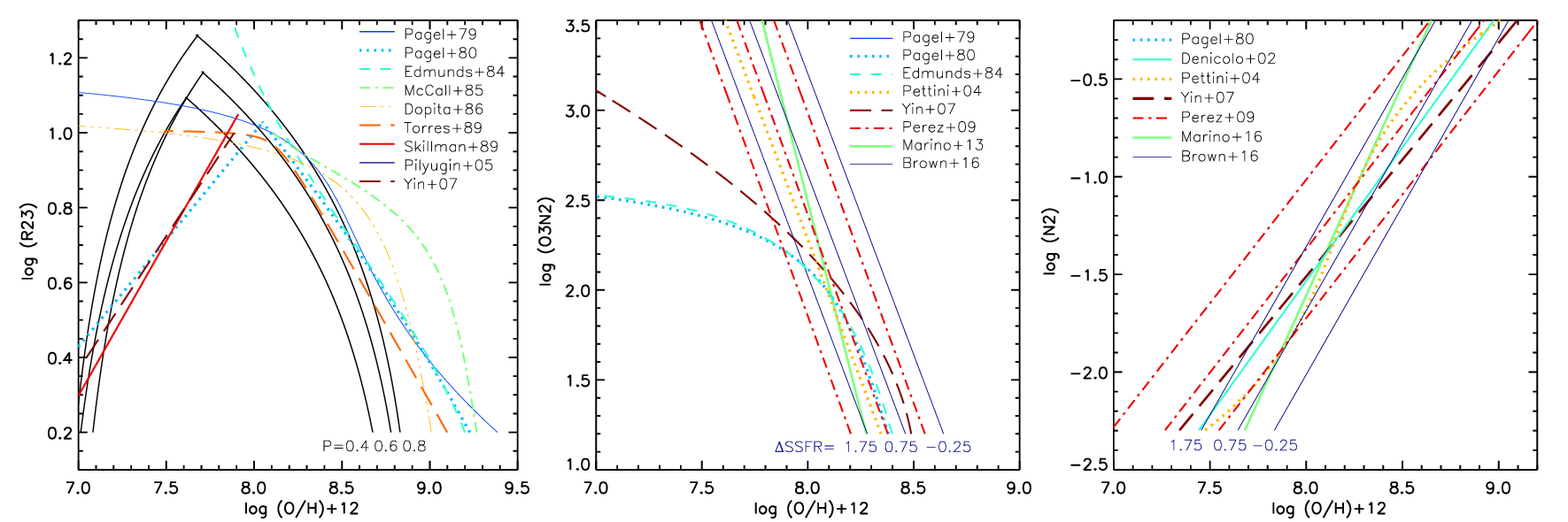}
\caption[Kewley_figure8.eps]{Empirical calibrations based on fits to \Te\ metallicities for samples of \HII\ regions or galaxies.  Empirical calibrations for the ratios of \R23\ (top), \OIIIHb$/$\NIIHa (middle), and \NIIHa\ (bottom) are shown.  Some calibrations include corrections for the ionization state of the gas which is represented by a parameter $P$ in the \citet{Pilyugin05} method, and an offset in specific star formation rate, $\Delta SSFR$ in the \citet{Brown16} method.  Large discrepancies exist amongst the many empirical calibrations.}
\label{Te_conversion_comparison}
\end{figure*}

Many Auroral-strong-line calibrations exist in the literature.  Figure~\ref{Te_conversion_comparison} compares the Auroral-strong-line calibrations for three commonly used metallicity-sensitive line ratios: \R23, \OIIIHb$/$\NIIHa\ (known as O3N2), and \NIIHa.  These calibrations were derived by the original authors using simple curve fits to Auroral metallicities of either \HII\ region spectra or integrated spectra of galaxies.   

The differences among the Auroral-strong-line calibrations is dramatic; estimates vary up to 1 dex in metallicity.  The earlier calibrations were based on very small numbers of \HII\ regions which led to unusual shapes in the lower metallicity ends of the \R23\ and \NIIHa\ calibrations.  For calibrations developed after 1989, the discrepancies are up to 0.6~dex in \R23, 0.5~dex in O3N2, and more than 0.9~dex in \NIIHa.  The magnitude of these discrepancies has not improved in the past decade, despite significantly larger samples of \HII\ regions being used.  The causes of these discrepancies are difficult to dissect, and could be a product of many effects including all of the potential limitations with the Auroral method described in Section~\ref{Direct_method}, as well as sample bias.

Sample bias enters from at least two assumptions: (1) that \HII\ regions will always show the Auroral lines if observed for a sufficient amount of time, and (2) that \HII\ regions with Auroral lines are representative of all \HII\ regions and galaxies.  In the past, the number of \HII\ regions with measurable \Te\ abundances was small, and authors often combined \Te\ observations of \HII\ regions from different galaxies (including galaxies of different types), and \HII\ regions observed with different methods.  In the 1970s-1990s, some \HII\ region samples were observed with objective prism surveys, which are known to be biased towards strong \OIII~$\lambda 4959, 5007$.  Recently, integral field spectroscopy has made it possible to derive  \Te\ metallicities for large numbers of galaxies \citep{Marino13,Curti17}, which allows sample bias to be understood and largely overcome.

Recently, stacking has been used on large galaxy samples to increase the S/N of the Auroral lines in galaxies \citep{Liang07,Andrews13,Brown16}.  \citet{Andrews13} show that Auroral metallicities have a factor of 2-3 times larger dependence on the star formation rate than strong-line metallicity methods, while \citet{Brown16} show that the \HII\ regions and stacked data used in Auroral line samples and stacked galaxy data do not span the full range of line ratios seen in the full sample of individual SDSS galaxies.  These studies emphasize the need to analyse all of the \HII\ regions in samples, {\it both with and without} Auroral line detections.  A simple exercise to test for sample bias is to compare the strong-line ratio distributions (i.e. \OIIIOII, \NIIHa, \R23) of the \HII\ regions with and without Auroral line detections.  An Auroral line sub-sample is only representative of all \HII\ regions if it can span the full range of line ratios seen in all \HII\ regions. 
       
Some empirical calibrations attempt to overcome the sample biases of the empirical methods by including theoretical model fits to individual \HII\ region data at the high metallicity end \citep[e.g.,][]{Pettini04,Maiolino08}.  One of the first metallicity calibrations, by \citet{Pagel79}, is based on \HII\ regions with Auroral metallicities at low abundances, and on photoionization model fits to a metal-rich \HII\ region in the galaxy M101.  Such model fits rely on assumptions about the stellar radiation field within the \HII\ region, and the resulting calibrations are still dependent on the choice of \HII\ regions across the entire metallicity range. These mixed calibrations typically give metallicities that are in-between Auroral metallicities and pure photoionization methods.  

Depletion of oxygen onto dust grains has traditionally not been taken into account in Auroral line methods.  Dust depletion measurements in nearby \HII\ regions require a 0.08 - 0.11 dex correction to the oxygen abundance measurements \citep{Mesa10,Peimbert10}.  \citet{Peimbert10} recommend dust corrections of +0.11 dex to the measured \OH\ for \HII\ regions with \OH$>8.3$, +0.10 dex for $8.3 > $\OH$ > 7.8$, and +0.09 dex for \HII\ regions with $7.8 > $\OH.

\subsection{Recombination Line Methods}

Recombination line fluxes are almost all proportional to the electron density and inversely proportional to the electron temperature.  These relationships hold across an order of magnitude change in electron temperature, and for densities up to $10^{10} {\rm cm}^{-3}$.  As a result, an error in the measured electron temperature or density has negligible effect on the resulting metallicity estimates, and recombination lines are often used as a "gold standard" for comparison to other, more problematic metallicity estimation techniques.  \citet[][]{Peimbert17} gives an excellent tutorial and review of the use of recombination lines and Auroral lines to determine metallicities.  Aside from the helium lines, the \CII~$\lambda 4267$ and \OII~$\lambda 4650$ doublets are most commonly used to determine metallicities.  

The main disadvantage of using recombination lines for deriving metallicities is that aside from hydrogen and helium, the recombination lines are very weak, and are only seen in nearby \HII\ regions and planetary nebulae.  In these regions, the recombination lines have been successfully used to measure carbon and oxygen abundances in the Milky Way and in some nearby extragalactic \HII\ regions \citep[][and references therein]{Peimbert13}.
Recombination lines have been used to determine the steep metallicity gradient in our Milky Way \citep{Esteban05}, as well as a flattening in the gradient at the outskirts of the Milky Way \citep{Esteban13}.

In \HII\ regions, recombination lines consistently give larger metallicities than Auroral methods \citep{French83,Mathis99,Peimbert93}.   The electron temperatures of recombination lines are larger by 
$16.1\pm0.01$\% than the electron temperatures of \HII\ regions measured with the Auroral lines \citep[e.g.,][]{Peimbert13}.  The scatter in this offset for \HII\ regions is remarkably small.  For planetary nebulae, the offset is in the same direction, and is larger on average ($27\pm0.03$\%).   The logarithm of the difference between the recombination line and Auroral metallicities is known as the Abundance Discrepancy Factor (ADF).  The ADF is around $\sim 0.2$~dex in many Milky Way and extragalactic \HII\ regions \citep{Garcia05,Garcia-Rojas07,Esteban09,Garcia-Rojas14}, but is as large as 0.59~dex in some cases \citep{Liu00,Liu06,Tsamis04,Tsamis08,Mesa10}.  The ADF for C/O and N/O abundances appear to be significantly smaller \citep{Zuckerman86}. 

\citet{Peimbert67} proposed that electron temperature fluctuations in \HII\ regions cause the abundance discrepancy factor.  Auroral lines are significantly more sensitive to the electron temperature of the gas than recombination lines; Auroral lines depend exponentially on a single assumed electron temperature, whereas recombination methods scale with electron temperature through a power law.   \citet{Peimbert13} show that the pressure within the regions where the Auroral lines are produced is larger than the pressure within the recombination line regions, suggesting that the Auroral lines are produced in higher pressure, higher temperature regions.  These high pressure regions may result from the dissipation of turbulent energy in shocks, or from magnetic reconnection.  

Electron temperature correction methods assume that the ADF is entirely due to electron temperature discrepancies, and that the interstellar medium in chemically homogeneous.  However, \citet{Tsamis05} argue that small scale chemical inhomogeneities from incomplete small-scale mixing are responsible for the abundance discrepancy.  They produced tailored photoionization models to the UV, optical, IR and radio spectra of 30 Doradus.  Their models reproduce the recombination lines and nebular lines only if chemical inhomogeneities exist in the nebula. 

High S/N spatially-resolved spectra of individual \HII\ regions can quantify dust depletion, chemical inhomogeneities and temperature fluctuations, which can then be taken into account in tailored photoionization modelling.  When the thermal and ionization structure of the nebula and dust depletion is taken into account, the systematic temperature differences from different methods can be resolved in some cases \citep{Pena12}.  

Recently, \citet{Nicholls12} proposed that there may be departures from thermal equilibrium in the
electron energies, described by a parameter $\kappa$. This distribution is capable of
explaining the discrepancy in a physically plausible way. The equilibrium Maxwell-Boltzmann distribution is the limiting
case of a $\kappa$ distribution where $\kappa$ = infinity. \citet{Livadiotis18} demonstrated that $\kappa$ distributions can arise 
naturally in astrophysical plasmas.  However, \citet{Draine18} argue that $\kappa$ distributions do not arise in \HII\ regions or planetary nebulae.  

The full cause of the ADF remains unknown. Metallicity calibration discrepancies can mimic or hide metallicity evolution with redshift, and may significantly change metallicity gradient estimates.  Until the abundance discrepancy is conclusively resolved, we recommend (1) the application of a consistent calibration or calibrations across all samples, or all \HII\ regions being studied, and (2) the use of several metallicity calibrations to quantify the effect of different metallicity calibrations on the scientific conclusions.

\subsection{Theoretical Metallicity Diagnostics}  \label{theoretical_metal_calibs}

Purely theoretical metallicity calibrations were developed to overcome the systematic differences among the Auroral-strong-line calibrations as well as the discrepancy between the Auroral and recombination line metallicities.  Theoretical calibrations are calculated by combining stellar population synthesis and photoionization models. \citet{McGaugh91} created the first theoretical metallicity calibrations using stellar evolution tracks to calculate the ionizing radiation field corresponding to a zero-age burst.  He calculated theoretical calibrations between line ratios and metallicity that take into account the relationship between radiation field and temperature.  Theoretical calibrations were subsequently computed with updated photoionization models \citep[e.g.,][and references therein]{Charlot01,Kewley02,Kobulnicky04,Dors11,Nagao11,Morisset16,Byler18}.  

One of the major benefits of theoretical calibrations is that a large range of parameter space can be explored, allowing one to understand the effects of galaxy properties.  In this Section, we illustrate the effect of ionization parameter and ISM pressure on the diagnostic line ratios, and provide new calibrations that take these parameters into account, where needed.  The main disadvantages of theoretical calibrations are the model assumptions, which include a plane parallel or spherical geometry, uncertain stellar atmosphere models, reliance on current atomic data, and (sometimes) a simplified temperature and density structure.  We discuss these further in Sections~\ref{Models} and \ref{Challenges}.

Bayesian methods have recently been developed that use theoretical models to fit the observed nebular emission-lines  \citep{Brinchmann04,Tremonti04,Blanc15,Chevallard16,Thomas18}.  These methods are primarily used to derive metallicity, but may also be used to calculate the ISM pressure, ionization parameter, and other galaxy properties \citep[e.g.,][]{Thomas16}.  With Bayesian methods, the number of parameters being solved need to be equal to or less than the number of constraints provided by the emission-lines and continuum.  Bayesian methods are very powerful, but without priors, they don't take into account the regimes over which specific lines are not valid diagnostics.  For example, the \NII\ and \OII\ lines may be used in Bayesian methods as the main constraint of metallicity in the low metallicity regime where the \NIIOII\ ratio is in fact insensitive to metallicity.  An understanding of the sensitivity of the diagnostic lines to other galaxy parameters and the limitations of sets of emission lines is essential for interpreting the estimates derived from Bayesian methods.  The use of priors is recommended to ensure that emission-lines are only used when they provide useful estimates of the parameter of interest.

\begin{textbox}[h]
Students should first apply individual strong-line calibrations to gain an understanding of the diagnostics being used, prior to applying Bayesian methods.  Being based on the same models, Bayesian methods suffer from many of the same limitations as the individual calibrations, and unreliable estimates can be difficult to identify without a physical understanding of the limitations of each set of emission-lines.
\end{textbox}

\subsection{Theoretical UV Metallicity Diagnostics}

\begin{figure*}[!t]
\includegraphics[width=12cm]{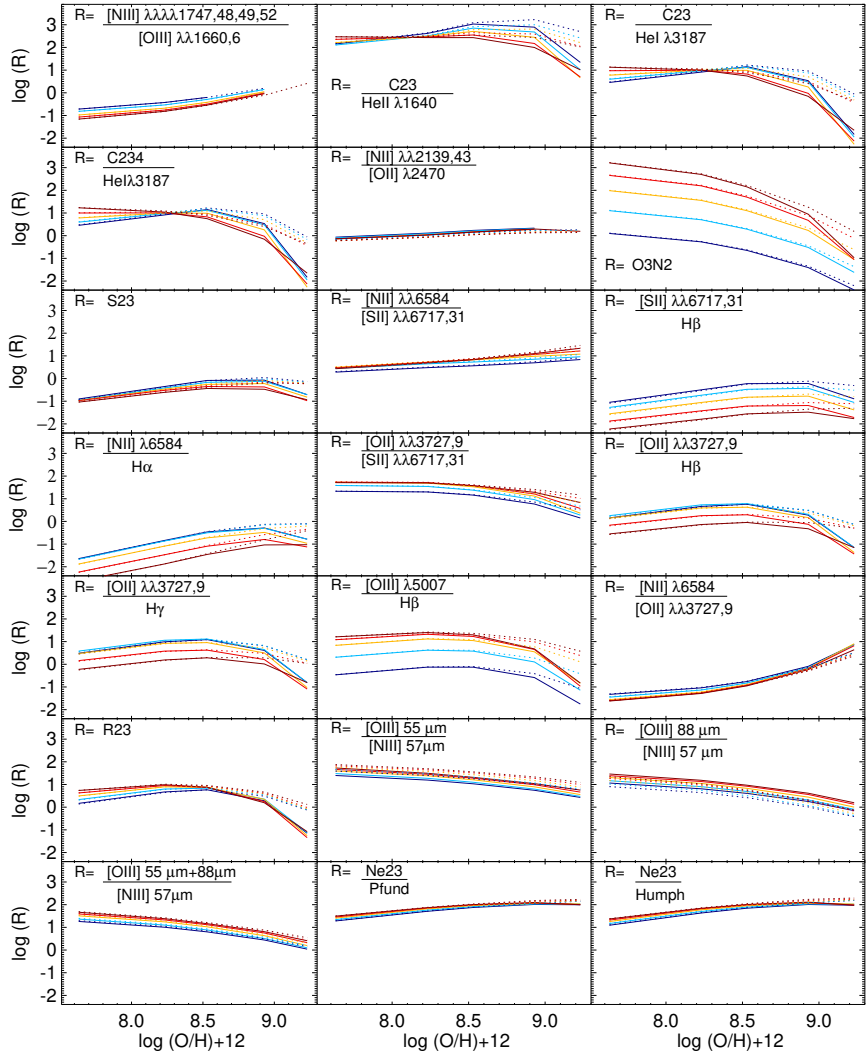}   
\caption[Kewley_figure9.eps]{UV-optical-IR metallicity diagnostic ratios versus metallicity in units of log(O/H)+12. Colors blue through red correspond to ionization parameter $\log(U)=[-3.98,-3.73,-3.48,-3.23,-2.98,-2.73,-2.48,-2.23,-1.98]$, respectively.  Solid and dotted lines give ISM pressures of $\log({\rm P/k})=5$ and $\log({\rm P/k})=7$, respectively. }
\label{Metallicity_Diagnostics}
\end{figure*}

UV metallicity diagnostics are scarce and were originally developed to interpret IUE or COS data of \HII\ regions and small samples of galaxies.  Recently, new UV diagnostics have been developed that will be key tools for interpreting spectra on the next generation space and ground-based telescopes \citep{Vidal-Garcia17,Byler18}.  \citet{Byler18} combined Starburst99 stellar evolution models with Cloudy photoionization models to investigate the diagnostic potential of the UV lines, focusing on [\ion{C}{3}]~$\lambda 1907$, \ion{C}{3}]~$\lambda 1909$, \ion{O}{3}]~$\lambda \lambda 1661,66$, \ion{Si}{3}]~$\lambda \lambda 1883,92$, \ion{C}{4}~$\lambda \lambda 1548,1551$, \ion{N}{2}]~$\lambda \lambda 1750,52$ and \ion{Mg}{2}~$\lambda 2796$.  They provide a useful suite of emission-line and equivalent width diagnostic diagrams for the separation and identification of metallicity and ionization parameter, as well as a figure giving key diagnostic UV emission-lines and their observability with NIRSpec on the James Webb Space Telescope.

Figure~\ref{Metallicity_Diagnostics} shows the most promising UV metallicity diagnostics, where colors correspond to ionization parameter, and solid and dotted lines give ISM pressures of log(P/k)=5 and log(P/k)=7, respectively.  Table~\ref{Metallicity_Coeffs} gives the coefficient fits for the diagnostics in terms of the ionization parameter and ISM pressure.

Hydrogen lines are often used as baselines for metallicity diagnostics.  There are no useful Hydrogen lines in the UV, so we use the \HeI~$\lambda 3187$ and the \HeII~$\lambda 1640$ lines as proxies.  The \HeI~$\lambda 3187$ line is an ideal proxy because it is $10 \times$ stronger than \HeII\ in a typical spectrum, and it is less sensitive to uncertainties in the EUV spectrum through the stellar atmosphere models.  It is important to keep in mind that the ionization potential of the \HeII\ line (54.4 eV) is in a region of the EUV spectrum where stellar atmosphere uncertainties are large, and the number of ionizations into He$^{+}$ is very sensitive to the ionization parameter.  

The UV contains many carbon lines which may be used in metallicity diagnostics.  The C23/\HeI\ ratio is  useful as a metallicity diagnostic in metal rich regimes (\OH$> 8.5$), where the dependence on ionization parameter is significantly reduced.  The \CIII~$\lambda \lambda 1906,08$ lines are usually strong in star-forming galaxies and trace the C$^{++}$ zone, while the \CII~$\lambda 2325$ blend (i.e. the sum of 2323.50\AA, 2324.69\AA, 2325.40\AA, 2326.93\AA, and 2328.12\AA) traces the C$^{+}$ zone.  Ideally, the full suite of \CII, \CIII, and \CIV\ lines could be used to measure metallicity via the C234/\HeI\ ratio, but the \CIV\ line can be contaminated by line-driven winds from massive stars, affecting the ratio by 0.1 - 0.3 dex \citep{Byler18}.  The C23 ratios are double-valued with metallicity for most ionization parameters because (like all collisionally excited lines), at low metallicities, the nebula is hot, and the ionized carbon emission scales with carbon abundance.  At high metallicity, the low temperatures prevent much of the collisional excitations of the carbon ions.  In this domain, the mid-infrared fine structure lines contribute significantly to the cooling of the nebula.  Additional line ratios are needed to determine whether the C23 metallicity lies on the upper or lower branch.  The C23 ratio is strongly dependent on both ionization parameter and ISM pressure and C23 diagnostics need to include a correction for these factors.  However, at the large ionization parameters typical of high redshift galaxies, the C23/HeI and C324/HeI ratios provide useful monotonic diagnostics of metallicity that vary little with ionization parameter.

The Si23/\HeI\ and Si23/\HeII\ ratios using the \SiIII~$\lambda \lambda 1883,92$ and \SiII~$\lambda 1808$ lines are sensitive metallicity diagnostics, but should be used with caution because silicon can be heavily depleted out of gas phase and onto dust grains.  The amount of depletion varies from galaxy to galaxy, and silicon may be returned to the gas phase by grain destruction by supernova shocks, which affects Mg, Si, and Fe \citep[see eg.,][]{Jones00,Byler18b}.  If dust depletion is known, and similar to the depletion used in the photoionization models, the Si23 diagnostics can be used at metallicities \OH$>8.5$ where they depend little on ionization parameter.  
 
The ideal UV metallicity diagnostic is the \NIIOII\ ratio based on the \NII~$\lambda 2139,43$ and \OII~$\lambda 2470$ lines, as long as the $\log($\NIIOII) ratio can be determined within an accuracy of $\sim 0.05$~dex.
The \NII\ line has both a primary and a secondary nucleosynthetic origin \citep[e.g.,][]{Considere00}. In 
primary nucleosynthesis, nitrogen is created for the first time in the current generation of stars.  In secondary nucleosynthesis, nitrogen created through fusion processes using elements that pre-existed prior to the current generation of stars.  For most metallicities (\OH$>8.2$), secondary nucleosynthesis is the dominant production mechanism of nitrogen, making nitrogen highly sensitive to the overall metallicity of a galaxy.  When combined with the temperature-sensitive \OII\ line, the \NIIOII\ ratio is an extremely strong function of metallicity, and is unchanged for ISM pressures typical of star-forming galaxies ($4<\log(P/k)<6$).   Because \NII\ and \OII\ have similar ionization potentials, the \NIIOII\ ratio is not sensitive to ionization parameter, and it is relatively insensitive to the presence of a hard radiation field from an AGN or shocks \citep{Kewley08}.  

The UV \NIIIOIII\ ratio is a useful UV metallicity diagnostic, but with a larger dependence on the ionization parameter than \NIIOII\ due to the ionization potentials of N$^{++}$ and O$^{++}$.  This ratio uses the \NIII\ complex of 5 lines at 1747, 1748, 1749,1752, and 1754~\AA\ and the \OIII\ doublet at 1660, 1666\AA.  The \NIIIOIII\ ratio is insensitive to ISM pressure variations.

With limited UV wavelength coverage, many galaxies only have absorption-lines available in their spectrum.   Starburst galaxies produce strong interstellar absorption lines, stellar photospheric and wind lines, some of which are sensitive to the metallicity.  \citet{Rix04}, \citet{Leitherer11}, and \citet{Zetterlund15}  provide useful diagnostics for interpreting the absorption lines in a UV spectrum.

\subsection{Theoretical Optical Metallicity Diagnostics}

The most commonly used theoretical optical metallicity diagnostic is the oxygen \R23\ ratio (\R23$= $\{\OII$\lambda \lambda 3727,9 + $\OIII$\lambda \lambda 4959,5007\}/$\Hb), first proposed by \citet{Pagel79}.  
Many theoretical calibrations of this ratio now exist \citep[e.g.,][]{McGaugh91,Oey00,Kewley02,Kobulnicky04}.  \citet{Skillman89b} was first to show that the relationship between \R23\ and metallicity depends on ionization parameter.  Most subsequent theoretical calibrations included a correction for ionization parameter.   The \R23\ ratio is relatively insensitive to the ionization parameter at metallicities above \OH$>8.5$ because the \OIII\ and \OII\ lines have opposite dependencies on ionization parameter, which are roughly cancelled out.  The \R23\ ratio is sensitive to the ISM pressure for high metallicities (\OH$>8.5$), and should be used in conjunction with an ISM pressure diagnostic.

The \NIIOII\ ratio, based on the \NII$\lambda 6584$ and \OII$\lambda \lambda 3727,9$ lines, is by far the most reliable metallicity diagnostic in the optical, with no dependence on ionization parameter, and only marginal dependence on the ISM pressure for $4\leq \log(P/k) \leq8$. Several simple optical \NIIOII\ metallicity calibrations have been developed \citep{Jensen76,Dopita86,Kewley02}.  The \NIIOII\ ratio is highly sensitive to metallicity thanks to the primary and secondary nature of nitrogen, plus the temperature sensitivity of the \OII\ line.  The optical \NIIOII\ ratio is insensitive to ionization parameter, and depends on ISM pressure only at the highest metallicities (\OH$\geq 9.23$) and at the highest ISM pressures ($\log(P/k)>10^8$).  The \NIIOII\ ratio is also the least sensitive optical diagnostic to the presence of an AGN or diffuse ionized gas \citep{Kewley06b,Zhang17}.  

\citet{Pettini04} proposed a calibration called O3N2 that uses the \OIIIHb\ and \NIIHa\ ratios (O3N2$=$\OIII$\lambda 5007/$\Hb)/(\NII$\lambda 6584$/\Ha).  This calibration is based on a combination of Auroral metallicities and photoionization models, and was developed for spectra of high redshift galaxies where flux calibration and extinction correction is difficult.  Figure~\ref{Metallicity_Diagnostics} shows that the O3N2 ratio has a strong dependence on ionization parameter; varying by 2-3 orders of magnitude across the entire metallicity range. We do not recommend the use of the O3N2 ratio, especially at high redshift where the ionization parameter is usually significantly larger than in nearby \HII\ regions.  Instead, if only the red spectrum is available, we recommend the \NII$\lambda 6584/$Ha\ ratio, which has a smaller (although not negligible) ionization parameter dependence; the \NIIHa\ metallicity varies by $\sim 1$~dex with ionization parameter.  Several \NIIHa\ calibrations are available \citep{Storchi94,Denicolo02,Pettini04}.  These calibrations are primarily based on the relationship between Auroral metallicities and the \NIIHa\ line ratio in local \HII\ regions, and include a hidden assumption that the ionization parameter of the sample that the calibration is being applied to is the same as the mean ionization parameter of the \HII\ region sample that was used to derive the calibration.  For high redshift galaxies, we recommend the use of a \NIIHa\ calibration that includes a correction for ionization parameter, such as in Table~\ref{Metallicity_Coeffs}.

To overcome the ionization parameter problem, \citet[][; hereafter D16]{Dopita16} proposed a composite diagnostic based on the \NII$\lambda 6574/$\SII$\lambda \lambda 6717,31$ and \NIIHa\ ratios:
\begin{equation}
\log(O/H)+12=8.77+y + 0.45(y+0.3)^5
\end{equation}

where $y=\log$\NIISII$+0.264\log$\NIIHa.  The \NIISII\ ratio alone has a larger dependence on ionization parameter than the \NIIOII\ ratio, and is independent of ISM pressure for $4\leq \log(P/k) \leq7$, although it has been used in the past as a metallicity diagnostic \citep{Jensen76}.  However, when combined with the \NIIHa\ ratio, the D16 function becomes relatively insensitive to ionization parameter and ISM pressure over $4\leq \log(P/k)\leq 7$, and is a promising diagnostic for high redshift galaxies where only the red spectrum may be available.

\begin{textbox}[b] 
Nitrogen diagnostics are sensitive to the elemental N/O ratio.  Our models take into account the fact that the N/O ratio changes with metallicity due to the primary and secondary nature of nitrogen.  If the N/O ratio is expected to follow the standard primary and secondary curves, then these theoretical nitrogen metallicity diagnostics can be applied.  However, if the N/O ratio is expected to be elevated, then tailored models with a given input N/O ratio should be used instead.
\end{textbox}

The \SIIHa\ ratio (\SII$\lambda \lambda 6717,31$/\Ha) has been calibrated by \citet{Denicolo02} and \citet{Kewley02}.  The \SII\ lines are produced in a partially ionized zone at the edge of \HII\ regions.  The length of this partially ionized zone is extremely sensitive to the ionization parameter.  Therefore, the \SIIHa\ ratio 
is strongly dependent on the ionization parameter and is not a useful metallicity diagnostic unless the ionization parameter is determined to an accuracy of 0.05 dex in $\log(q)$.  This ionization parameter dependence can be removed by including the \SIII\ line on the numerator, called S23 (S23$=\{$\SII$\lambda \lambda 6717,31$+\SIII$\lambda \lambda 9069,9532$\}/\Ha) \citep{Vilchez96,Oey00,Diaz00,Kewley02}.   \citet{Oey00}  suggest also including the \SIV\ $10.5\mu$ line on the numerator to create a diagnostic known as S234.  Figure~\ref{Metallicity_Diagnostics} shows that the S23 and S234 ratios are sensitive metallicity diagnostics, particularly at high pressure, when they become almost linear functions of metallicity, because large pressures strengthen \SIII\ and \SIV\ at high metallicity.  

The \OII~$\lambda \lambda 3727,9$/\SII~$\lambda \lambda 6717,31$ ratio was proposed by \citet{Dopita86} as a metallicity diagnostic.  This ratio is less sensitive to the ionization parameter than the \SIIHa\ ratio, and is useful for metallicities \OH$>8.2$.   However, if the full optical spectrum is available to calculate this ratio, we recommend the use of the \NIIOII\ ratio instead because the \NIIOII\ ratio is considerably less sensitive to ionization parameter and ISM pressure.

The ratios of \OIII$\lambda 5007$/Hb, \OIII$\lambda 5007$/\NII$\lambda 6584$, \ArIII$\lambda 7135$/OIII$\lambda 5007$, \SIII$\lambda 9069$/OIII$\lambda 5007$, {\NeIII~$\lambda 3967$+\OII~$\lambda \lambda 3727,9$\}/H$\gamma$), \NeIII~$\lambda 3967$/\OII~$\lambda \lambda 3727,9$, and \OIII~$\lambda 5007$/\NeIII~$\lambda 3967$ have been proposed as metallicity diagnostics when limited sets of emission-lines are available \citep{Alloin79,Dopita86,
Charlot01,Stasinska06,Nagao06,Perez-Montero07,Maiolino08}.  Unfortunately, these ratios are relatively insensitive to metallicity, except in the very metal-rich regime (\OH$>8.5$), and are much more sensitive to the ionization parameter than to metallicity.  If only a small portion of the spectrum around \OII\ is available (i.e. no \OIII\ or \Hb), we recommend the use of \OII/H$\gamma$ as a metallicity diagnostic, which has a similar relationship to metallicity as \R23, but with a larger ionization parameter dependence.

\subsection{Theoretical Infrared Metallicity Diagnostics}

Infrared emission-lines were first used to study the abundance variations within the Milky Way, thus avoiding the effects of dust \citep[see][and references therein]{Simpson04}.  Most strong lines in the infrared spectrum are high ionization species and are unable to provide reliable diagnostics of metallicity without a significant dependence on the ionization parameter \citep[see e.g.,][]{Afflerbach97}.  Fortunately, there are two neon lines (\NeII~$12\mu$ and \NeIII~$15\mu$) that, when combined with the nearby Hydrogen lines (either at $7.46\,\mu$m or $12.37\,\mu$m) give robust metallicity diagnostics that are independent of ionization parameter and ISM pressure.  We give calibrations for both Hydrogen lines because while the Humphrey-$\alpha$ line at $12.37\,\mu$m is closest in wavelength to the neon lines, the Pfund-$\alpha$ line at $7.46\,\mu$m is approximately 10$\times$ stronger in flux and is more observable.

\citet{Nagao11} used photoionization models to calibrate the (\OIII~$57\mu$m+\OIII~$88\mu$m)/[\ion{N}{3}]~$57\mu$m, \OIII~$51\mu$m/[\ion{N}{3}]~$57\mu$m and \OIII~$88\mu$m/[\ion{N}{3}]~$57\mu$m ratios, taking into account the gas density and ionization parameter.  These ratios provide useful diagnostics of metallicity, providing the ionization parameter is taken into account.  The (\OIII~$51\mu$m+\OIII~$88\mu$m)/[\ion{N}{3}]~$57\mu$m ratio is the most robust to changes in ISM pressure, and is independent of ISM pressure for most of the normal range ($4\leq \log(P/k) \leq 7$).  Calibrations for these metallicities are given in Table~\ref{Metallicity_Coeffs}.

The far-IR  \NII~$205\mu$m / \CII~$158\mu$m ratio has also been used as a metallicity diagnostic \citep{Nagao13}.  Unfortunately, this ratio is degenerate with metallicity (sometimes being triple-valued), with a 0.3 dex variation in metallicity from the ionization parameter.  In addition, \CII\ may also originate from photo-dissociation regions (PDRs), which are not included in our photoionization model, and would need to be removed prior to the application of a metallicity calibration.

\begin{textbox}[h] 
{\bf Caution:} All emission-line metallicity calibrations suffer from systematic discrepancies of up to 0.7 dex in \OH\ \citep{Kewley08}.  Fortunately, relative metallicities calculated with the same calibration are usually (though not always) accurate to within 0.03~dex on average.  Nevertheless, we recommend the use of three or more metallicity calibrations (preferably with different line ratios and by different authors), where possible, to check how the science results are influenced by the metallicity calibrations used.
\end{textbox}

\section{Excitation Sources in Galaxy Evolution Studies}

Understanding the relative fraction of star formation, AGN, and shocks in galaxies has been a hot topic of research for several decades.  We now know that many AGN in local galaxies are often surrounded by circumnuclear regions of star formation; around half of nearby, optically-selected Seyfert 2 galaxies host a nuclear starburst \citep[e.g.,][]{Cid01,Veilleux03,Veilleux05}.  Star formation may be related to AGN fuelling;  AGN activity is observed systematically after stellar winds and supernova in the nuclear regions have subsided \citep{Wild10}, and AGN-driven winds may also quench star formation \citep[e.g.,][]{Gabor10}.  

Shock excitation in galaxies can be produced by many phenomena, including galactic-scale outflows, galaxy interactions, RAM pressure stripping, and AGN related activity such as jets.  Mergers can produce widespread shocks throughout galaxies which significantly affect the emission-line spectrum of a galaxy at both  kpc-scales and sub-kpc scales within a galaxy \citep{Medling15}.  

Many galaxies show evidence for wind-induced shocks. Our Milky Way has strong shocks near the Galactic centre from stellar winds associated with Wolf Rayet Stars \citep{Simpson07}.  Some galaxies have massive winds with outflow velocities of up to several hundred km/s \citep[e.g.,][]{Strickland04,Veilleux05,Sharp10,Ho16}.  These ÒsuperwindsÓ are produced by either (1) a combination of supernovae and wind-blown bubbles from massive stars (starburst driven winds), or (2) outflows associated with the accretion of material onto AGN (AGN driven winds) \citep[see][for a review]{Veilleux05}.  Starburst driven winds may eject gas into the galactic halo, which can subsequently fall back onto the disk, like a galactic fountain.  Powerful starburst winds may also drive gas outside galaxies and into the CGM \citep{Heckman17}.  Integral field spectroscopy shows that starburst driven winds are common in local disk galaxies of all stellar masses \citep[][]{Ho16}, although winds appear to be more common in galaxies with large star formation surface densities and in galaxies that have experienced recent bursts of star formation \citep{Heckman03}.  

Galactic scale winds are readily observed with spectroscopy in edge-on galaxies because winds often produce conical structures, with limb-brightening along the edges of expanding X-ray bubbles that entrain ambient cold gas. The wind bubbles expand supersonically and shock waves excite emission lines from the UV through to the infrared.  Fast shocks have a photoionizing precursor that produces strong high ionization lines, while the hard radiation field from the shock front itself produces an extended partially ionized zone where low ionization lines such as \OI, \NI, and \SII\ are observed.  Shocked regions may also have higher electron densities than the surrounding non-shocked medium \citep{Ho14}.  

Starburst driven winds are commonly observed at high redshift ($z >1$) because star formation rates (1-100 \Msun$/yr$) are higher than in the local universe.  Winds are observable at high redshift through either absorption or emission-lines \citep[e.g.,][]{Weiner09,Steidel10,Chisolm18,Rigby18,Davies18}.  \citet{Davies18} used emission-lines to show that faster outflows with larger mass loading factors are associated with higher star formation surface densities at $2.0<z<2.6$, similar to local galaxies.  
  
The field of excitation source diagnostics is currently in flux (no pun intended!).  Many emission-line diagnostics have been developed to determine the dominant excitation source in galaxies, but these diagnostics were primarily developed for single aperture spectroscopy with relatively small ($\sim 1$~kpc) apertures.  Recently, wide area integral field spectroscopy, and high spatial resolution integral field spectroscopy have revolutionized our ability to identify and separate multiple excitation sources in the same galaxy.  In particular, the MUSE and Keck Cosmic Web Imager instruments are likely to play a major role in our understanding of the power sources across nearby galaxies in the near future. 

Theoretical models of excitation sources have been essential for understanding the relationship between specific line ratios and the fundamental properties of excitation sources.  Major advances are currently being made in modelling the spectra from star forming galaxies, AGN, and shocks.  It is now possible to model the spectra from galaxies that contain a mixture of excitation sources, although the models are not yet fully self-consistent.

Given these ongoing and exciting changes, we briefly review the current diagnostics in use for single aperture spectroscopy in the UV, optical and infrared.  We then describe some of the most promising diagnostics for use with integral field spectroscopy.  We end with an overview of the recent advances in excitation source modelling.  Because these advances are occurring rapidly, and we expect further advances to be made in the coming 2-3 years, we refrain from providing new excitation class diagnostics at this time. 

\subsection{Excitation sources using single aperture spectroscopy}

The most common method to determine the excitation sources of galaxies is using ratios of strong optical emission-lines.  The excitation diagnostic diagram of \OIII~$\lambda5007/{\rm H}\beta$ vs NII~$\lambda6583/{\rm H}\alpha$ is known as the BPT diagram, after \citet{Baldwin81}, who proposed the use of the \OIII~$\lambda5007/{\rm H}\beta$, \NII~$\lambda6583/{\rm H}\alpha$, and \OI~$\lambda6300/{\rm H}\alpha$ ratios for distinguishing among normal \HII\ regions, planetary nebulae, and objects photoionized by a harder radiation field.  \HII\ regions and star forming galaxies form a clean sequence on the BPT diagram, from low metallicity to high metallicity.  This sequence is known as the ``\HII\ region abundance sequence" or the ``star forming galaxy abundance sequence" (Figure~\ref{BPT}).  The position of this sequence can be affected by the metallicity of the ionized gas, the ISM pressure, the hardness of the ionizing radiation field, and the ionization parameter, as illustrated in Figure~\ref{BPT} (right panel).  Due to these changes, the position of the star forming abundance sequence on the BPT diagram is expected to change with redshift \citep{Kewley13b}.

Seyfert galaxies and shocks lie at large \OIIIHb\ ratios on the BPT diagram.  It is well known that AGN reside in the most massive and most metal-rich galaxies \citep[e.g.,][]{Thomas18}.  Therefore, an AGN contribution raises the \NIIHa\ ratio above the star forming abundance sequence, allowing clean separation of galaxies containing an AGN.  
The position of AGN on the BPT diagram is extremely sensitive to metallicity, and will move towards smaller \NIIHa\ ratios at low metallicity.   Low metallicity AGN are extremely rare in local galaxies \citep{Groves06} but this is likely to change at high redshift, causing the AGN region to move towards, or even overlap with the star forming abundance sequence \citep{Kewley13b}.  The series of galaxies joining the star forming galaxy abundance sequence and the AGN region is commonly referred to as a ``mixing sequence".  Galaxies that lie along the mixing sequence, are commonly referred to as ``composites" or ``transition" objects.  Composites may contain a mixture of star formation, shock excitation, and/or AGN activity. 

\begin{figure*}[!t]
 \centering
  \begin{minipage}[b]{0.63\textwidth}
    \includegraphics[width=\textwidth]{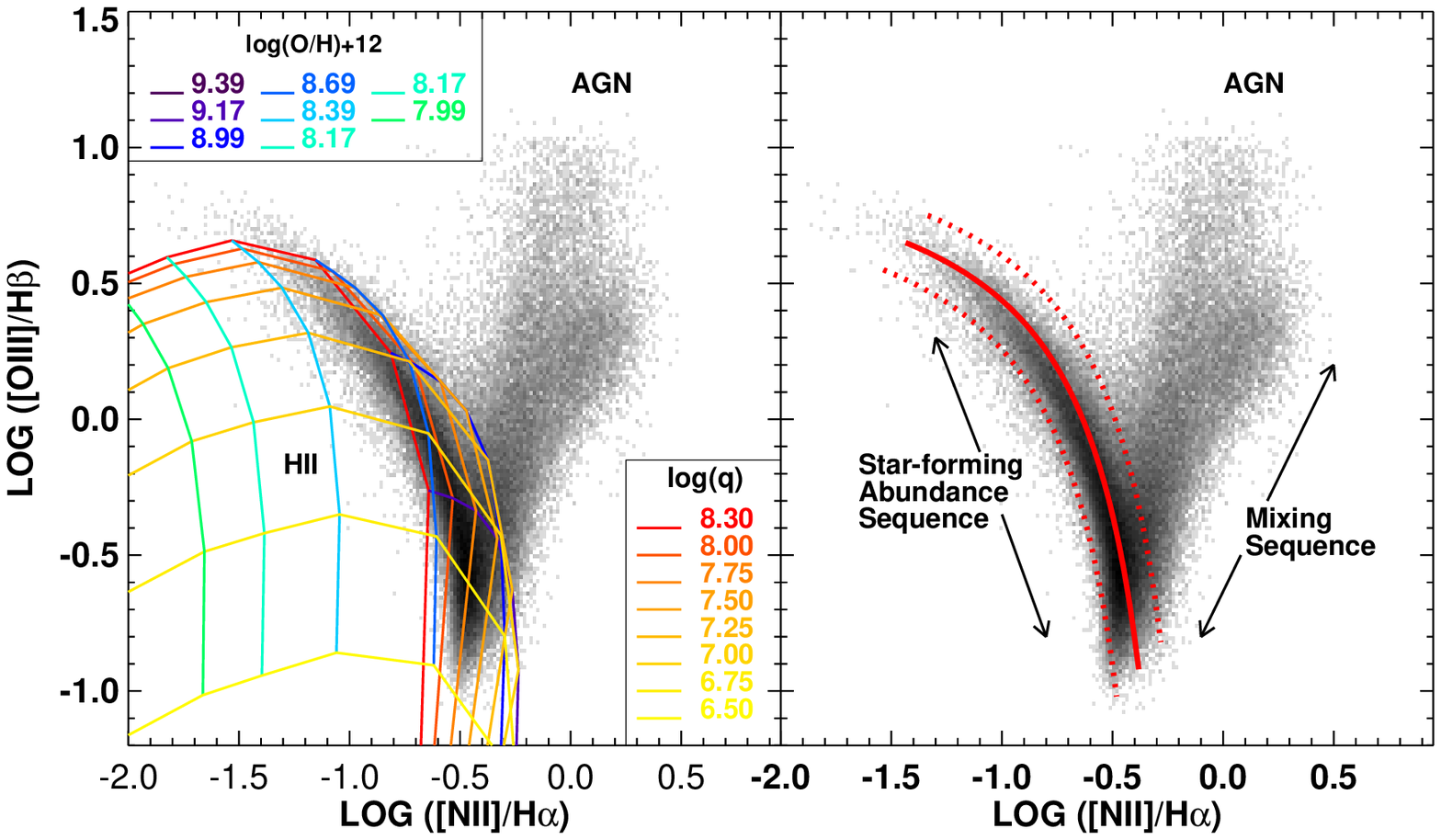}
  \end{minipage}
  \begin{minipage}[b]{0.35\textwidth}
    \includegraphics[width=\textwidth]{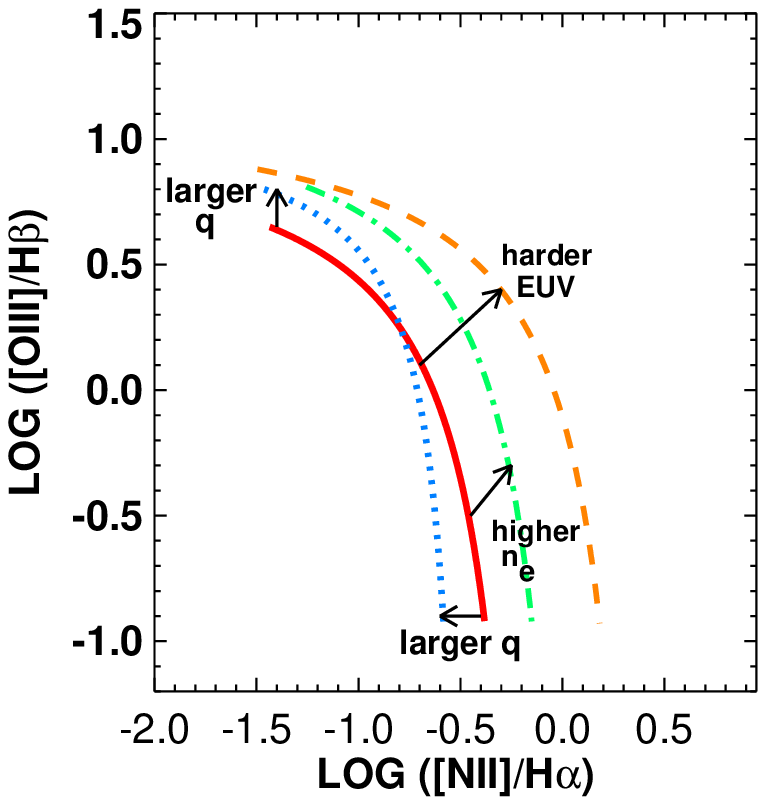}
  \end{minipage}
\caption[Kewley_figure10a.eps,Kewley_figure10b.eps]{The \NIIHa\ versus \OIIIHb\ optical diagnostic diagram for the Sloan Digital Sky Survey galaxies analyzed by \citet{Kewley06a}.  Left: The colored curves show theoretical photoionization model fits to the star forming abundance sequence.  Middle: The red solid curve shows the mean star-forming sequence for local galaxies.  The shape of the red solid curve is defined by our theoretical photoionization models, while the position is defined by the best-fit to the SDSS galaxies.  The $\pm 0.1$ dex curves (dashed lines) represent our model errors and contain 91\% of the SDSS star-forming galaxies.  Right: An illustration of the effect of varying different galaxy parameters on the star-forming galaxy abundance sequence in the \NIIHa\ versus \OIIIHb\ diagnostic diagram.  }
\label{BPT}
\end{figure*}

\citet{Kennicutt84} and \citet{Keel83} extended the BPT classification scheme to include the \SII~$\lambda\lambda6716,6731/{\rm H}\alpha$\ line ratio, which is also sensitive to the hardness of the ionizing radiation field.   \citet{osterbrock85} and \citet[][hereafter VO87]{Veilleux87} derived the first semi-empirical classification lines to be used with the standard optical diagnostic diagrams.  The full suite of  ``standard optical diagnostic diagrams" based on the \OIIIHb, \NIIHa, \SIIHa, and \OIHa\ line ratios are commonly known as VO87 diagrams.   

The first purely theoretical optical classification scheme was developed by \citet{Kewley01a} (hereafter Ke01), using a combination of stellar population synthesis, photoionization, and shock models to derive a ``maximum starburst line" as an upper limit for star-forming galaxies on the BPT diagrams.  Galaxies that lie above this line cannot be explained by any combination of pure starburst models, and must contain a significant fraction (30-50\%) contribution from an AGN or shock excitation.  To produce samples containing purely star-forming galaxies, \citet{Kauffmann03b} shifted the Kewley et al. model shape to fit a boundary to the SDSS star-forming galaxy sequence.  \citet{Stasinska06b} subsequently used photoionization models to derive a new theoretical classification line for classifying local star-forming galaxies. 

\begin{textbox}[b]
{\bf Caution:} Classification lines derived for local galaxies may not be applicable to to high redshift galaxies, or spatially resolved regions within galaxies because the ionization parameter, the radiation field shape (due to the age of the stellar population), and/or the ISM pressure may be different than in the nuclear spectra of SDSS galaxies.
\end{textbox}
 
 The SDSS sample enabled a significant advance  in the optical classification of galaxies because the large number of galaxies ($\sim 45,000$) revealed clearly formed branches on the \SIIHa\ versus \OIIIHb\ and \OIHa\ versus \OIIIHb\ diagrams, for the first time \citep{Kewley06a}.  On these two diagrams, one branch joins the star-forming galaxy abundance sequence to the Seyfert region, and the other branch joins the star-forming galaxy abundance sequence to the region where Low Ionization Nuclear Emission Regions (LINER) are located.  The two branches are not observed on the BPT diagram, because the \NIIHa\ ratio does not provide sufficient separation between the two branches, which overlap.
   
LINER emission can be produced by different emission mechanisms \citep[see e.g.,][for an overview]{Ho08,Yan12}.  Some LINER emission is associated with Low Luminosity AGN (LLAGN).  These LINERs are identified with small ($<1$~kpc) nuclear apertures \citep{Heckman80,Filippenko85,Ho93a}, and have properties consistent with gas ionized by the radiation from an inefficiently accreting, low luminosity supermassive black hole \citep{Ho96,Constantin06}. LINERS in the SDSS observed through $\sim 1$~kpc apertures follow the same relation in Eddington ratio as Seyfert galaxies, but with a harder ionizing radiation field and a lower ionization parameter \citep{Kewley06a}.  Further evidence of AGN in LINERS is found in the radio.  More than 50\% of nuclear LINERs have pc-scale radio nuclei, with implied brightness temperatures $\gtrsim 10^7$~K, and sub-parsec jets \citep{Heckman80,Nagar05}.  High spatial resolution spectra and X-ray studies suggest that the ionizing radiation from the accretion onto the low luminosity black hole produces insufficient ionizing radiation to produce the line ratios observed in LINERS \citep{Flohic06}.  Shocks by jets or other outflows may be required to power this LINER emission, in addition to the emission from the AGN accretion disk \citep{Molina18}.   

LINER-type emission is also seen in the extended gas outside the nuclear regions (i.e. $> 1$~kpc)\citep{Phillips86,Goudfrooij94}. Extended LINER emission can be produced by shocked regions produced by galactic-scale outflows \citep{Dopita95,Ho14,Ho16}.  An evolved stellar population, such as post-AGB stars can also produce extended LINER emission \citep[e.g.,][]{Binette94,Sarzi10,Yan12,Singh13,Hsieh17}.  The term LINER is not appropriate for these extended regions, because LINERs were originally defined as being "nuclear" on very small scales.  An alternative name, LIER (low ionisation emission-line regions) has been suggested by some authors \citep{Belfiore16}, but this definition can include both extended shocks or evolved stellar populations as power sources. In particular, the spatial and spectral resolution of many large IFU surveys is insufficient to unequivocally rule out shocks from galactic-scale winds as the power source for extended LINER emission.

To avoid continued confusion in the literature regarding the power source in LINERs, we suggest that authors avoid the use of terms like ``LINER" and ``LIER".  We advocate the use of more descriptive power source terms with resolution scales, such as ``low luminosity AGN over $< 200~h^{-1}$pc scales", ``galactic-scale wind shocks over a 1~kpc scales", ``post-AGB emission-line regions over 1~kpc scales".  Where aperture size or spatial resolution cannot distinguish between potential power sources, this should be made clear.  Furthermore, more than one of these power sources may exist in a galaxy.  Aperture size, spectral resolution, and emission-line signal-to-noise ratios influence which excitation source is deemed responsible for the low ionization emission.

\begin{figure*}[!t]
\includegraphics[width=12cm]{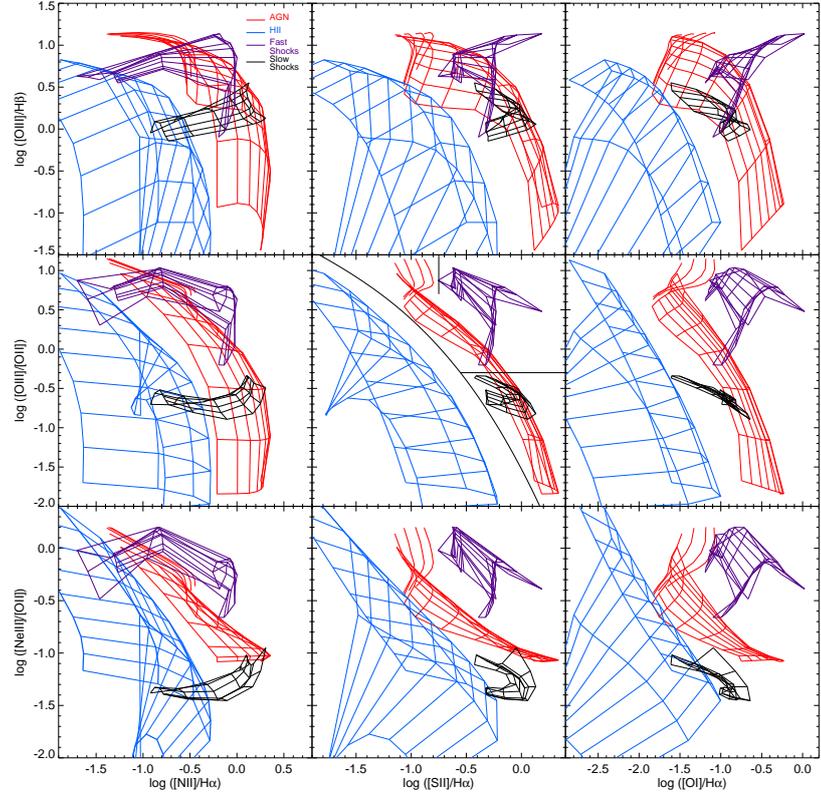}
\caption[BPT_diags_logP_5_mega_multi.pdf]{(Top row) The standard optical diagnostic diagrams showing the location of our Pressure stellar photoionization models (blue), AGN photoionization models (red), fast shock models (purple) and slow shock models (black).  (Middle row) The standard optical diagnostic diagrams but with \OIIIHb\ replaced with \OIII~$\lambda 5007/$\OII~$\lambda \lambda 3727,9$. (Lower row) The standard optical diagnostic diagrams but with \OIIIHb\ replaced with \NeIII~$\lambda 3869/$\OII~$\lambda \lambda 3727,9$.}
\label{Optical_diagnostic}
\end{figure*}

Optical diagnostics are not ideal for separating excitation sources if shock emission is suspected.  Figure~\ref{Optical_diagnostic} (top row) shows the location of our Pressure models, the fast shock models of \citet{Allen08}, the slow shock models of \citet{Rich11}, and the AGN models of \citet{Thomas16,Thomas18b} on the three standard optical diagnostic diagrams.  Although the stellar photoionization models and the AGN models are separated, the fast and slow shock models lie in the same region as the AGN models. Mixing sequences between star formation and shocks are likely coincide with mixing sequences between star formation and AGN.  The shock models lie predominantly away from the pure star forming regions (blue grid), but there is some overlap on the \NIIHa\ vs \OIIIHb\ diagram of shock models with the star forming sequence.  Replacing the \OIIIHb\ ratio with \OIIIOII\ or \NeIIIOII\ improves the situation a little (Figure~\ref{Optical_diagnostic}, middle and lower rows), but there is still overlap amongst the power sources.

Some diagnostics in the UV are more promising.  UV excitation source diagnostics were first developed to understand the UV spectra from high-z radio galaxies, where both shocks and AGN may contribute to the spectrum \citep{Villar-Martin96,Villar-Martin97}.  Unlike the optical, the UV contains a large suite of high ionization emission-lines that are stronger in shocked regions than in photoionized regions.  Shocks drive high temperature gas which produce large quantities of high ionization lines such as \CIV~$\lambda 1549$ \citep[see][for an overview]{Groves04c}.  Many UV diagnostic diagrams have been proposed to either (1) separate AGN from shocks, (2) to separate AGN from star formation, or (3) to separate star formation from shocks \citep{Allen98,Best00,Moy02,Groves04b,Jaskot16,Feltre16}.  In particular, \citet{Feltre16} investigates many UV diagnostic diagrams with current models.  

The \CIII/\HeII\ ratio from the \CIII~$\lambda 1909$ and  \HeII~$\lambda 1640$ lines is one of the most useful discriminants between star formation and harder ionizing sources.  The \HeII~$\lambda 1640$ line is weak in most star-forming galaxies, but even a \CIII/\HeII\ lower limit is useful for identifying star formation dominated galaxies. Ratios of $\log({\rm CIII}]/{\rm HeII}) > 1$ are unambiguously produced by purely star forming models, Shock and AGN models usually produce significantly smaller \CIII/\HeII\ ratios because harder radiation fields produce more \HeII, as well as more ionizations from C++ into C+++.  The \CIII/\HeII\ ratio is insensitive to changes in the ISM pressure. 

The highest ionization UV lines, like \NV~$\lambda 1240$ and \NeV~$\lambda 3426$, are not typically produced in \HII\
regions and the presence of these lines are useful for diagnosing a harder ionizing radiation field than pure star formation.  However, the ratio of \NeV\ to the other strong UV lines produces significant overlap between the AGN and shock models and are not useful diagnostics to separate AGN from shocks.

Resonance lines like \CIV\ are commonly used in the UV, but radiative transfer of resonance lines is approximated with an escape probability that takes into account the opacity effect of dust, and these assumptions introduce additional uncertainties into the resulting model fluxes of \CIV (e.g., Byler et al. 2018, in prep).

A major disadvantage to the UV lines, and sometimes the optical lines is that galaxies containing dust-obscured AGN can be misclassified as purely star forming galaxies.  Infrared emission-lines can identify AGN in the presence of dust.   
Many useful line ratios are now in use for discriminating AGN from star formation in the infrared \citep[e.g.,][]{Genzel98,Fernandez16}.  \citet{Lutz14} discusses use of the infrared diagnostics for galaxy evolution.  

The ratio of high ionization to low ionization fine structure lines such as the \NeV~$14.3,24.3 \mu {\rm m}/$\NeII~$12.8 \mu{\rm m}$ and \OIV~$25.9 \mu {\rm m}/$\NeII~$12.8 \mu{\rm m}$ line ratios discriminate between starbursts and AGN, thanks to the harder radiation field from an AGN \citep{Genzel98}.  The \NeV\ and the \OIV\ lines have ionization potentials of $>54$eV and are produced in the high pressure narrow-line regions surrounding the AGN \citep[see][for a discussion]{Fernandez16}.  

The infrared continuum and PAH features can also provide useful diagnostics for distinguishing the excitation source.  PAHs are poly-aromatic hydrocarbons which are destroyed by the harder radiation field associated with an AGN.  Many IR diagnostics use PAH features, including\citet{Genzel98}, \citet{Laurent00}, \citet{Spoon07}, and \citet{Armus07}. 

Some IR excitation diagnostics can simultaneously trace other properties of the galaxies.  For example, \citet{Spoon07} combined the $6.2\mu$m PAH emission feature and the strength of the $9.7\mu$m silicate absorption line to produce a sophisticated diagram that includes the effect of dust obscuration and geometry as well as excitation source.  The \OIV~$25.9 \mu {\rm m}/$\OIII~$88 \mu {\rm m}$ versus \NeIII~$15.6 \mu {\rm m}/$\NeII~$12.8 \mu {\rm m}$ diagram proposed by \citet{Fernandez16} is sensitive to both the excitation source and the metallicity of the gas through the neon lines.  

Overall, the current suite of UV-infrared excitation diagnostics is extensive and useful for diagnosing the power source in galaxies that contain a single power source.  However, many galaxies contain multiple power sources, and single aperture spectroscopy is unable to unambiguously identify multiple power sources within an aperture, nor estimate the relative contribution of the excitation sources to each emission-line.  Integral field spectroscopy overcomes this problem.

\subsection{Excitation sources using integral field spectroscopy}

Different power sources can be unambiguously indentified using high spectral and spatial resolution optical integral field spectroscopy, as long as the effects of beam smearing (emission from one pixel contributing to emission in another pixel) have been taken into account.  The standard optical diagnostic diagrams can be used in conjunction with photoionization models to determine the fractional contribution from an AGN and star formation to each emission-line, for every spaxel with sufficient S/N.  
If a galaxy displays a clean mixing sequence from the star forming galaxy sequence to the AGN region on the BPT diagram, the relative distance along the sequence gives the fractional contribution from star formation and AGN.  If this fractional contribution is used to combine star formation and AGN photoionization models, the relative contribution to each emission-line can be determined.  Examples can be found in \citet{Davies14a}, \citet{Davies14b}, \citet{Davies16b}, and \citet{D'Agostino18}.  \citet{Davies14b} showed that using this method, the star formation rate can be successfully derived from the \Ha\ line after the AGN fractional contribution has been removed, and that the bolometric luminosity of the AGN can be successfully estimated after the star formation contribution has been removed from the \OIII\ emission line.  This method holds substantial promise for future work investigating the starburst-AGN fraction as a function of redshift and environment, particularly if a similar method can be applied in the infrared to overcome the effects of dust.

\begin{textbox}[b]
\noindent
With high spectral resolution integral field spectroscopy, shocked regions can be unambiguously identified when the following criteria are met: 
\begin{enumerate}
\item The velocity dispersion distribution is bimodal.
\item The velocity dispersion of the broad component is greater than 80~kms/s.
\item The velocity dispersions correlate with shock-sensitive emission-line ratios, such as \NIIHa, \SIIHa, or \OIHa.  
\item The line ratios are consistent with the predictions from shock models.
\end{enumerate}
\end{textbox}

Separating shocks from star forming regions can also be accomplished using high spectral resolution integral field spectroscopy.  This method takes advantage of the impact of shocks on the gas kinematics.   Gas excited by thermal shocks produces emission-lines with velocity dispersions (i.e. emission-line widths in velocity units) around the mean shock velocity.  These shocked components often present as separate kinematic components in integral field data, and can be seen as broad lines, usually underlying the narrow lines typical of star forming regions.  Both galactic wind shocks and merger-induced shocks are known to produce separate kinematic components to the emission-lines which have velocity dispersions of $150 - 500$~km/s, significantly larger than the velocity width of \HII\ regions or gas ionized by evolved stellar populations (both typically $< 40$~kms/s).  With this method, shocks have been successfully separated from star forming regions in merging galaxies \citep{Rich10,Rich11,Rich15}, and in isolated spiral galaxies \citep{Ho14,Ho16}.


The correlation between velocity dispersion and line ratios such as \NIIHa\ or \SIIHa\ \citep[see][]{Rich10} is critical for unambiguous identification of  shocks because neither beam smearing nor aged stellar populations can produce correlations between velocity dispersion and emission-line ratios.  High spectral resolution of 30-50~km/s at \Ha\ is typically required to identify shocks with this method.
Unfortunately, the spectral and spatial resolution of many current large IFU surveys are insufficient to separate evolved stellar populations from shock emission using velocity dispersion.  

Separating shocks from both star formation and AGN is significantly more difficult because the AGN narrow line region also produces a kinematic component to the emission-line profiles that may have velocity dispersions similar to (or larger than) shocks.
However, it appears that separation of all three power sources is now be possible, using the spatial information afforded by integral field spectroscopy, in conjunction with the emission-line ratios and velocity dispersion information.  A 3-D diagnostic diagram that uses radius as the x-axis, velocity dispersion as the y-axis, and a combined line ratio as the z-axis is able to separate all three power sources, as long as two of the power sources are extended beyond the nucleus (D'Agostino et al. in prep).  Any of the BPT or VO87 diagnostic line ratios can be used in this 3D diagram, and we recommend the use of multiple diagnostic line ratios for verification.  Figure~\ref{Josh_3D} gives an example of the 3D diagram to separate shocks, AGN, and star formation from D'Agostino et al. This diagram utilizes the fact that gas excited by an AGN can be identified by its central position within galaxies as well as its large velocity dispersion, which is typically $>300$~km/s.  Additional multi-wavelength information such as hard-X-ray emission or a compact steep spectrum radio source can help to confirm the AGN identification.   In Figure~\ref{Josh_3D}, at low line ratios and low velocity dispersions, star formation dominates.  At large line ratios, large velocity dispersions, and large radii, shock emission dominates, while at large line ratios, large velocity dispersion and small radii, AGN dominates, producing a double peaked fork shape that is two mixing sequences: (1) a star formation - shock mixing sequence (indicated by the blue line), and (2) a star formation - AGN mixing sequence (indicated by the red line).  If this 3D diagram is combined with theoretical photoionization and shock models, one could potentially derive the fractional contribution from star formation, shocks, and AGN to each spaxel and each emission-line, as long as the spatial resolution allows clean separation between the star forming, shock, and AGN kinematic components.   

\begin{figure*}[!t]
\includegraphics[width=\textwidth]{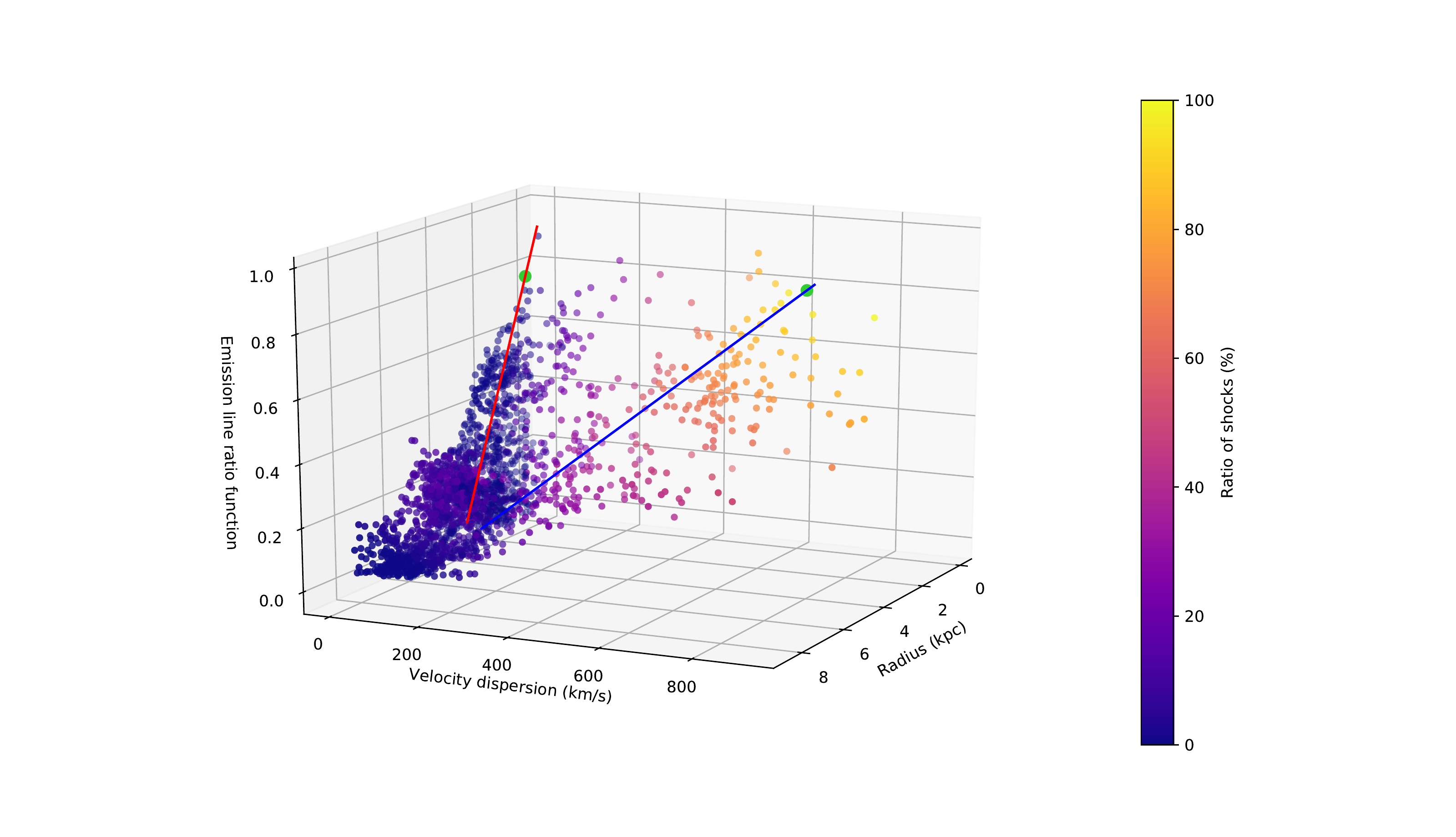} 
\caption[Kewley_figure12.pdf]{A 3D diagnostic diagram showing how star formation, AGN, and shocks can be separated using a combination of emission-line ratios, velocity dispersion, and radial information.  The colored circles are spaxels in NGC 1068 from \citet{D'Agostino18}, color coded according to shock fraction as shown in the legend. The red and blue lines indicate the direction of the starburst-AGN and starburst-shock mixing sequences, respectively.}
\label{Josh_3D}
\end{figure*}

\subsection{Excitation Source Modelling}\label{Models}

Modern excitation source diagnostics rely on stellar evolution synthesis, photoionization, and shock models both to (1) predict the regions of the diagrams where each power source is likely to be located, and (2) understand the impact of star formation, AGN, and shock properties on the predicted locations of the power sources on the diagnostic diagrams.  

Most stellar evolutionary synthesis models are based on single stars.  Single star models all produce a relatively soft ionising radiation field, with extremely similar (to within $\sim 0.15$~dex) emission-line predictions for the key optical diagnostic ratios (D'Agostino et al. 2018).  Single star stellar evolution models contain too few ionising photons above $>40$~eV to reproduce the observed strength of key UV, optical, and IR emission-lines, particularly at low metallicity.  This problem was identified through comparisons between models and observations for the optical \SIIHa, \OIHa\ ratios in local galaxies \citep{Kewley01b}, the \NeIII\ line strength \citep[e.g.,][]{Sellmaier96}, and the IR \OIV$25.9 \mu$m/\OIII$88 \mu$m ratio \citep{Fernandez16}.  The lack of ionising photons has been referred to as the  ``[Ne III] problem", but it affects a substantially larger number of lines than just \NeIII.   

Several methods have been proposed to produce more ionising photons above $>40$~eV, including stellar rotation and binary star models. Stellar tracks calculated with stellar rotation produce a harder ionising radiation field, that remains hard for longer timescales than stellar tracks without rotation included \citep{Levesque12}.  There are multiple reasons for this effect.  Rotation causes stars to spend more of their lifetime on the main sequence because Helium enhancement causes rotating stars to have both hotter effective temperatures and larger luminosities than non-rotating stars \citep{Leitherer08,Leitherer14}.   Rotational mixing also extends the W-R phase by causing more stellar mass loss, lowering the minimum mass for a WR star \citep{Georgy12}.  An entire population of rotating stars will contain a larger fraction of W-R stars than a non-rotating population, and these W-R stars will be longer lived.   
Stellar evolution synthesis models now include rotation, although a full suite of rotation models across the full metallicity range is still lacking.

Binary star models produce similar effects on the ionizing spectrum as models with rotation. Binary star models can produce a harder ionizing radiation field particularly at low metallicity \citep{Xiao18}.  Binary stars remove the hydrogen envelope of some red supergiants to form Wolf-Rayet stars, which make a significant contribution to the ionizing radiation field, resulting in a radiation field that remains harder for a longer period of time than single star models \citep{Eldridge09}.  To reproduce the observed strengths of UV lines in high redshift galaxies, binary star models require a sub-solar metallicity and a depleted carbon-to-oxygen ratio \citep{Eldridge12}.  These advances, in combination with advances in stellar atmosphere modelling, may resolve the Ò[Ne III]Ó problem and provide more robust predictions of the UV-infrared spectral lines for the next generation of excitation source diagnostics.

Shock models have also made significant advances recently.  Fast shocks ($v>500\, {\rm km s}^{-1}$) produce strong high ionization lines such as \OIII\ \citep{Allen98,Allen08}, while slow shocks produce relatively weak high ionization lines, but strong low ionization species such as \SII\ and \NII\ \citep{Rich11,Rich15}.  This difference occurs because fast shocks produce a photoionizing precursor which produces significant ionization in front of the shock.  Slow shocks are unable to drive a  precursor \citet{Sutherland17}.  Previously, shocks and precursors were run separately, with several iterations of shock and precursor models to ensure that the ionization state of the precursor gas is reliable.  The latest shock models include self-consistent treatment of shocks and their photoionizing precursor, the latest atomic data, and the shock models now span a significantly larger range of velocity \citet{Sutherland17}.  The new generation of shock models is currently only available for solar metallicity.  When available, the full suite of metallicities will enable one to derive shock properties, such as the shock velocity and the metallicity in the shocked region using emission-line diagnostics.

To derive reliable excitation source diagnostics, theoretical modelling of pAGB stars is required to identify where pAGB stars contribute to the emission line ratios.  Theoretical modeling of pAGB star emission assumes that the radiation field produced by the pAGB stars ionizes nearby gas with a close to 100\% covering fraction.  Under these assumptions, pAGB star models can produce LINER emission spectra, although \citet{Yan12} suggest that a larger ionization parameter is required than current pAGB models are able to produce.  We expect further advances in pAGB star modelling in the near future, with the combination of new high spatial and spectral resolution integral field spectroscopy and pAGB models.

\section{Diffuse Ionized Gas} 
 
The ionizing radiation produced by the star clusters is only partially absorbed by the \HII\ regions.  Superbubble models suggest that the combined effects of supernovae, stellar winds, and large-scale ionization by OB associations create a complex density and ionization structure that can be porous to ionizing radiation, allowing some radiation to escape \citep{Shields90}.  The escaped radiation ionizes the diffuse gas outside \HII\ regions, both in the disk and up to several kiloparsecs above the disk. The gas that receives leaked ionizing radiation from \HII\ regions is referred to as Diffuse Ionized Gas (DIG).    

The DIG has been studied extensively in the Milky Way \citep{Madsen06,Minter97,Rand98} (referred to as the Warm Ionized Medium, or WIM), and more recently in nearby galaxies \citep[e.g.,][and references therein]{Zhang17,Poetrodjojo18}.  These studies suggest that the radiation in the DIG may come from a variety of sources.  \citet{Martin97} showed that the DIG has a radial gradient that is consistent with the dilution of radiation from a centralized source, indicating that the dominant excitation mechanism of the DIG is photoionization by the radiation from massive stars.  However, in some galaxies, 30-50\% of the DIG emission may come from shock excitation \citep{Martin97,Ramirez-Ballinas14}, and a very minor component ($<20$\% of \Ha) may come from dust scattered radiation \citep{Barnes15,Ascasibar16}.

The absorption by the \HII\ region produces a hard residual ionizing radiation field producing strong diffuse gas emission in low-ionization species, such as \OI, \NII, \CII, and \SII, and weak emission in high ionization species such as \OIII\ \citep[see reviews by][]{Rand98,Mathis00,Haffner09}.   The DIG may contribute between 10-50\% of the \Ha\ emission, but may contaminate the \NII, \SII\ and other low ionization lines up to 2-3$\times$ more \citep{Madsen06,Oey07,Blanc09}.  As shown by \citet{Zhang17}, the \NIIOII\ ratio is relatively unaffected by the DIG, on average.  The \OIIIOII\ ratios are also likely to be relatively unaffected by the DIG.  

The line ratios of the DIG have important implications for estimates of the electron density, ISM pressure, ionization parameter, metallicity, and excitation source in galaxies.  In spectra where \HII\ regions are unresolved, estimates of the electron density or ISM pressure using the \SII\ doublet are likely to be strongly contaminated by the DIG and are likely to yield density and pressure estimates that are larger than the true values.  The DIG is likely to introduce scatter in estimates of the ionization parameter using \OIIIOII\, but the average ionization parameter will be close to the true value.  The DIG will cause the metallicity to be overestimated if the \NIIHa\ and \SIIHa\ (or \NIISII) ratios are used, but the metallicity is likely to be close to the true (strong-line) value if the \NIIOII\ or \R23\ ratios are used.  When the standard optical diagnostic diagrams are used to classify excitation source, strong contamination from the DIG may cause galaxies to be misclassified as LINERs.  

One of the most serious consequences of the DIG is its affect on metallicity gradient measurements.  In face-on spiral galaxies, the DIG contributes more at larger radii than at smaller radii, causing metallicity gradients to appear significantly flatter than the true gradient.  Spatial resolution of the scales of the \HII\ regions (40-100 pc) is required to overcome this problem.  This has important implications for studies of metallicity gradients with large IFU surveys or at high redshift, where the spatial resolution is typically $\sim 1$~kpc.

\section{Challenges}\label{Challenges}

Galaxy evolution studies rely on reliable emission-line diagnostics.  The current challenges to theoretical modeling of galaxy spectra and observations of galaxies are opportunities to improve our estimates of fundamental galaxy properties in future.  In this Section, we describe some areas where improved theoretical models or observations will make a significant impact on galaxy evolution studies.\\
 
\noindent
{\bf Stellar tracks:} Current stellar population synthesis and photoionization models are limited to a coarse grid of five metallicities that is determined by the metallicities of the stellar tracks and opacity tables.  Stellar tracks track how stars of a given mass evolve on the Hertzspring-Russell diagram, and are computed from stellar evolution theory and large observational stellar libraries.  Stellar tracks require a set of elemental abundance ratios as an input parameter, and have usually been calculated by scaling the relative abundance ratios in \citet{Anders89}.  There are two problems with this.  First, five sets of metallicities limits the resolution at which theoretical strong line metallicity diagnostics can be calculated and the stellar evolution models and/or photoionization models are currently being interpolated.  Ideally, stellar tracks (and opacity tables) would be calculated at intervals of 0.2-0.5 dex in $\log({\rm O/H})$.  Secondly, it is now known that the \citet{Anders89} relative abundance ratios do not match observed abundance ratios in \HII\ regions in the Milky Way, LMC, or SMC, at a given metallicity.  Stellar tracks and opacity tables that represent the observed abundance ratios in the Milky Way and nearby galaxies \citep[see][]{Nieva12,Nicholls17} are critically needed, and could be computed with current stellar evolution models and datasets.  These tracks would allow consistent abundance ratios to be used in stellar population synthesis and photoionization models to derive high resolution metallicity diagnostics, for the first time.\\

\noindent
{\bf Stellar population synthesis models:} The theoretical prediction of emission-lines depends on the ionisation potential of each emission-line, and is therefore sensitive to the shape of the ionising EUV radiation field at the wavelength of the ionisation potential.  The EUV radiation field is produced solely by the stellar evolution synthesis models, and different models produce remarkably different SED shapes, depending on the input physics. 

Despite significant advances in stellar population synthesis models, the cause(s) of the discrepancy between observed and predicted emission-lines such as (but not limited to) \NeIII\ remain unknown.  Detailed multilevel NLTE treatment including radiation-driven wind theory and a consistent calculation of NLTE line blocking opacities for all contributing ionization stages is critical \citep{Sellmaier96,Giveon02,Sellmaier96,Weber15}, as is understanding the contribution from scattered light to \HII\ region spectra \citep{Simon11}.  Stellar evolution models also need to include both the effects of stellar rotation and binary stars, rather than these effects being taken into account in different stellar evolution models.  Ideally, both members of binary star systems would be evolved simultaneously, with NLTE atmospheres that include metal opacities for all contributing ionization stages.\\

\noindent
{\bf Ionization structure of HII regions:} High spatial resolution, high signal-to-noise observations with broad wavelength coverage across nearby \HII\ regions are needed to understand the ionization structure in \HII\ regions. Observations of multiple ionization states of different species is needed.  Single star \HII\ regions are useful for such studies because the effective temperature of the ionizing star is known \citep{Zastrow13}, but star clusters must be modeled also.  With the detailed ionization structure known, it may be possible to reverse-engineer the detailed shape of the ionizing radiation field required to produce such an ionization structure through photoionization models.  These observations would provide important constraints on the shape of the ionizing radiation field required from the stellar population synthesis models.\\

\noindent
 {\bf Super star clusters:} Spectra of high redshift galaxies are a luminosity-weighted average of the star-forming regions within the spectral aperture.  If super star clusters are common in high redshift star-forming galaxies, then a larger proportion of super star clusters may dominate the emission-line ratios observed.  Surface brightness dimming may exacerbate this effect.  
The ionization parameter is large in local super star clusters $-1.5<\log(U)<-2.3$ \citep{Smith06,Snijders07}.  As seen in this review, the ionization parameter can significantly affect the measurement of fundamental galaxy properties such as the ISM pressure and the metallicity.  To estimate the importance of this effect, systematic studies are needed on the properties of nearby super star clusters, as well as the fraction of super star clusters in galaxies as a function of redshift. \\
 
\noindent
{\bf HI region structure:} Most photoionization models are calculated for radiation-bounded \HII\ regions, in which the nebula ends where hydrogen is completely recombined.  However, \citet{Nakajima12} suggest that Lyman-$\alpha$ emitters at high redshift contain density-bounded \HII\ regions.   In a density-bounded nebula, the density is sufficiently low that the stars can ionize the entire nebula.  While the higher ionization lines like \OIII\ are closer to the star cluster and are unaffected by how the nebula is bounded,  the hydrogen recombination zone may be shortened in density-bounded nebulae.  The \OIIIHb\ ratio observed will then be larger in density-bounded nebulae than in a radiation-bounded nebula.  The outer partially ionized zone is shortened in density-bounded nebulae, reducing the emission of \SII, \OI\ and possibly even \NII.  It is currently unclear whether density-bounded nebulae are common in normal star-forming galaxies, either locally or at high redshift.  Detailed ionization mapping of the \HII\ regions in nearby galaxies would constrain the fraction of \HII\ regions that are radiation bounded and density bounded.   A mixture of radiation-bounded and density-bounded \HII\ regions have been observed in the local group \citep{Pellegrini12}. 

A better understanding of the density structure of local \HII\ regions is required.  This area would benefit from high spatial resolution observations of nearby \HII\ regions in a broad range of environments.  In concert with these observations, complex theoretical modelling of \HII\ regions needs to evolve from simple plane parallel or spherical geometries to full Monte-Carlo radiative transfer codes in order to take complex geometries and density structures into account \citep[see][for a review]{Steinacker13}.  Initial Monte-Carlo radiative transfer or dust models with ray-tracing techniques have been developed \citep{Gordon01,Ercolano08,Popescu13,Vandenbroucke18}. These 3D models offer much promise for future modeling of \HII\ regions and galaxies.  For example, \citet{deLooze14} and \citet{Law11} use 3D radiative transfer to model the young stellar populations and dust heating processes in nearby star-forming galaxies.  De Looze et al. show that 3D models can reproduce the far-infrared and sub-mm observations of M51, and they calculate the relative contributions of stars and dust. \citet{Popescu17} use their 3D models to solve for the UV-sub-mm interstellar radiation fields of the Milky Way, a notoriously difficult problem.  We anticipate that the next major advance in the galaxy spectral modeling field will be the development of fully self consistent 3D radiative transfer models, that will allow detailed dust and gas distributions to be embedded within the photoionized nebula with arbitrary temperature, density and dust distributions.  Promising work in this direction includes \citet{KLaw18}.  \\

\noindent
{\bf Identifying shocks:} 
Shock excitation can contaminate both low ionization line ratios and high ionization line ratios, depending on the shock velocity. High resolution integral field spectroscopy allows the shock component to be resolved and separated from the \HII\ region component of optical emission-lines through the combination of morphological information, velocity maps, velocity dispersion, and emission-line ratios \citep[see][for examples of how shocks can be identified at low and high redshift, respectively]{Ho14,Yuan12}.  We recommend that these tools be used to identify galaxies containing shocks prior to the application of emission-line diagnostics.  In cases, where shocks and star forming regions can be separated, it is feasible to apply pressure, ionization parameter and metallicity diagnostics to the \HII\ region component of the spectra.  

Three-dimensional diagnostics using velocity dispersion, radius and line ratio information are exceptionally promising ways to separate and quantify the contribution from shocks, star formation and AGN across galaxies (D'Agostino et al., in prep).  Currently, 3-D diagnostics only exist for the optical.  Extensions of such diagnostics to the UV and the infrared, as well as comprehensive studies of the location of theoretical models on these diagrams are needed to fully exploit the potential of these diagnostics.\\

\noindent
{\bf Diffuse ionized gas modelling:} Theoretical models are needed to help remove the DIG component from spectra.  Unfortunately, current photoionization models can only partially reproduce the DIG line ratios \citep[e.g.,][]{Bland-Hawthorn97,Zhang17}, leading to additional ionizing sources being proposed.  For example, \citet{Martin96} used a combination of photo-ionized gas with shock-ionized gas using various shock speeds, and turbulent mixing layers of various temperatures to model the DIG in irregular galaxies.  \citet{Zhang17} showed that photoionization models fail to reproduce the strength of the low ionization optical lines, and suggests that a radiation field from evolved stellar populations may resolve this discrepancy.  High spatial resolution ($10-100$~pc) integral field spectroscopy of large samples of face-on galaxies is required to characterize the conditions within the DIG to help inform DIG models.  With such information, DIG models might allow reliable properties to be derived for the global spectra of galaxies, or coarse ($\sim 1$~kpc) resolution integral field spectroscopy by providing an estimate of, and correction for, the contamination from the DIG to each emission-line across a spectrum.  In particular, it is necessary to understand the contamination from the DIG to the UV diagnostic line ratios, because it is currently difficult, and in many cases impossible to observe the contribution from the DIG in the UV.\\

\noindent
{\bf Understanding global spectra:} Understanding the properties derived for global spectra of galaxies (or for large spaxels within galaxies where \HII\ regions are unresolved) is non-trivial, but is critical for studies of the highest redshift galaxies.  Previous work in understanding global spectral properties uses single slit spectroscopy of \HII\ regions or integral field spectroscopy where \HII\ regions are unresolved \citep[e.g.,][]{Kobulnicky99a,Moustakas06,Gavazzi18}.  Much may be learned by repeating these studies using significantly higher spatial resolution observations that resolve the temperature and density structure within \HII\ regions.
High spatial resolution observations of nearby galaxies could be used to simulate surface brightness dimming of high redshift galaxies and then be convolved with the response functions of current and future instruments to understand how derived global properties of high redshift galaxies relate to the actual properties within galaxies.  For example, the global metallicity of a high redshift galaxy may not be the true mean metallicity, but may be weighted towards specific \HII\ regions with certain sets of properties.  Note that because high redshift galaxies are not the same as local galaxies, theoretical simulations that place \HII\ regions with complex temperature and density structures within a model galaxy from cosmological zoom-in simulations at a given redshift may help inform observational studies in future.  Recent theoretical work in this area indicates that there are large effects on metallicity gradients at high redshift from low signal-to-noise and beam smearing \citep{Accharya18}.  The inclusion of simulations of the diffuse ionized gas would significantly improve these simulations.  \\

\noindent
{\bf Limited sets of emission-lines:} Some samples contain only limited sets of emission-lines, necessitating the use of only two or three emission-lines for diagnosing fundamental galaxy properties.  Common examples include spectra of high redshift galaxies obtained using infrared bandpasses.  
Care must be taken not to derive more parameters than there are constraints from the emission-lines.  Each unique emission-line ratio provides a single constraint, unless additional information is included, such as equivalent widths (where the continuum provides an additional constraint), or stellar masses.  Theoretical models should be used to investigate the dependence of sets of line ratios on relevant galaxy properties prior to the application of the line ratios.  If a specific line is to be used regularly in a given sample, we strongly recommend undertaking a comprehensive study of the properties of that line and potential caveats, like in \citet{Jaskot16}.  Jaskot et al. use models with varying C/O abundances, dust content, gas density, nebular geometry and optical depth to understand the impact of each of these properties on the \CIII\ line.  Other lines that warrant such a study include, but are not limited to, the \CIV~$\lambda 1549$ line, the \SiIIIf\ and \SiIII\ ($\lambda \lambda 1883, 92$) lines, and the UV and optical nitrogen lines.
Tracking individual elements through cosmological simulations is also needed. \citet{Naiman18} presents such a study of Magnesium and Europium.  Further studies like these, extended to other elements, will be essential to inform galaxy evolution studies using emission-lines with future extremely large telescopes.\\

\section{Acknowledgements}

The authors would like to thank the editor, Luis Ho, for his feedback and support of this manuscript.  L.J.K. would like to acknowledge the staff and residents of {\it Varuna the writer's house} and the {\it UNSW ASTRO 3D Writing Retreat} for providing tranquility, excellent food, and a supportive writing atmosphere.  Parts of this research were supported by the Australian Research Council Centre of Excellence for All Sky Astrophysics in 3 Dimensions (ASTRO 3D), through project number CE170100013.  L.J.K. gratefully acknowledges the support of an ARC Laureate Fellowship (FL150100113). This research has made extensive use of NASA's Astrophysics Data System Bibliographic Services.

\begin{table}
\caption{Ionization parameter calibrations for the diagnostic line ratios shown in Figure~\ref{ionization_diags}.}
\begin{center}
    \begin{tabular}{lrrrrrrrr} \hline \hline
    \multicolumn{9}{l}{Bi-Cubic Surface fits: $z = A + Bx + Cy + Dxy + Ex{^2} + Fy{^2} + Gxy{^2} + Hyx{^2} + Ix{^3} + Jy{^3}$ }  \\
    \multicolumn{9}{l}{where $x=$log(line ratio), $y=\log(O/H)+12$, $z=log(U)$}  \\
    \hline
R & $\frac{\rm CIII]}{\rm [CII]}^{\rm b}$ & $\frac{\rm SiIII]}{\rm [SiII]}^{\rm b}$ & $\frac{\rm [AlIII]}{\rm [AlII]}^{\rm b}$  & $\frac{\rm [OIII]}{\rm [OII]}^{\rm b}$ & $\frac{\rm [SIII]}{\rm [SII]}^{\rm b}$ & $\frac{\rm [NeIII]}{\rm [NeII]}^{\rm c}$ & $\frac{\rm [SIV]}{\rm [SIII]}^{\rm c}$ & $\frac{\rm [NIII]}{\rm [NII]}^{\rm c}$ \\
    \hline
\multicolumn{9}{c}{Fits for $\log({\rm P/k})=5.0^{\rm a}$ }\\ \hline    
$Z_{min}$           & 7.63          & 7.63 & 7.63       & 7.63      & 7.63      & 7.63       & 7.63      & 7.63 \\ 
$Z_{max}$          & 8.93    & 8.93       & 8.93       & 8.93      & 9.23      & 9.23       & 9.23      & 9.23 \\ 
$\log(U_{min})$  &   -3.98 &.     -3.98 & -3.98     & -3.98     & -3.98    & -3.98    & -3.98     & -3.98 \\ 
$\log(U_{max})$ &  -2.98 &       -2.73 & -2.73     & -2.98      & -2.48    & -2.48     & -2.48      & -2.48 \\ \hline
$A$                    & -354.86 & -271.55 & -56.999 &  13.768 &  90.017 &  -454.99 & -236.35 & -189.71 \\
$B$                   & 62.164 &    68.254 &  17.186 &  9.4940 &   21.934 &   19.251 & 10.170 & 13.207 \\
$C$                   & 137.35 &    102.23 &  22.538 & -4.3223 & -34.095 &   168.78 & 88.399 & 69.119 \\
$D$                   & -15.604 & -16.983 & -3.4525 &  -2.3531 & -5.0818 & -4.6574 & -2.3369 & -3.1714 \\
$E$                   & -1.5532 & -2.7208 &  0.7664 &  -0.5769 & -1.4762 &  0.0451 & 0.0849  & 0.1490 \\
$F$                  & -17.901 &  -13.023 & -3.0032 &   0.2794 &  4.1343 & -21.058 & -11.140 & -8.5599 \\
$G$                  & 0.9936 &    1.0647 & 0.1910 &    0.1574 &  0.3096 &  0.2962 & 0.1467 & 0.1990 \\
$H$                 & 0.2248 &     0.3931 & 0.0000 &    0.0890 &  0.1786 &  0.0000 & 0.0000 & 0.0000 \\
$I$                   & 0.0000 &     0.0000 & 0.3419 &    0.0311 &  0.1959 &  0.0000 & 0.0000 & 0.0000 \\
$J$                  & 0.7778 &     0.5531 & 0.1300 &    0.0000 & -0.1668 &  0.8765 & 0.4683 & 0.3535 \\ \hline
 RMS err (\%)     & 1.69  &        2.71  &    3.01  &       1.35  &     1.96  &  1.92     & 2.05     & 2.94 \\ \hline \hline
\multicolumn{9}{c}{Fits for $\log({\rm P/k})=7.0^{\rm a}$ } \\ \hline    
$Z_{min}        $     & 7.63     & 7.63      & 7.63       & 7.63      & 7.63        & 7.63         & 7.63      & 7.63 \\ 
$Z_{max}$           & 8.93      & 8.93      & 8.93       & 8.93      & 9.23         & 9.23        & 9.23      & 9.23 \\ 
$\log(U_{min})$   &  -3.98    & -3.98     & -3.98      & -3.98     & -3.98       & -3.98       & -3.98     & -3.98 \\ 
$\log(U_{max})$  & -2.48     & -2.48     & -2.48      & -2.48     & -1.98       & -1.98       & -1.98     & -1.98 \\ \hline
$A$                     & -415.93  & 190.14  & 13.393   & -48.953 & 22.308    & -445.65   & -213.24 & -164.64 \\
$B$                     & 49.915   & 30.405  & 29.853   & 6.076     & 10.928   & 26.720    & 11.204   & 10.994 \\
$C$                     & 157.61   & -67.224 & -3.2608  & 18.139   & -9.5322  & 164.91   & 80.309   & 59.247 \\
$D$                     & -12.538  & -7.1308 & -6.5926  & -1.4759  & -2.3701  & -6.4297 & -2.6648   & -2.6271 \\
$E$                     & -0.7865  & -4.1040 & 0.8510   & -0.4753  & -0.7150  & -0.4018  & -0.1224  & 0.1438 \\
$F$                     & -20.097  & 7.6694  & 0.1452    & -2.3925  & 1.1679  & -20.531  & -10.206   & -7.2715 \\
$G$                    & 0.8021   & 0.4308   & 0.3855   & 0.1010    & 0.1432  & 0.4011   & 0.1721    & 0.1647 \\
$H$                    & 0.1180   & 0.5029   & 0.0000   & 0.0758    & 0.0793  & 0.0541   & 0.0325    & 0.0000 \\
$I$                     & -0.0560  & 0.1883   & 0.3998   & 0.0332    & 0.1904  & 0.0000   & 0.0159    & 0.0000 \\
$J$                    & 0.8551   & -0.2878   & 0.0021  & 0.1055    & -0.0476 & 0.8529   & 0.4329    & 0.2976 \\
    \hline
    RMS err (\%) & 2.43 & 2.59 & 1.62 & 1.48 & 1.28 & 1.42 & 1.71 & 2.52 \\ \hline \hline
  \end{tabular}%
 \end{center}
 \begin{tabnote}
 $^{\rm a}$Valid over $Z_{min}\leq \log({\rm O/H})+12 \leq Z_{max}$ and $ \log(U_{min})\leq \log(U) \leq \log(U_{max})$.\\
 $^{\rm b}$ Wavelengths in \AA: \ion{C}{3}]/[\ion{C}{2}]$=\frac{1907+1908}{2323.50+2324.69+2325.40+2326.93+2328.12}$, 
 \ion{Si}{3}]/[\ion{Si}{2}]$= \frac{1883+1892}{1808}$, [\ion{Al}{3}]/[\ion{Al}{2}]$=\frac{1856+1862}{1670}$, 
 [\ion{O}{3}]/[\ion{O}{2}]$=\frac{5007}{3727+3729}$, [\ion{S}{3}]/[\ion{S}{2}]$=\frac{9069+9531}{6717+6731}$\\
 $^{\rm c}$ Wavelengths in $\mu$m: [\ion{Ne}{3}]/[\ion{Ne}{2}]$=\frac{15}{12}$, [\ion{S}{4}]/[\ion{S}{3}]$=\frac{10.51}{18.71}$, [\ion{N}{3}]/[\ion{N}{2}]$=\frac{57}{122}$
 \end{tabnote}
\label{pres_diag_Pk5}
\end{table}

\begin{table}
  \caption{Metallicity diagnostic calibrations for UV and optical line ratios described in Section 4.5 - 4.7.}
  \begin{center}
    \begin{tabular}{lrrrrrrr}    \hline    \hline
    \multicolumn{8}{l}{Bi-Cubic Surface fit: $z = A + Bx + Cy + Dxy + Ex{^2} + Fy{^2} + Gxy{^2} + Hyx{^2} + Ix{^3} + Jy{^3}$ }  \\
    \multicolumn{8}{l}{where $x=log(R)$, $y=log(U)$, $z=log({\rm O/H})+12$, and $\log({\rm P/k})=5.0^{\rm a}$}  \\
\hline    
R  & $\frac{\rm [NIII]}{\rm [OIII]}^b$ & $\frac{\rm C23}{\rm HeII}^b$ & $\frac{\rm C23}{\rm HeI}^b$ & $\frac{C234}{\rm HeI}^b$ &   $\frac{\rm [NII]}{\rm [OII]}^b$  \\ 
\hline 
$Z_{min}$  &     7.63 &      8.53  &       8.53 & 8.53        &  7.63  \\ 
$Z_{max}$ &     8.93 &      9.23  &       9.23 & 9.23        &   8.93 \\  \hline
$A$            & 8.0527 & 10.8288 & 10.3120 & 10.3407  &   12.3718 \\  
$B$            & 2.9475 &   0.1049 & -0.6145 & -0.5974    &   17.6449 \\ 
$C$           & -1.0845 &  2.1818 &   1.7227 & 1.7496     &    -1.7778\\ 
$D$           & 1.3608  &   0.5483 & -0.1752 & -0.1654    &  0.8281 \\  
$E$           & 2.1376  &   0.4194 & -0.1032 & -0.0986    &  17.8471  \\ 
$F$           & -0.3984 &   0.9858 &  0.6505 & 0.6585     &   -0.6297 \\ 
$G$          & 0.2190  &    0.1893 & -0.0224 & -0.0211    &   0.0962 \\ 
$H$          & 0.5947  &    0.0918 &  0.0205 & 0.0217     & 0.2310 \\ 
$I$           & 0.8740   &  -0.1302 &  -0.0342 & -0.0337   &  6.9436 \\  
$J$          & -0.0427  &   0.1455 &    0.0733 & 0.0741    &   -0.0658  \\ 
RMS err (\%) & 1.86  &     2.95  &       0.35  & 0.37        &  2.11 \\  \hline \hline
R & O3N2$^b$ & S23$^b$   & $\frac{\rm [NII]}{\rm [SII]}^b$ & $\frac{\rm [SII]}{\rm H\alpha}^b$ & $\frac{\rm [NII]}{\rm H \alpha}^b$ \\ \hline
$Z_{min}$    & 8.23      & 7.63     & 7.63      & 7.63     & 7.63\\ 
$Z_{max}$   & 8.93      &  8.53.   & 8.53      &  8.53    &  8.53 \\ \hline
$A$              & 10.312  & 11.033 & 5.8892  & 23.370 & 10.526 \\
$B$              & -1.6575 & 0.9907 & 3.1688  & 11.700 & 1.9958 \\
$C$              & 2.2525  & 1.5789 & -3.5991 & 7.2562 & -0.6741 \\
$D$              & -1.3594 & 0.4233 & 1.6394  & 4.3320 & 0.2892 \\
$E$              & 0.4764  & -3.1663 & -2.3939 & 3.1564 & 0.5712 \\
$F$              & 1.1730  & 0.3666  & -1.6764 & 1.0361 & -0.6597 \\
$G$             & -0.2968 & 0.0654   & 0.4455 & 0.4315  & 0.0101 \\
$H$             & 0.1974   & -0.2146 & -0.9302 & 0.6576 & 0.0800 \\
$I$               & -0.0544 & -1.7045 & -0.0966 & 0.3319 & 0.0782 \\
$J$               & 0.1891 & 0.0316   & -0.2490 & 0.0336 & -0.0982 \\ \hline
RMS err (\%) & 2.97    & 0.55       & 1.19      & 0.92     & 0.67 \\
\hline
   \hline
    \end{tabular}%
    \end{center}
 \begin{tabnote}
 $^{\rm a}$ Valid over $Z_{min}\leq \log({\rm O/H})+12 \leq Z_{max}$ and $-3.98 \leq \log(U) \leq -1.98$.\\
 $^{\rm b}$ Wavelengths in \AA: [NIII]/[OIII]=$\frac{1747,48,49,52}{1660,66}$, C23/HeII=$\frac{1906,08 + 2325b^c}{1640}$,  C23/HeI=$\frac{1906,08 + 2325b^c}{3187}$,  C234/HeI=$\frac{1906,08 + 2325b^c + 1548,51}{3187}$, O3N2 $= \frac{\rm [OIII]/H\beta}{\rm [NII]/H\alpha} = \frac{5007/4861}{6584/6563}$, S23 = $\frac{\rm [SII]+[SIII]}{\rm H\alpha}=\frac{6717,31+9069+9531}{6563}$, \NIISII $= \frac{6584}{6717,31}$, \SIIHa $= \frac{6717,31}{6563}$,  \NIIHa $= \frac{6584}{6563}$.\\
 $^{\rm c}$ [\ion{C}{2}] blend (2325b) = [\ion{C}{2}]~$\lambda 2323$ + [\ion{C}{2}]~$\lambda 2324$ + [\ion{C}{2}]~$\lambda 2325$ + [\ion{C}{2}]~$\lambda 2326$ + [\ion{C}{2}]~$\lambda 2328$
 \end{tabnote}
\label{Metallicity_Coeffs}%
\end{table}%
\clearpage

\begin{table}
  \caption{Metallicity diagnostic calibrations for optical and IR line ratios described in Section 4.5 - 4.7 (Table 2 continued)}
  \begin{center}
    \begin{tabular}{lrrrrrrr}    \hline    \hline
    \multicolumn{8}{l}{Bi-Cubic Surface fit: $z = A + Bx + Cy + Dxy + Ex{^2} + Fy{^2} + Gxy{^2} + Hyx{^2} + Ix{^3} + Jy{^3}$ }  \\
    \multicolumn{8}{l}{where $x=log(R)$, $y=log(U)$, $z=log({\rm O/H})+12$, and $\log({\rm P/k})=5.0^{\rm a}$}  \\
\hline    
R  & $\frac{\rm [OII]}{\rm [SII]}^b$ & $\frac{\rm [OII]}{\rm H \beta}^b$  & $\frac{\rm [NII]}{\rm [OII]}^b$ & R23$^b$ \\
\hline
$Z_{min}$  & 8.23.     & 8.53             & 7.63     & 8.53  & \\ 
$Z_{max}$ & 9.23      & 9.23             & 9.23      & 9.23  & \\  \hline
$A$           & 12.4894   & 6.2084      & 9.4772  &  9.7757 & \\
$B$           & -3.2646  & -4.0513     & 1.1797  &  -0.5059 & \\
$C$           & 3.2581   & -1.4847     & 0.5085  &  0.9707& \\
$D$           & -2.0544  & -1.9125    & 0.6879  &   0.1744 & \\
$E$           & 0.5282   & -1.0071     & 0.2807  &   -0.0255 & \\
$F$           & 1.0730   & -0.1275     & 0.1612  &   0.3838 & \\
$G$          & -0.3445  & -0.2471     & 0.1187  &   -0.0378 & \\
$H$          & 0.2130    & -0.1872    & 0.1200  &   0.0806 & \\
$I$           & -0.3047   & -0.1052    & 0.2293  &   -0.0852 & \\
$J$           & 0.1209   & 0.0173     & 0.0164  &  0.0462 & \\
RMS err (\%) & 2.52  & 2.49         & 2.65       & 0.42 & \\
\hline
R  & $\frac{\rm [OIII] 88 \mu}{\rm [NIII] 57\mu}$ & $\frac{\rm [OIII] 55 \mu}{\rm [NIII] 57\mu}$ & $\frac{\rm [OIII] 55+ 88 \mu}{\rm [NIII] 57\mu}$ & $\frac{\rm Ne23}{\rm Pfund}^c$ & $\frac{\rm Ne23}{\rm Humph}^c$  \\ \hline
$Z_{min}$  & 7.63      & 7.63       & 7.63         & 7.63       & 7.63  \\ 
$Z_{max}$ & 9.23      & 9.23       & 9.23         & 8.93       & 8.93  \\ \hline
$A$            & 8.8463 & 8.8593   & 9.7206     & -2.3231  & -18.7919   \\
$B$            & -2.6206 & -2.6454 & -3.4595    & 23.2584  & 45.1168  \\
$C$            & -0.8838 & -0.6418 & -0.4186    & -0.2016   & -1.5507   \\
$D$            & -1.1524 & -1.2794 & -1.4070    & 1.9163    & 2.9812   \\
$E$            & 1.3261  & 1.1269  & 1.6997     & -16.559   & -24.953   \\
$F$            & -0.3990 & -0.3004 & -0.3098    & 0.3813    & 0.3725   \\
$G$           & -0.1721  & -0.2098 & -0.1853  & 0.0005     & 0.0071   \\
$H$            & 0.3436  & 0.3968  & 0.3625     & -0.7283   & -0.8470   \\
$I$              & -0.5433 & -0.5161 & -0.5318.  & 4.1178    & 4.7163   \\
$J$               & -0.0489 & -0.0360 & -0.0444  & 0.0395   & 0.0396  \\
RMS err (\%) & 1.70    & 1.69       & 1.69       & 1.57        & 1.81    \\
   \hline
       \hline
    \end{tabular}%
    \end{center}
 \begin{tabnote}
 $^{\rm a}$ Valid over $Z_{min}\leq \log({\rm O/H})+12 \leq Z_{max}$ and $-3.98 \leq \log(U) \leq -1.98$.\\
 $^{\rm b}$ Wavelengths in \AA: $\frac{\rm [OII]}{\rm [SII]} = \frac{3727,9}{6717,31}$, $\frac{\rm [OII]}{\rm H \beta} = \frac{3727,9}{4861}$, $\frac{\rm [OIII]}{\rm H \beta} = \frac{5007}{4861}$, $\frac{\rm [NII]}{\rm [OII]}=\frac{6584}{3727,9}$, R23$=\frac{4959+5007+3727,9}{4861}$\\
  $^{\rm c}$ Wavelengths in $\mu$: $\frac{\rm Ne23}{\rm Pfund}= \frac{23}{7.46}$, $\frac{\rm Ne23}{\rm Humph}=\frac{23}{12.37}$ 
 \end{tabnote}
\end{table}%


\bibliography{library}

\end{document}